\documentclass[12pt]{article}
\usepackage{bbm}
\usepackage{graphics}
\usepackage{color}
\usepackage{authblk}
\usepackage{latexsym}
\usepackage{epsfig,amssymb,euscript, mathrsfs,cite}
\usepackage{amsmath}
\usepackage{slashed}
\usepackage{soul}
\usepackage{cancel}
\usepackage{verbatim}
\usepackage{mathrsfs} 
\usepackage[normalem]{ulem}
\usepackage{enumerate}
\definecolor{MyDarkBlue}{rgb}{0.15,0.15,0.45}
\usepackage[linktocpage=true]{hyperref}
\hypersetup{
colorlinks=true,
citecolor=MyDarkBlue,
linkcolor=MyDarkBlue,
urlcolor=MyDarkBlue,
pdfauthor={},
pdftitle={},
pdfsubject={hep-th}
}
\newsavebox{\ns}
\newsavebox{\dbrane}
\newsavebox{\dbshort}

\def\be{\begin{equation}}
\def\ee{\end{equation}}
\def\bea{\begin{eqnarray}}
\def\eea{\end{eqnarray}}

\newcommand{\nn}{\nonumber\\}

\newcommand{\hook}{\mathbin{\rule[.2ex]{.4em}{.03em}\rule[.2ex]{.03em}{.9ex}}}

\newcommand\cF{\mathcal{F}}

\newcommand\cL{\mathcal{L}}

\newcommand\cV{\mathcal{V}}
\newcommand\cW{\mathcal{W}}

\newcommand{\Z}{\mathbb{Z}}

\newcommand{\vol}{\mathrm{vol}}

\newcommand{\rd}{{\rm{d}}}

\newlength{\sswidth}

\numberwithin{equation}{section}       % equation numbers in each section

\newcommand{\identity}{\mathbbm{1}}
\newcommand{\hcf}{\textrm{hcf}}

\newcommand{\nkk}{n_{\rm KK}}

\usepackage{color}

\begin{document}

\pagestyle{plain}
\setcounter{page}{1}
\newcounter{bean}
\baselineskip18pt

\begin{titlepage}

\vfill

\begin{flushright}
CCTP-2026-10\\
ITCP-2026-10
\end{flushright}

%\vfill

\begin{center}
   \baselineskip=16pt
   {\Large\bf 
Spindle solutions with hyperscalars\\ in $D=4$ gauged supergravity}
  \vskip 1cm
  %\vskip 1.5cm
Igal Arav$^1$,  Jerome P. Gauntlett$^2$, Jaeha Park$^2$ \\
Matthew M. Roberts$^{3}$ and Christopher Rosen$^4$\\
     \vskip 1cm     
                                                    \begin{small}
                                \textit{$^1$Instituut voor Theoretische Fysica, KU Leuven,\\
                                Celestijnenlaan 200D, B-3001 Leuven, Belgium}
        \end{small}\\
        \begin{small}\vskip .3cm
      \textit{$^2$Abdus Salam Centre for Theoretical Physics, 
  Imperial College\\ Prince Consort Rd., London, SW7 2AZ, U.K.}
        \end{small}\\
                \begin{small}\vskip .3cm
      \textit{$^3$    Science Institute, University of Iceland\\
    Dunhaga 3, 107 Reykjav\'ik, Iceland }
        \end{small}\\
              \begin{small}\vskip .3cm
           \textit{$^4$Crete Center for Theoretical Physics, Department of Physics, University of Crete,\\
71003 Heraklion, Greece}
        \end{small}\\
                       \end{center}
\vfill

\begin{center}
\textbf{Abstract}
\end{center}

\begin{quote}
We construct new classes of supersymmetric $AdS_2\times \Sigma$ solutions, 
where $\Sigma=\Sigma(n_N,n_S)$ is a spindle. 
Such solutions can arise as the near horizon limit of supersymmetric, accelerating black holes.
The solutions are constructed using $D=4$ STU $U(1)^4$ gauged supergravity theory coupled to a charged hyperscalar, and can be uplifted to obtain
smooth, supersymmetric $AdS_2\times Y_9$ solutions of $D=11$ supergravity. 
We allow $(n_N,n_S)$ to be non-coprime integers, 
including orbifolds of the round $S^2$. We also allow the hyperscalar to vanish at the poles.
The $AdS_2$ solutions with non-vanishing hyperscalar can naturally arise as the endpoint of 
holographic RG flows,
triggered by relevant hyperscalar deformations of the $AdS_2$ solutions of the STU model.
\end{quote}

\vfill

\end{titlepage}

\tableofcontents

\newpage

\section{Introduction}

Rich classes of examples of the AdS/CFT correspondence can arise from branes wrapping
manifolds while preserving supersymmetry. This was first pointed out in the pioneering paper
of \cite{Maldacena:2000mw}, which initiated a broad and ongoing research programme exploring such constructions.
More recently, it has been appreciated that one can also consider  
branes wrapping orbifolds. In particular, starting with \cite{Ferrero:2020laf},
there is a wide variety of supersymmetric solutions involving two-dimensional orbifolds known
as spindles \cite{Ferrero:2020twa,Hosseini:2021fge,Boido:2021szx,Ferrero:2021wvk,Cassani:2021dwa,Ferrero:2021ovq,Couzens:2021rlk,Faedo:2021nub,Ferrero:2021etw,Couzens:2021cpk,Giri:2021xta,Cheung:2022ilc,Suh:2022olh,Arav:2022lzo,Couzens:2022yiv,Couzens:2022aki,Couzens:2022lvg,Faedo:2022rqx,Boido:2022mbe,Suh:2022pkg,Suh:2023xse,Amariti:2023mpg,Hristov:2023rel,BenettiGenolini:2023yfe,Amariti:2023gcx,Inglese:2023tyc,Faedo:2024upq,Macpherson:2024frt,Boisvert:2024jrl,Ferrero:2024vmz,Hristov:2024qiy,Bomans:2024mrf,Suh:2024twt,Suh:2024fru,Crisafio:2024fyc,Kim:2025ziz,Arav:2025jee,Conti:2025rfd}. 

For supersymmetric $AdS\times \Sigma$ solutions of gauged supergravity, where $\Sigma$ is a spindle, it has
been shown that supersymmetry can be preserved 
via a \emph{twist}, generalising the standard topological twist when branes wrap Riemann surfaces, as well as the novel
 possibility of an \emph{anti-twist} \cite{Ferrero:2021etw}.
 Another notable feature is that after uplifting these solutions to $D=10$ or $D=11$ supergravity, one can
 obtain solutions that are completely free of any orbifold singularities. 
 In the case of $AdS_2\times \Sigma$ solutions of $D=4$ gauged supergravity, which is the focus of this paper, such solutions also naturally arise as the near horizon limit of accelerating black holes \cite{Ferrero:2020twa,Cassani:2021dwa,Ferrero:2021ovq}. 

We extend previous investigations on $AdS_2\times \Sigma$ solutions in several ways.
Earlier constructions considered \emph{coprime} spindles which have orbifold singularities $\mathbb{R}/\mathbb{Z}_{n_N}$
and $\mathbb{R}/\mathbb{Z}_{n_S}$ at the two poles of the spindle, with $n_N,n_S$ coprime integers.
Here we also consider non-coprime spindles, focussing on constructions that uplift on $S^7$ to give smooth
and supersymmetric solutions of $D=11$ supergravity. 
As in \cite{Arav:2025jee}, for non-coprime spindles there can be several inequivalent uplifted solutions
for given $n_{N,S}$ and specified magnetic fluxes through the spindle.
We construct the new solutions within the $D=4$ STU gauged supergravity model
coupled to a complex hyperscalar field which is charged with respect to a gauge field $A_B$ \cite{Bobev:2018uxk}. 
In previous work, solutions with the hyperscalar non-vanishing at both poles were analysed \cite{Suh:2022pkg,BenettiGenolini:2024kyy}, which necessarily have vanishing 
magnetic flux of $F_B=\rd A_B$ through the spindle. Here we investigate the possibility of solutions in which the
hyperscalar can vanish at the poles of the spindle, which allows for non-vanishing $F_B$ flux, and again focus on cases that uplift to give smooth and supersymmetric solutions of $D=11$ supergravity. 
The work is complementary to \cite{Arav:2025jee}, where similar investigations were carried out for $AdS_3\times \Sigma$ solutions
of $D=5$ gauged supergravity. We obtain analogous results to those in \cite{Arav:2025jee}, but there are some differences in the details.

As in \cite{Arav:2025jee}, by analysing the BPS equations we are able to derive algebraic conditions that determine
the values of the metric functions and the neutral scalars at the poles of the spindle, as well as an expression for $S_{BH}$,
which would be the Bekenstein-Hawking entropy of the black hole with the $AdS_2\times\Sigma$ solution as the near horizon limit, provided the solution exists. In the case of
solutions to the STU model, i.e. solutions with vanishing hyperscalar, we managed to find closed form expressions for these values, 
extending the results of \cite{Couzens:2021cpk} who directly analysed the known analytic solutions \cite{Ferrero:2021etw}. Similar algebraic conditions
also arise when the hyperscalar is non-vanishing; now, however, solving these constraints subject to some
obvious necessary conditions is not sufficient for the existence of a hyperscalar solution. To actually determine whether or not 
the hyperscalar spindle solution exists one needs to solve the BPS equations numerically, which we do in this paper.

In practice, we only find hyperscalar
solutions in the anti-twist class and moreover, the hyperscalar can only vanish at one of the poles (and necessarily
with an even power of a locally defined radial coordinate).
In addition, we find substantial evidence for the claim that
a necessary and sufficient condition for the existence of these solutions is that
there is an STU solution, with the same magnetic fluxes through the spindle, that admits a relevant hyperscalar mode.
This then gives rise to what we call the \emph{RG scenario}: the hyperscalar solutions precisely exist when there can be a holographic RG flow
from an STU solution to the hyperscalar solution, driven by a relevant deformation. It would be interesting to construct such RG flow solutions, but as this requires solving partial differential equations, we leave this for future work.
In all cases, we find that $S_{BH}$ for the STU solution is always greater than  $S_{BH}$ for the hyperscalar solution, in the large $N$ limit that we use throughout this paper.
Thus, if such RG flows do exist then  $S_{BH}$ would necessarily decrease along the RG flow. This is noteworthy, as we are 
unaware of a general monotonicity argument\footnote{In the $AdS_3\times \Sigma$ solutions of \cite{Arav:2025jee}
there was a similar RG scenario and there it was the central charge, $c$, of the dual SCFT that was decreasing, 
consistent with known monotonicity of $c$ under RG flow.}
for supersymmetric flows from one $AdS_2$ solution to another one that, \emph{a priori}, implies that this should happen.

A complementary analysis of $AdS_2\times \Sigma$ solutions of the STU model coupled to the hyperscalar was made 
in \cite{BenettiGenolini:2024kyy} using equivariant localization in supergravity \cite{BenettiGenolini:2023kxp}. There the case of coprime spindles and the hyperscalar non-vanishing at both poles was considered. In particular, it was shown how
to obtain the values of the neutral scalars and the metric functions at the poles of the spindle 
by extremising an off-shell entropy function. 
Here we show how this approach can be simply generalised to allow for solutions where
the hyperscalar vanishes at the poles. In addition we clarify how one can obtain an on-shell result for
the entropy by extremizing the off-shell entropy function or, alternatively, by utilising 
some conserved quantities implied by the gauge equations of motion.

The plan of the rest of the paper is as follows. In section \ref{sec:gensetup} we introduce the $D=4$ supergravity model and the $AdS_2$ ansatz that we study, as well as 
summarising aspects of the BPS equations coming from \cite{Suh:2022pkg,Arav:2024wyg}. We also summarise some of the analysis of
\cite{Arav:2025jee}, regarding the smoothness conditions for $S^1$ bundles over spindles as well as the conditions required for
well-defined spinors, generalising \cite{Ferrero:2021etw,Arav:2022lzo}.
In section \ref{section:STU} we discuss solutions of the STU model: analytic STU solutions are known \cite{Ferrero:2021etw,Couzens:2021cpk}, but we focus on extracting results without having the explicit solution.
Section \ref{hyperfluc} investigates the spectrum of hyperscalar fluctuations about STU solutions, and determines when there
can be a relevant hyperscalar mode associated with the RG scenario. 
Section \ref{sec5} analyses the BPS equations for hyperscalar solutions and section \ref{sec:examplesofhssols} summarises many new representative examples. Section \ref{section:equivlocal} switches gears and makes the connection with equivariant localization. We conclude with some final comments in
section \ref{sec:finalcomments}.

The paper has six appendices. In appendix \ref{app:S2} we discuss the special case of $AdS_2\times S^2$ solutions, as in \cite{Benini:2015eyy}, as well as new constructions where we take a quotient of the $S^2$, $S^2/\mathbb{Z}_h$, with $h$ odd that have a regular and supersymmetric uplift. Appendices \ref{minimal}-\ref{3p1} 
present some further simplifications that are associated with
various subtruncations of the main supergravity model that we consider.
In appendix \ref{app:E} we prove a general result for STU solutions in the anti-twist class and finally
appendix \ref{plotssols} presents some numerical plots for some particular hyperscalar solutions.

\section{General setup}\label{sec:gensetup}

\subsection{The D=4 supergravity model}
Our analysis will be in the context of a $D=4$ gauged supergravity theory whose solutions can be uplifted on $S^7$ to obtain exact solutions of $D=11$ supergravity. 
It 
consists of the $U(1)^4$ STU model, which can be viewed as an $\mathcal{N}=2$ gauged supergravity coupled to three vector multiplets, 
coupled to an additional hypermultiplet \cite{Bobev:2018uxk}.
One of the two complex scalar fields in the hypermultiplet is consistently truncated to zero, and we restrict to solutions with $F^I \wedge F^J = 0$, such that the three complex scalars in the vector multiplets are real, in alignment with the 
model also studied in \cite{Suh:2022pkg,Arav:2024wyg,BenettiGenolini:2024kyy}, in a closely related context.
Our conventions agree with \cite{BenettiGenolini:2024kyy}. 

The bosonic part of the Lagrangian is given by\footnote{\label{footnote:conventions}To compare with \cite{Arav:2024wyg}, we have $g^{\textrm{there}} = \frac{1}{\sqrt{2}}$, $A^I_{\textrm{there}} = -\frac{1}{\sqrt{2}} A^I_{\textrm{here}}$, $\lambda_i^{\textrm{there}} = \frac{1}{2}\varphi^i_{\textrm{here}}$, $\varphi_{\textrm{there}} = \frac{1}{2}\rho_{\textrm{here}}$, and $\mathcal{P}_{\textrm{there}} = \mathcal{V}_{\textrm{here}}$.}
\begin{align}
\label{eqn:4daction}
    \mathcal{L} = \frac{1}{16\pi G_4} \sqrt{-g} \Big[ R & - \frac{1}{2} \sum_{i=1}^3 (\partial\varphi^i)^2 - \frac{1}{4} \sum_{I=0}^3 (X^I)^{-2} F_{\mu\nu}^I F^{I\mu\nu} - \mathcal{V}\nn
    & - \frac{1}{2} (\partial \rho)^2 - \frac{1}{2} \sinh^2\rho (D\theta)^2\Big]\,.
\end{align}
The four $U(1)$ gauge fields are denoted as $A^I$, with field strengths $F^I = \rd{}A^I$. As in \cite{BenettiGenolini:2024kyy}, we choose to parametrise the $X^I$ using three real scalars $\varphi^i$ in the vector multiplet via
\begin{equation}
\label{eqn:sections}
    X^0=e^{\frac{1}{2}(\varphi^1+\varphi^2+\varphi^3)}\,,\quad X^1 = e^{\frac{1}{2}(\varphi^1-\varphi^2-\varphi^3)}\,,\quad X^2=e^{\frac{1}{2} (-\varphi^1+\varphi^2-\varphi^3)}\,,\quad X^3 = e^{\frac{1}{2}(-\varphi^1-\varphi^2+\varphi^3)}\,,
\end{equation}
which satisfy the constraint
\begin{equation}
\label{eqn:sectionsconstraint}
    X^0X^1X^2X^3 = 1\,.
\end{equation}
The scalar potential $\mathcal{V}$ is
\begin{equation}
    \mathcal{V} = 2 \left( \frac{\partial \mathcal{W}}{\partial\rho}\right)^2 + 2 \sum_{i=1}^3\left( \frac{\partial \mathcal{W}}{\partial\varphi^i}\right)^2 - \frac{3}{2}\mathcal{W}^2\,,
\end{equation}
where the real superpotential $\mathcal{W}$ is given by
\begin{equation}
\label{eqn:superpotential}
    \mathcal{W} = -\frac{1}{2} \sum_{I=0}^3 X^I + \zeta_I X^I \sinh^2 \frac{\rho}{2}\,,
\end{equation}
with the FI parameters given by $\zeta_I = \frac{1}{2} (1,-1,-1,-1)$.
The hyperscalar $\rho e^{i\theta}$ is charged with respect to the ``broken symmetry" gauge field 
\begin{align}
A_B \equiv A^0 - A^1 - A^2 - A^3\,,
\end{align} 
and in (\ref{eqn:4daction}) we defined the gauge invariant quantity
\begin{equation}
\label{Dthetadef}
    D\theta \equiv \rd\theta + \zeta_I A^I =\rd\theta + \frac{1}{2}A_B \,.
\end{equation}
We also define the ``R-symmetry" gauge field $A_R$ as well as two ``flavour-symmetry" gauge fields $A_{F_1},A_{F_2}$
via
\begin{align}
\label{eqn:flavoursymm}
A_R& \equiv A^0+A^1+A^2+A^3\,,\nn
    A_{F_1}& \equiv A^1-A^2\,,\quad A_{F_2} = A^2-A^3\,, 
\end{align}
such that $A_R,A_B,A_{F_1},A_{F_2}$ form an alternative basis for $U(1)^4$.

A bosonic solution preserves supersymmetry if there exists a Dirac spinor $\epsilon$ that solves
the Killing spinor equations:
\begin{align}
\label{eqn:4dKSEs}
    \Big[\nabla_\mu - \frac{i}{2}Q_\mu - \frac{1}{4}\mathcal{W}\,\Gamma_\mu + \frac{i}{16} \sum_{I=0}^3 (X^I)^{-1} F_{\nu\rho}^I \Gamma^{\nu\rho}\Gamma_\mu \Big]\epsilon = 0\,,\nn
    \Big[\Gamma^\mu \partial_\mu \varphi^i + 2\partial_{\varphi^i} \mathcal{W} + \frac{i}{2} \sum_{I=0}^3 \partial_{\varphi^i}(X^I)^{-1}F_{\mu\nu}^I \Gamma^{\mu\nu} \Big]\epsilon = 0\,,\nn
    \left[ \Gamma^\mu \partial_\mu \rho + 2\partial_\rho \mathcal{W} + 2i\partial_\rho Q_\mu \Gamma^\mu \right]\epsilon = 0\,,
\end{align}
where the one-form $Q$ is defined via
\begin{equation}
\label{eqn:Q}
    Q \equiv \frac{1}{2} A_R - \frac{1}{2} (\cosh \rho-1) D \theta\,.
\end{equation}
In particular, we see that the Killing spinor is charged with respect to the R-symmetry.

The model extends the STU model and so it admits a vacuum $AdS_4$ solution with unit radius, 
$R_{\textrm {ABJM}}^2\equiv 1$
which is dual to ABJM theory after uplifting on $S^7$. We have
\begin{equation}
\label{eqn:G4N32}
    \frac{1}{G_4} = \frac{2\sqrt{2}}{3} N^{3/2}\,,
\end{equation}
where $N$ is the rank of the gauge group of ABJM. The free energy of ABJM theory on $S^3$ is given by
\begin{equation}
    F_{S^3} = \frac{\pi}{2G_4} = \frac{\pi \sqrt{2}}{3} N^{3/2}\,.
\end{equation}

The model admits another supersymmetric $AdS_4$ solution \cite{Warner:1983vz}, with radius squared equal to $R_{\textrm
{mABJM}}^2\equiv \frac{4}{3\sqrt{3}}$, $e^{\frac{1}{2}\varphi^i} = 3^{1/4}$, $\tanh\frac{\rho}{2} = \frac{1}{\sqrt{3}}$, and vanishing gauge fields. After uplifting on $S^7$, the solution preserves $SU(3)\times U(1)_R$ global symmetry. It is dual to the $d=3$, $\mathcal{N}=2$ mABJM SCFT, which arises as the IR fixed point of the RG flow of a mass deformation of ABJM theory \cite{Ahn:2000aq,Corrado:2001nv,Benna:2008zy,Klebanov:2008vq}. The free energy of this SCFT
on $S^3$ is given by
\begin{equation}\label{freemABJM}
    F_{S^3} = \frac{\pi}{2G_4} R_{\textrm{mABJM}}^2 = \frac{4\sqrt{2}\pi}{9\sqrt{3}} N^{3/2}\,.
\end{equation}

Associated with the above supersymmetric vacua there are correspondingly two truncations of our supergravity model to
minimal gauged supergravity, as explained in appendix \ref{minimal}. Solutions of minimal gauged supergravity can also be uplifted on Sasaki-Einstein spaces, not just $S^7$, to obtain solutions of $D=11$ supergravity \cite{Gauntlett:2007ma}.

We also consider a $2+2$ truncation, as discussed in appendix \ref{2p2}. This keeps
pairwise equal gauge fields, $A^0 = A^1$, $A ^2 = A^3$, a single neutral scalar, via $X^0=X^1$ and $X^2=X^3$,
and the hyperscalar. 
In addition we consider a $3+1$ truncation in appendix \ref{3p1}. This sets three gauge fields equal
$A^1 = A^2 = A^3$, keeps a single neutral scalar via $X^1=X^2=X^3$, as well as the hyperscalar. Note that
the $3+1$ truncation has both the ABJM and the $SU(3)\times U(1)_R$ invariant $AdS_4$ vacua.
We also note that the STU solutions of the 3+1 truncation with vanishing hyperscalar preserve $SU(3)\times U(1)^2$
symmetry.

The model also has non-supersymmetric $SU(4)^-$ vacua with
$\varphi^i=0$, $e^{\frac{\rho}{2}}=\sqrt{2}\pm 1$ 
and $R^2_{SU(4)^-}=\frac{3}{4}$.
However, they are unstable
\cite{Pope:1984bd,Bobev:2010ib} and so they won't be a focus in the sequel.

\subsection{The $AdS_2$ ansatz and BPS equations}
\label{subsection:BPSanalysis}
We consider the following ansatz
\begin{align}\label{eq:ansatz}
    {\rd} s_4^2 &= e^{2V} {\rd} s^2(AdS_2) + f^2 {\rd} y^2 + h^2 {\rd} z^2\,,\nn
     A^I &= a^I {\rd} z\,,
\end{align}
where ${\rd} s^2(AdS_2)$ has unit radius, $V,f,h,a^I$ are all functions of $y$, and $\Delta z=2\pi$.
We are interested in solutions in which $y,z$ parametrise a two-dimensional spindle denoted by $\Sigma$.
We assume that the three vector multiplet real scalars $\varphi^i$ are functions of $y$, and the complex hyperscalar is of the form $\rho(y)e^{i\bar{\theta}z}$ with constant $\bar{\theta}$. We utilize Poincar\'e coordinates on $AdS_2$ as well as the following $D=4$ orthonormal frame
\begin{equation}
\label{eqn:frame}
    e^0=e^V \frac{{\rd}t}{u}\,,\quad e^1 = e^V \frac{{\rd}u}{u}\,,\quad e^2=f \, {\rd}y\,,\quad e^3 = h \,{\rd}z\,,
\end{equation}
where $f,h>0$.
We also impose an ansatz on the Killing spinor $\epsilon$
\begin{equation}
    \epsilon = \vartheta \otimes \upsilon \,,
\end{equation}
where $\upsilon$ is a spinor on $\Sigma$, and $\vartheta$ is a Killing spinor on $AdS_2$.
Correspondingly, we decompose $\textrm{Cliff}(1,3)$ as
\begin{equation}
\label{eqn:Clifford}
    \Gamma_i = \beta_i \otimes \gamma_3\,,\quad \Gamma_{a+1} = \identity \otimes \gamma_a\,,
\end{equation}
where $\beta_i$, with $i=0,1$, generate $\textrm{Cliff}(1,1)$, and $\gamma_a$, with $a=1,2$, generate $\textrm{Cliff}(2)$. We define $\beta_3 \equiv -\beta_{01}$ and $\gamma_3 \equiv -i \gamma_{12}$. For concreteness, we 
take $\gamma_i$ be the Pauli matrices $\sigma_i$, to write explicit expressions for $\upsilon$ (see \eqref{eqn:spindlespinor}).

The $AdS_2$ Killing spinor $\vartheta$ satisfies
\begin{equation}
\label{eqn:AdS2spinor}
    D_i \vartheta = \frac{\kappa}{2} \beta_i \vartheta\,,
\end{equation}
with $\kappa=\pm1$. Concretely, we have two Poincar\'e Killing spinors given by
\begin{equation}
\label{eqn:AdS2poincaresusy}
    \vartheta_Q = u^{-1/2}\vartheta_0\,,\qquad \beta_1 \vartheta_0 = -\kappa\, \vartheta_0\,,
\end{equation}
where $\vartheta_0$ is a constant spinor.
Note there are also two superconformal Killing spinors given by
\begin{equation}
    \vartheta_S = u^{-1/2}(\beta_0t + \beta_1 u)\vartheta_0\,,\qquad \beta_1\vartheta_0 = +\kappa \,\vartheta_0\,.
\end{equation}
Substituting the above into the Killing spinor equations (\ref{eqn:4dKSEs}), one obtains a set of $D=2$ Killing spinor equations for $\upsilon$. These were derived and analysed in \cite{Suh:2022pkg}, and subsequently used in \cite{Arav:2024wyg}. We refer readers to these references for a detailed derivation of the relations below (recalling our footnote \ref{footnote:conventions}).

Analyzing in the frame (\ref{eqn:frame}), $\upsilon$ takes the form
\begin{equation}
\label{eqn:spindlespinor}
    \upsilon = e^{V/2}e^{i \bar{s}z/2}\begin{pmatrix}
        \sin \frac{\xi}{2}\\
        \cos \frac{\xi}{2}
    \end{pmatrix} \equiv e^{V/2} \zeta \,,
\end{equation}
where the constant $\bar{s}$ is the gauge-dependent charge of the spinor under $\partial_z$. Using the spinor $\zeta$, we can form the following real bilinears \cite{BenettiGenolini:2024kyy}: 
\begin{align}
\label{eqn:bilinears_v0}
    S &= \zeta^\dagger \zeta = 1\,,\nn
     P &= \zeta^\dagger \gamma_3 \zeta = - \cos \xi \,,\nn
      \xi^\mu \partial_\mu &= - i \zeta^\dagger \gamma^\mu\gamma_3\zeta \partial_\mu = {h^{-1}e^V \sin \xi} \partial_z \,.
\end{align}
From bilinear relations, one can show that $\xi^\mu \partial_\mu$ is a Killing vector \cite{BenettiGenolini:2024kyy}, which implies that $h^{-1} e^V \sin \xi$ is a constant. This can be seen explicitly from the BPS equations, as we explain below.

Note the angle $\xi$ in our conventions satisfies (cf. (B.7) in \cite{Suh:2022pkg})
\begin{equation}
    \mathcal{W}\sin\xi = 2f^{-1} V'\,,\qquad \mathcal{W} \cos \xi = -2\kappa e^{-V} + \frac{1}{2} \sum_{I=0}^3 (X^I)^{-1} F^I_{23}\,,
\end{equation}
where $F^I_{23}=f^{-1}h^{-1} (a^I)'$. We have
\begin{equation}
\label{eqn:proj}
    \left[ i \cos\xi \,\Gamma_{23} + \sin\xi \,\Gamma_2\right] \epsilon = \epsilon\,,\quad \left[ i \cos\xi + \sin\xi \,\Gamma_{3}\right] \epsilon = -\Gamma_{23} \, \epsilon\,.
\end{equation}
Recall that we decompose $\textrm{Cliff}(1,3)\cong\textrm{Cliff(1,1)}\otimes \textrm{Cliff(2)}$ via (\ref{eqn:Clifford}). For the Poincar\'e Killing spinors (\ref{eqn:AdS2poincaresusy}), we can write the projection condition on the $D=4$ spinors as
\begin{equation}
\label{eqn:Poinproj}
    \Gamma_1 \epsilon_Q = i \kappa \Gamma_{23} \epsilon_Q\,.
\end{equation}
Assuming $\sin\xi\neq0$, the complete BPS equations are given by
\begin{align}
\label{eqn:BPSeq}
    f^{-1}\xi' & =\mathcal{W}\cos\xi + \kappa e^{-V}\,,\nn
    f^{-1}V' & = \frac{1}{2}\mathcal{W}\sin\xi\,,\nn
    f^{-1}\varphi_i' & = -2 \partial_{\varphi_i} \mathcal{W}\, \sin\xi\,,\nn
    f^{-1}\rho' & = - \frac{2}{\sin\xi} \partial_\rho \mathcal{W}\,,\nn
    f^{-1}\frac{h'}{h}\sin\xi & = \kappa e^{-V}\cos \xi + \frac{1}{2}\mathcal{W}(1+\cos^2\xi)\,,
\end{align}
with the following constraints
\begin{align}
\label{eqn:BPSconstraints}
    (\bar{s}-Q_z) \sin\xi & = -\mathcal{W} \, h \cos\xi - \kappa h e^{-V}\,,\nn
    \partial_\rho \mathcal{W}\cos\xi & = \partial_\rho Q_z \sin\xi \, h^{-1}\,.
\end{align}
We also find that the field strengths can be expressed as
\begin{equation}\label{eq:fieldstrength}
    (X^I)^{-2}F^I_{23} = \frac{\kappa}{e^V X^I} - \cos\xi + 2\cos\xi \sinh^2\frac{\rho}{2} \zeta_I \,.
\end{equation}

An integral of the BPS equations is given by
\begin{equation}
\label{eqn:BPSint}
    h e^{-V} = k \sin \xi \,,
\end{equation}
where $k$ is a constant, 
as anticipated below \eqref{eqn:bilinears_v0}. 
This allows us to write $\xi^\mu \partial_\mu = \frac{1}{k} \partial_z$. The BPS equations can be used to express the field strengths in the form
\begin{equation}
\label{eqn:BPSFequiv}
    F^I_{yz} = (a^I)' = (\mathcal{I}^I)'\,,
\end{equation}
where we have defined
\begin{equation}
\label{eqn:mathcalI}
    \mathcal{I}^I \equiv k x^I\,,
\end{equation}
where the ``dressed scalars" $x^I$ are given by
\begin{equation}
\label{eqn:dressedscalar}
    x^I \equiv - \cos\xi \, e^V X^I \,.
\end{equation}
Notice that the constraint \eqref{eqn:sectionsconstraint} implies we can write
\begin{align}\label{efvcsxfth}
e^{4V}=(\cos\xi)^{-4} x^0 x^1 x^2 x^3\,.
\end{align}

Further information can be obtained from the gauge equations of motion and we find (no sum on $I$):
\begin{equation}
    \left( e^{2V} (X^I)^{-2} F^I_{23} \right)' = e^{2V} f h^{-1}\sinh^2\rho \, (D\theta)_z \zeta_I\,.
\end{equation}
For vanishing hyperscalar ($\rho=0$), all four of these equations can be integrated. 
Using \eqref{eq:fieldstrength}
it is straightforward to see that
\begin{equation}
\label{eqn:fluxconstraint4}
    \rho = 0 : \qquad \frac{1}{\kappa}\mathcal{E}^I =  - e^{2V}\cos\xi \left(\frac{1}{x^I} + \kappa \right)\,,
\end{equation}
where $\mathcal{E}^I$ are constants. For non-vanishing hyperscalar, only three linear combinations can be integrated. A natural choice is 
\begin{align}
\label{eqn:fluxconstraint3}
    \frac{1}{\kappa}\mathcal{E}_R & = -e^{2V}\cos\xi \left(\frac{3}{x^0} + \frac{1}{x^1} + \frac{1}{x^2} + \frac{1}{x^3} + 6 \kappa \right)\,,\nn
    \frac{1}{\kappa}\mathcal{E}_{F_1} & = -e^{2V} \cos\xi \left( \frac{1}{x^1} - \frac{1}{x^2} \right)\,,\nn
    \frac{1}{\kappa}\mathcal{E}_{F_2} & = -e^{2V} \cos\xi \left( \frac{1}{x^2} - \frac{1}{x^3} \right)\,,
\end{align}
where $\mathcal{E}_R = 3\mathcal{E}^0+\mathcal{E}^1+\mathcal{E}^2+\mathcal{E}^3$, $\mathcal{E}_{F_1} = \mathcal{E}^1-\mathcal{E}^2$ and $\mathcal{E}_{F_2} = \mathcal{E}^2-\mathcal{E}^3$ are constants.

\subsection{Boundary conditions and the Bekenstein-Hawking entropy of spindle black holes}
\label{subsection:BCs}
We now restrict to conformal gauge
\begin{equation}\label{confgauge}
    f=e^V\,.
\end{equation}
Using (\ref{eqn:BPSint}), the $D=4$ metric can be written as 
\begin{equation}
\label{eqn:AdS2ansatz}
    {\rd}s_4^2 = e^{2V} \left[ {\rd}s^2(AdS_2) + {\rd}y^2 + k^2 \sin^2\xi {\rd}z^2 \right] \,.
\end{equation}
We assume $y\in[y_N,y_S]$ with $y_S>y_N$, such that $y,z$ parametrise a compact spindle $\Sigma$. By modelling the neighbourhood of the two poles of $\Sigma$ as $\mathbb{R}^2/\mathbb{Z}_{n_{N}}$ and $\mathbb{R}^2/\mathbb{Z}_{n_{S}}$, we infer that
\begin{align}\label{eqn:sincosBC}
k\sin\xi &= \frac{1}{ n_{N}}(y-y_{N})+\dots\,,\qquad k\sin\xi = \frac{1}{ n_{S}}(y_S-y)+\dots\,,\nn
\cos\xi|_{N,S} &= (-1)^{t_{N,S}} \,, \qquad \text{$t_{N,S}=0,1$}\,.
\end{align} 
Here we have taken $k\sin\xi \geq 0$, without making any sign choices for $k$. This can be accomplished by taking
\begin{align}
\xi = t_N \pi + \frac{(-1)^{t_N}}{k n_N} (y-y_N) +\dots\,,\qquad
\xi = t_S \pi + \frac{(-1)^{t_S}}{k n_S} (y_S-y) +\dots\,,
\end{align}
at the two poles. Combining (\ref{eqn:BPSint}) with the first equation of (\ref{eqn:BPSeq}) and (\ref{eqn:BPSconstraints}), we deduce that
$ f^{-1}\xi'=k^{-1}e^{-V}(Q_z-\bar{s})$ and hence
\begin{align}\label{eqn:bdycond1}
(Q_z-\bar s)|_{N} = \frac{(-1)^{t_N}}{n_N}\,,\qquad
(Q_z-\bar s)|_{S} = \frac{(-1)^{t_S+1}}{n_S}\,,
\end{align}
near the poles. The first equation of (\ref{eqn:BPSconstraints}) then implies
\begin{equation}
\label{eqn:bdycond2}
   \frac{(-1)^{t_N}}{kn_N}  = \kappa + \left.(-1)^{t_N} e^V \mathcal{W} \right\vert_{N}\,,\qquad
     \frac{(-1)^{t_S+1}}{kn_S}  = \kappa + \left.(-1)^{t_S} e^V \mathcal{W} \right\vert_{S}
    \,.
\end{equation}

The values of $x^I$ at the poles determine the vector multiplet scalars via 
$X^I=-e^{-V}x^I/\cos\xi$, and from \eqref{efvcsxfth} the warp factor at the poles is given by $e^{4V}|_{N,S}=x^0x^1x^2x^3|_{N,S}$. Moreover, recalling the prepotential for the STU model is 
\begin{align}\label{prepotstu}
\mathcal{F}(X^I) = -2 i \sqrt{X^0X^1X^2X^3}\,,
\end{align} 
we have
\begin{equation}
   i \mathcal{F}(x^I) = 2  e^{2V}\cos^2\xi\,.
\end{equation}
Viewing the metric (\ref{eqn:AdS2ansatz}) as that arising from a near-horizon limit of accelerating black holes, we may obtain expressions for the Bekenstein-Hawking entropy and the magnetic charges. From $S_{\textrm{BH}}=\frac{A}{4G_4}$, we have
\begin{equation}\label{expressionforsbh}
    S_{\textrm{BH}} \equiv  \frac{\Delta z}{4G_4} \int_N^S f h \, {\rd}y = \frac{\pi\sqrt{2}}{3} N^{3/2} \frac{\Delta z}{2\pi} \int_N^S f h \, {\rd}y\,,
\end{equation}
where the second equality comes from the relation between the vacuum $AdS_4$ solution and ABJM theory (\ref{eqn:G4N32}). Recall that $\Delta z = 2\pi$. Using the BPS equations (\ref{eqn:BPSeq}), it is straightforward to show
\begin{equation}
    fh=-\kappa k (e^{2V}\cos\xi)'\,.
\end{equation}
It follows that we can write
\begin{equation}\label{BHentropy}
    S_{\textrm{BH}} = -(-1)^{t_S} \frac{\pi\sqrt{2}}{3} N^{3/2} \frac{k}{2}\kappa \left[ i \mathcal{F}(x_S^I) - \sigma i \mathcal{F}(x_N^I) \right]\,,
\end{equation}
where 
\begin{align}
\sigma \equiv (-1)^{t_N-t_S}\,,
\end{align}
with $\sigma=\pm 1$ corresponds to supersymmetry being preserved by a twist/anti-twist, respectively, which are the only possibilities \cite{Ferrero:2021etw}. Next, from (\ref{eqn:BPSFequiv}),  (\ref{eqn:mathcalI}), the fluxes of the four gauge fields are given by
\begin{equation}
\label{eqn:BPS_flux}
    \frac{p^I}{n_Nn_S} \equiv \frac{1}{4 \pi} \int_{\Sigma} F^I = \frac{k}{2}(x^I_S - x^I_N)\,,
\end{equation}
where, as explained in section \ref{sec:smooth}, the $p^I$ will be taken to be integers in order to have a smooth uplift on $S^7$.

The above results have direct analogues in the analysis of section \ref{section:equivlocal}, that uses equivariant localisation, generalising \cite{BenettiGenolini:2024kyy}. In that case one is led to an off-shell entropy function, which can be extremised to get the on-shell result. In the following sections, we instead demonstrate how the gauge equations of motion allows us to obtain the on-shell expression directly, without utilising an extremization principle. To do so, we exploit additional information specified by integrating the gauge equations of motion.
For vanishing hyperscalar, (\ref{eqn:fluxconstraint4}) provides four additional constraints on $x^I_{N,S}$
\begin{equation}
\label{eqn:STUcharges}
    \rho=0\,:\quad\frac{1}{\kappa}\mathcal{E}^I = \frac{1}{2} (-1)^{t_N+1} i \mathcal{F}(x^I_N) \left(\frac{1}{x_N^I} + \kappa \right) = \frac{1}{2}(-1)^{t_S+1} i \mathcal{F}(x^I_S) \left(\frac{1}{x_S^I} + \kappa \right)\,,
\end{equation}
coming from the conserved charges $\mathcal{E}^I$ having the same value at the two poles. For the associated STU model solutions the Bekenstein-Hawking entropy is then expressed in terms of $n_{N,S}$, $t_{N,S}$, and the fluxes $p^I$. Similarly, for non-vanishing hyperscalar, we have three additional constraints from (\ref{eqn:fluxconstraint3}). The Bekenstein-Hawking entropy of the corresponding solutions then also depends on the behaviour of the hyperscalar at the poles of the spindle.
This is not immediately obvious from \eqref{BHentropy}; however, we will see later that $x^I_{N,S}$ are constrained by the behaviour of
the hyperscalar at the poles (see \eqref{eqn:zetaxNS}
or \eqref{eqn:brokensymmconstraint}).

\subsection{Smooth orbibundles}
\label{sec:smooth}

We are interested in obtaining smooth solutions of $D=11$ supergravity after uplifting on $S^7$. The $D=11$ solutions will be of the form $AdS_2 \times Y_9$, where $Y_9$ is an $S^7$ bundle over $\Sigma$. This bundle can be constructed by considering smooth circle bundles over $\Sigma$, associated with the four $U(1)$'s in the $D=4$ gauged supergravity. The $S^7$ can then be embedded in the $\mathbb{C}^4$ fibre of the associated complex line bundles.

In this section we summarise the regularity conditions arising from gauge fields, Killing spinors, and the hyperscalar as sections of bundles over $\Sigma \cong \mathbb{WCP}^1_{[n_N,n_S]}$, considering cases where $(n_N, n_S)$ are both coprime and non-coprime. 
The essential analysis was systematically performed in \cite{Arav:2025jee}, generalizing \cite{Ferrero:2021etw,Arav:2022lzo}, and we refer the reader to those references for more details.

\subsubsection{Circle fibrations over spindles}\label{2pt4pt1sec}
We consider two patches for the $U(1)^4$ orbibundle, covering the north and south poles of $\Sigma$, with $\psi^I_{N,S}$ coordinates on the $S^1$ fibres in each patch with $\Delta \psi^I_{N,S} = 2\pi$. For each circle fibre, the covering space on each patch is $S^1 \hookrightarrow \mathbb{C}\times S^1/\mathbb{Z}_n \rightarrow \mathbb{C} / \mathbb{Z}_n$, with local metrics\footnote{Note the factor of $\frac{1}{2}$ here and in subsequent formulae is such that when the solutions are uplifted on $S^7$ we get a well-defined solution of $D=11$ supergravity \cite{Cvetic:1999xp}. This also applies to solutions with non-zero hyperscalar; the fluxes $p^I$ defined in \eqref{eq:defpI} should be quantized (c.f. \eqref{eqn:fluxmIcond}), such that the $S^7$ fibration over $\Sigma$ is smooth.}
\begin{align}
    {\rd} s^2_{\mathbb{C}\times S^1/\mathbb{Z}_{n_N}} &  = \left( {\rd} \psi^I_N + \frac{1}{2} A_N^I \right)^2 + {\rd} s^2_{\mathbb{C}/\mathbb{Z}_{n_N}} \, ,\nn
    {\rd} s^2_{\mathbb{C}\times S^1/\mathbb{Z}_{n_S}} &  = \left( {\rd} \psi^I_S + \frac{1}{2} A_S^I \right)^2 + {\rd} s^2_{\mathbb{C}/\mathbb{Z}_{n_S}} \, .
\end{align}
On the total space of the orbibundle, $({\rd} \psi^I + \frac{1}{2} A^I)$ are global one-forms, where we use a gauge such that the connection one-forms are given by
\begin{align}
\label{eq:circleconnection}
    \frac{1}{2} A^I_{N} = \frac{1}{2} A^I_{(0)N} + \frac{m^I_{N}}{n_{N}} {\rd} z \,,\qquad 
        \frac{1}{2} A^I_{S} = \frac{1}{2} A^I_{(0)S} + \frac{m^I_{S}}{n_{S}} {\rd} z \,.
\end{align}
Here, $A^I_{(0)}$ are arbitrary, regular one-forms on $\mathbb{C}$ (i.e. they vanish at the poles). 
As the coordinate $z$ is not defined at the origin of $\mathbb{C}$, the gauge fields in each patch are defined only on $\mathbb{C} \backslash \{ 0\}$.
The bundle data are specified by $m^I_N \in \mathbb{Z}_{n_N}$ and $m^I_S \in \mathbb{Z}_{n_S}$, that appear in the flat connection pieces.
For the total space of each circle orbibundle to be smooth, it is necessary and sufficient that
\begin{equation}
\label{eqn:smoothcond}
    \hcf(m^I_N,n_N) = 1 \, , \qquad \hcf(m^I_S,n_S) = 1 \,,
\end{equation}
accordingly in each patch. One then constructs an orbibundle over $\Sigma$ by gluing the two local models, noting that $A_N^I$ and $A_S^I$ may differ by a $U(1)$ gauge transformation $g: S^1_{\textrm{glue}} \rightarrow U(1)$, of the form $g^I = \exp (i \gamma^I z)$. On the overlap, we then have  
\begin{equation}
\label{eq:circlegdg}
    \frac{1}{2} A_N^I = \frac{1}{2} A_S^I + \gamma^I {\rd}z \, ,
\end{equation}
with $\gamma^I \in \mathbb{Z}$.
Recalling that
\begin{equation}
\label{eq:defpI}
    \frac{p^I}{n_N n_S} \equiv \frac{1}{4 \pi} \int_{\Sigma} F^I \,,
\end{equation}
Stokes' theorem implies that
\begin{equation}
\label{eqn:fluxmIcond}
    p^I = n_Nm_S^I - n_S m_N^I + \gamma^I n_N n_S \in \mathbb{Z} \, .
\end{equation}
Furthermore, for each $I$, we immediately have
\begin{equation}
\label{eqn:hcfcond}
    \hcf(p^I,n_N) = \hcf(p^I,n_S)=\hcf(n_N,n_S) \,.
\end{equation}

Note that in place of the condition \eqref{eqn:fluxmIcond}, with $m_N^I \in \Z_{n_N}$, $m_S^I \in \Z_{n_S}$ satisfying \eqref{eqn:smoothcond}, we can instead demand that $m_N^I \in \Z$ and $m_S^I \in \Z$ satisfying \eqref{eqn:smoothcond}, with
\begin{equation}\label{eqn:fluxmIZcond}
	p^I = n_N m_S^I - n_S m_N^I \,.
\end{equation}
We will use this form when studying explicit examples in the following sections, though we continue here with $m_N^I \in \Z_{n_N}$, $m_S^I \in \Z_{n_S}$.

For \emph{coprime spindles} with $\hcf(n_N,n_S)=1$, smoothness is equivalent to the condition that 
\begin{align}\label{pcoprconds}
\hcf( p^I,n_N)=\hcf(p^I,n_S)=1\,.
\end{align}
If one specifies $(n_N,n_S,p^I)$, then $m_{N,S}^I \in \mathbb{Z}_{n_{N,S}}$ satisfying \eqref{eqn:fluxmIcond} are uniquely fixed. Further, for each circle orbibundle the total space is a uniquely determined Lens space. The orbibundles are thus classified by Chern numbers $p^I$; constructing a direct sum of the line bundles $O(p^0,p^1,p^2,p^3) = O(p^0) \oplus O(p^1) \oplus O(p^2) \oplus O(p^3)$, we may then form the $S^7$ bundle over $\Sigma$ by using the same $U(1)^4$ transition functions. 

We next consider \emph{non-coprime spindles}, by writing $h \equiv \hcf(n_N,n_S) \neq 1$. Given \eqref{eqn:hcfcond}, we must have
\begin{equation}
\label{eq:hfoldreduc}
    (n_N,n_S,p^I) = h (\hat{n}_N,\hat{n}_S,\hat{p}^I) \,,
\end{equation}
with $(\hat{n}_N,\hat{n}_S,\hat{p}^I)$ specifying a smooth orbibundle for a \textit{coprime} spindle. The total space of the orbibundle over the non-coprime spindle is then referred to as the ``flux quotient" of that of the coprime spindle, since
\begin{equation}
    \frac{1}{4 \pi} \int_\Sigma F^I = \frac{p^I}{n_N n_S} = \frac{1}{h} \frac{\hat{p}^I}{\hat{n}_N \hat{n}_S} \, .
\end{equation}
Importantly, for given $(n_N,n_S,p^I)$, the discrete orbibundle data $m_N^I \in \mathbb{Z}_{n_N}$ and $m_S^I \in \mathbb{Z}_{n_S}$ satisfying \eqref{eqn:fluxmIcond} are no longer uniquely fixed. In other words, the orbibundles are specified not just by the Chern number $p^I$, but also by the data $m_N^I,m_S^I$. For each circle orbibundle the total space is still a Lens space, determined by $(n_N,m_N^I,n_S,m_S^I)$. The associated complex line bundles might be denoted $O((n_N,m_N);(n_S,m_S))$. By constructing a direct sum, we may form the $S^7$ bundle as above, with the smoothness condition \eqref{eqn:smoothcond}.

Note that not every $h$-fold flux quotient of a smooth orbibundle for a coprime spindle will be smooth. Given $(n_N,n_S,p^I) = h(\hat{n}_N,\hat{n}_S,\hat{p}^I)$, the case with $\hat{n}_N,\hat{n}_S$ and any one of $p^I$ odd and $h$ even is obstructed \cite{Arav:2025jee} and it is believed that this is the only case that is obstructed. 
To have well defined spinors, there are additional conditions that need to be
satisfied.

\subsubsection{Spinors over spindles}
In order to have good supersymmetric solutions, we need to have well-defined spinors. From
the KSE \eqref{eqn:4dKSEs} and \eqref{eqn:Q}, we see that the Killing spinor has charge $-1/2$ with respect to $\frac{1}{2}A_R=\frac{1}{2}(A^0+A^1+A^2+A^3)$. From the results of \cite{Ferrero:2021etw} we deduce that
supersymmetry on $\Sigma$ is preserved via the ``twist" or the ``anti-twist" \cite{Ferrero:2021etw}. The spinors are then necessarily chiral at the poles, which implies that the scalar bilinears defined in \eqref{eqn:bilinears_v0} satisfy $S = \pm P = \pm \cos \xi$. For the twist case, the spinors have the same chirality at the poles and so $(t_N,t_S) = (0,0)$ or $(1,1)$ in \eqref{eqn:sincosBC}, 
whereas for the anti-twist case we have opposite chirality at the two poles, i.e. $(t_N,t_S) = (0,1)$ or $(1,0)$.

There is a constraint on the R-symmetry flux $p_R \equiv p^0+p^1+p^2+p^3$, given by
\begin{equation}
\label{eq:Rsymmconst}
	\frac{p_R}{n_Nn_S} = \frac{(-1)^{t_N+1}}{n_N} + \frac{(-1)^{t_S+1}}{n_S} \,,
\end{equation}
which also follows from the BPS equations as we show in later sections.

At the poles, the one-form $Q=\frac{1}{2}A_R$: when $\rho=0$ this is obvious and when $\rho\ne 0$ we should demand
$D\theta=0$ at the poles for regularity of the hyperscalar, as discussed below. By examining the spinor in a north and south pole patch, regularity requires \cite{Arav:2025jee}
\begin{equation}
\label{eq:spinorregularity}
    \frac{m_N^R}{n_N} - \bar{s}_N = \frac{(-1)^{t_N}}{n_N} \,,\qquad  \frac{m_S^R}{n_S} - \bar{s}_S = - \frac{(-1)^{t_S}}{n_S} \,,
\end{equation}
where $m^R \equiv \sum_I m^I$ and $\bar{s}_{N,S}\in\mathbb{Z}$. 

In fact, for coprime spindles, given \eqref{eq:Rsymmconst} is satisfied, the condition \eqref{eq:spinorregularity} is then automatically satisfied.
However, for non-coprime spindles it is an extra condition that needs to be imposed to have well-defined spinors.

A general result is that smooth, supersymmetric solutions necessarily have\footnote{This is in contrast to the $AdS_3\times \Sigma$ solutions of \cite{Arav:2025jee}.} 
\begin{align}\label{thensareodd}
n_N, n_S\qquad \text{are both odd}\,.
\end{align}
This essentially follows from the fact that we have four gauge fields: if we assume that $n_N$ is even, then from \eqref{eqn:smoothcond} we have $m_N^I$ are all odd. Then
$m^R_N$ is even and hence we can't satisfy \eqref{eq:spinorregularity}
 to have well-defined spinors. 
The same argument applies at the south pole.
An immediate corollary is that for non-coprime spindles, the $h$-fold flux quotients
necessarily have 
\begin{align}\label{hodd}
\text{$h$ is odd}\,.
\end{align}

\subsubsection{Non-trivial hyperscalars}\label{hyperscalar}
We now analyse regularity of the hyperscalar following \cite{Arav:2025jee} (see also \cite{Arav:2022lzo,BenettiGenolini:2024kyy}).
The hyperscalar has charge $1/2$ with respect to $A_B= 2\zeta_I A^I=A^0 - A^1 - A^2 - A^3$, as we see from \eqref{Dthetadef}.
We demand that the hyperscalar is a well defined section of a line orbibundle over the spindle. As usual, one analyses this by
introducing north and south pole patches. In addition we also want to satisfy the BPS conditions which impose further
restrictions. In particular, we observe that from $ \partial_\rho \mathcal{W} = \frac{1}{2} \zeta_I X^I \sinh \rho$ and
(B.25) of \cite{BenettiGenolini:2024kyy}, we have 
$S \, {\rd} \rho = \left( e^V \zeta_IX^I \, \xi\hook \vol - P \ast D\theta \right) \sinh \rho$.
Recall that the spinor bilinear $S=1$. Now at the poles $  \vol = \frac{1}{n_{N}} (y-y_{N}) {\rd} y \wedge {\rd} z$
and $  \vol = \frac{1}{n_{S}} (y_S-y) {\rd} y \wedge {\rd} z$, so only the second term in the previous expression is relevant and
we deduce 
\begin{align}\label{extrahypbpscond}
\rho'|_N=(-1)^{t_N}\frac{n_ND\theta_z}{y-y_N}\sinh\rho|_N\,,\qquad
\rho'|_S=(-1)^{t_S}\frac{n_SD\theta_z}{y_S-y}\sinh\rho|_S\,.
\end{align}
Notice that if $\rho$ is constant at a pole then $D\theta_z$ vanishes at the pole.

We 
assume that near the poles the hyperscalar behaves like
\begin{align}
\rho=C_N(y-y_N)^{r_N}+\mathcal{O}(y-y_N)^{r_N+1}\,,\qquad
\rho=C_S(y_S -y)^{r_S}+\mathcal{O}(y_S-y)^{r_S+1}\,,
\end{align}
for constants $C_{N,S}>0$ and with 
\begin{align}
r_{N,S}\in\mathbb{Z}_{\ge 0}\,.
\end{align}
In particular if $r=0$ then the scalar is non-vanishing at the pole, otherwise it vanishes at the pole.
Then \eqref{extrahypbpscond} implies
\begin{equation}
\label{Dthetasigma}
	D\theta\rvert_N =(-1)^{t_N} \frac{r_N}{n_N} {d}z \,,\qquad D\theta\rvert_S = (-1)^{t_S+1} \frac{r_S}{n_S} {d}z \,.
\end{equation}

The orbifold identifications for the hyperscalar line bundle also imply
\begin{equation}
\label{eq:Dthetahyper}
    D\theta_z\rvert_N = \bar{\theta}_N + \frac{2\zeta_I m_N^I}{n_N} \,,\qquad D\theta_z\rvert_S = \bar{\theta}_S + \frac{2\zeta_I m_S^I}{n_S} \, ,
\end{equation}
in the gauge \eqref{eq:circleconnection}, so that
\begin{align}
\label{eq:rNrSconstfirst}
    (-1)^{t_N} r_N = n_N \bar{\theta}_N + 2 \zeta_I m_N^I\,,\qquad
    (-1)^{t_S+1}r_S = n_S \bar{\theta}_S + 2 \zeta_I m_S^I \,,
\end{align}
with $\bar{\theta}_{N,S}\in \mathbb{Z}$ and
\begin{equation}
\label{eq:thetabarNS}
	\bar{\theta}_N = \bar{\theta}_S - 2 \zeta_I \gamma^I \,,
\end{equation}
where $\gamma^I$ is as in \eqref{eq:circlegdg}.
The broken flux $p_B \equiv p^0 - p^1 - p^2 - p^3$ through $\Sigma$ is given by
\eqref{eqn:fluxmIcond}. 
Using \eqref{eq:thetabarNS}, we deduce that $p_B$ is constrained:
\begin{equation}
\label{eq:brokenflux_v2}
	\frac{p_B}{n_N n_S} = (-1)^{t_N+1} \frac{r_N}{n_N} + (-1)^{t_S+1} \frac{r_S}{n_S} \,.
\end{equation}
Note that given $p_B$ satisfies \eqref{eq:brokenflux_v2}, then for coprime spindles the condition
\eqref{eq:rNrSconstfirst} is automatically satisfied, but for non-coprime spindles it is an extra condition that needs to be satisfied
to obtain smooth, supersymmetric uplifted solutions. Specifically, different choices of the bundle data $m_N^I$, $m_S^I$,
when they exist, are associated with different values of $r_N$ and $r_S$.

\section{STU solutions} \label{section:STU}
Solutions to the STU model are obtained by setting the hyperscalar to zero, $\rho=0$. Both twist and anti-twist class solutions in the STU model were studied in \cite{Ferrero:2021etw,Couzens:2021cpk}. In these papers the explicit analytic solutions were analysed.
Here we show that by examining the BPS equations we can obtain the Bekenstein-Hawking entropy without the explicit solution. Furthermore, we are also able to give explicit expressions for the values of scalars at the two poles in terms of the spindle data 
$n_{N,S}$, $t_{N,S}$ and the fluxes $p^I$. In addition, we comment on new non-coprime spindle solutions.
We also discuss $AdS_2\times S^2$ solutions which can be obtained, with care, as limits of twist solutions
by setting $n_N=n_S=1$, and we also consider quotients of the $S^2$.

\subsection{Analytically solving for boundary conditions}
We first note that 
\begin{equation}
    \mathcal{W}_{\textrm{STU}} = \frac{1}{2e^{V} \cos\xi} \, \sum_{I=0}^3 x^I\,,
\end{equation}
with $\rho=0$. From (\ref{eqn:bdycond2}), we then find
\begin{equation}
\label{eqn:xIconstraint}
    \sum_I x_N^I = -2\kappa + (-1)^{t_N} \frac{2}{kn_N}\,,\qquad \sum_I x_S^I = -2\kappa - (-1)^{t_S} \frac{2}{kn_S}\,.
\end{equation}
Together with (\ref{eqn:BPS_flux}), we confirm that the R-symmetry flux can be written as (c.f. \eqref{eq:Rsymmconst}):
\begin{equation}
\label{eqn:BPS_Rsymmconstraint}
    \frac{p_R}{n_N n_S} = \frac{(-1)^{t_N+1}}{n_N} + \frac{(-1)^{t_S+1}}{n_S}\,.
\end{equation}

We first consider coprime spindles. For given $\kappa \in \{\pm1 \}$, the spindle data for STU solutions can be taken to be $n_{N,S}$, $t_{N,S}$, and the fluxes $p^I$, constrained by \eqref{eqn:BPS_Rsymmconstraint}. 
In order that the uplift on $S^7$ is smooth, the fluxes are all coprime to both $n_N$ and $n_S$ as in \eqref{pcoprconds}.
In the coprime case, the discrete orbibundle data $m_{N,S}^I \in \mathbb{Z}_{n_{N,S}}$, coprime to $n_N,n_S$, are uniquely specified by
\eqref{eqn:fluxmIcond} and automatically satisfy \eqref{eq:spinorregularity}, so the uplifted solutions are smooth and
with well-defined spinors.

We then have an algebraic system of 9 variables: $x^I_N$, $x^I_S$, and $k$. The $x^I_S$ can be eliminated using \eqref{eqn:BPS_flux}, while \eqref{eqn:xIconstraint} provides an additional constraint on $x^I_N$ and $k$, leaving four undetermined parameters. By utilising the expression for the conserved charges \eqref{eqn:STUcharges}, we have four additional constraints, which allows us to solve for $x^I_N$, $x^I_S$, and $k$, and write the Bekenstein-Hawking entropy \eqref{BHentropy} without solving the BPS equations, just assuming the solutions exist. 
Additional necessary conditions for an STU solution to exist are given by
\begin{equation}\label{STUexistcond}
    S_{\textrm{BH}} > 0\,,\quad \left.(-1)^{t+1} x^I\right\vert_{N,S} > 0\,.
\end{equation}

For any given specific values of the spindle data, one can solve this algebraic system\footnote{An alternative approach is 
the equivariant localization computation of \cite{BenettiGenolini:2024kyy}, where an extremisation principle is utilised instead, as we discuss in section \ref{section:equivlocal}.}
explicitly and moreover we have found that in all cases
we have checked
the necessary conditions \eqref{STUexistcond} are in fact sufficient for the existence of an analytic STU solution as in \cite{Ferrero:2021etw,Couzens:2021cpk}.
By trial and error we have also obtained a solution for $x_N^I, x_S^I$ and $k$ that is expressed in terms of general spindle data; 
this agrees with all the explicit solutions to the constraint equations that we have found, for particular values of the spindle data, for both
the twist and the anti-twist class. 
Moreover, for the anti-twist class we can prove that the expressions satisfy all of
the relevant algebraic conditions and inequalities, assuming that $\kappa p^I>0$ (see section \ref{subsecstusols} and \eqref{anti-twist_bountiful}),
as we explain below and in appendix \ref{app:E}.

Following \cite{Couzens:2021cpk}, it is useful to define the following symmetric polynomials in the fluxes
\begin{equation}\label{couzens_Psymm}
    \hat{P}^{(2)} \equiv \sum_{0\leq I<J\leq3} p^I p^J \,, \qquad \hat{P}^{(3)} = \sum_{I=0}^3 \prod_{J \neq I} p^J \,,\qquad \hat{P}^{(4)} \equiv \prod_{I=0}^3 p^I \,,
\end{equation}
as well as
\begin{equation}\label{STU_defn_s}
	s \equiv \sqrt{(\hat{P}^{(2)} - \sigma n_N n_S)^2 - 4 \hat{P}^{(4)}} \,,
\end{equation}
where we recall $\sigma=(-1)^{t_N-t_S}=\pm1$, for twist and anti-twist, respectively.
We then find the following results for the dressed scalars at the poles:
\begin{align}\label{STU_xI}
	x_N^I &= -\frac{\kappa}{2} - \frac{\kappa \sigma}{2s}
	\Bigg[ \hat{P}^{(2)} - \sigma n_N n_S +2 ( (-1)^{t_S} n_N + p^I ) p^I              \Bigg] \,, \nn
 \nn
 	x_S^I &= -\frac{\kappa}{2} - \frac{\kappa \sigma}{2s}
	\Bigg[   \hat{P}^{(2)} - \sigma n_N n_S +2 ( (-1)^{t_N} n_S + p^I ) p^I        \Bigg] \,,
\end{align}
where the fluxes satisfy the constraint \eqref{eqn:BPS_Rsymmconstraint}, and
\begin{align}\label{STU_k}
k=\frac{2 \kappa s}{n_N n_S((-1)^{t_N}n_N-(-1)^{t_S} n_S)} \,.
\end{align} One can analytically check
that these expressions satisfy \eqref{eqn:xIconstraint} and \eqref{eqn:BPS_flux}.
It is straightforward to analytically check that the square of the conserved charge conditions
\eqref{eqn:STUcharges} are satisfied i.e.
\begin{equation}
\label{eqn:STUcharges2}
x_N^0x_N^1x_N^2 x_N^3\left(\frac{1}{x_N^I} + \kappa \right)^2 = 
x_S^0x_S^1x_S^2 x_S^3\left(\frac{1}{x_S^I} + \kappa \right)^2\,.
\end{equation}
It is more challenging to verify that the conserved charge conditions \eqref{eqn:STUcharges} themselves are satisfied in general, 
but we have managed to do this for the anti-twist class (assuming $\kappa p^I>0$; see appendix \ref{app:E}).

A compact, analytic expression for the Bekenstein-Hawking entropy of both twist and anti-twist solutions was given in \cite{Couzens:2021cpk},
\begin{equation}
\label{eq:Couzens_BH}
    S_{\textrm{BH}} = \frac{2\pi}{3n_N n_S} N^{3/2} \sqrt{\hat{P}^{(2)}- \sigma n_N n_S  + \sigma s} \,.
\end{equation}
The results of \cite{BenettiGenolini:2024kyy} imply that we should reproduce \eqref{eq:Couzens_BH} from \eqref{BHentropy} \emph{provided
that we have a solution that indeed satisfies \eqref{STUexistcond}}.\footnote{In appendix \ref{2p2} for the simpler $2+2$ truncation, we make some clarifying remarks regarding the validity of the general solutions above.} 
In \cite{BenettiGenolini:2024kyy} this was shown to be the case for anti-twist solutions in minimal gauged supergravity.
Using  \eqref{STU_xI} we have checked for specific solutions that the statement indeed holds for both twist and anti-twist solutions with unequal fluxes. Furthermore, for the anti-twist class we have proven it in general (assuming $\kappa p^I>0$; see appendix \ref{app:E}).

To summarise, we believe that \eqref{STU_xI}, \eqref{STU_k} and \eqref{eq:Couzens_BH} are true in general for both twist
and anti-twist STU \emph{bona fide} solutions, which must satisfy \eqref{eqn:xIconstraint}, \eqref{eqn:BPS_Rsymmconstraint} and \eqref{STUexistcond}. 
For the anti-twist class we do have a proof of this assuming $\kappa p^I>0$, which is in fact believed\footnote{This can be proven for pairwise equal fluxes; see appendix \ref{2p2}.} to hold
for all STU solutions in the anti-twist class (see section \ref{subsecstusols}).

For non-coprime spindles, the spindle data is given by $n_{N,S}$, $t_{N,S}$, fluxes $p^I$ as in
\eqref{eq:hfoldreduc}, with $p_R$ constrained by \eqref{eqn:BPS_Rsymmconstraint},
and we now also need to
specify the discrete orbibundle data $m_{N,S}^I \in \mathbb{Z}_{n_{N,S}}$. These $m_{N,S}^I$ should be coprime to $n_{N,S}$, respectively, and also satisfy both
\eqref{eqn:fluxmIcond} and 
\eqref{eq:spinorregularity}. In general, for given $n_{N,S}$, $t_{N,S}$, $p^I$ the 
$m_{N,S}^I \in \mathbb{Z}_{n_{N,S}}$ may no longer be unique and they will uplift
 to different solutions of $D=11$ supergravity. One way in which this is manifested is that the scaling dimension of operators dual to the hyperscalar will in general be different, as we discuss in later sections.
Solutions with the same $n_{N,S}$, $t_{N,S}$, and $p^I$, but different $m_{N,S}^I$ will have the same Bekenstein-Hawking entropy
(in the large $N$ limit). 
For a non-coprime spindle we have $(n_N,n_S,p^I) = h(\hat{n}_N,\hat{n}_S,\hat{p}^I)$, with $(\hat{n}_N,\hat{n}_S,\hat{p}^I)$ specifying a coprime spindle and $h\in \mathbb{Z}_{>0}$.
From the analytic expression, it is straightforward to see that
\begin{equation}\label{entropyfluxred}
    S_{\textrm{BH}}(n_N,n_S,p^I) = \frac{1}{h} S_{\textrm{BH}}(\hat{n}_N,\hat{n}_S,\hat{p}^I) \,.
\end{equation}

Generically, the uplifted STU solutions preserve $U(1)^4$ symmetry, but there are certain choices of fluxes with enhanced 
symmetry. For example, there are solutions within the $2+2$ truncation (see appendix \ref{2p2}), which have pairwise equal 
gauge fields which preserve 
$SU(2)_F \times SU(2)_F \times U(1)^2$ symmetry. Similarly, there are
STU solutions within the $3+1$ truncation (see appendix \ref{3p1}), which have $A^1=A^2=A^3$ 
which preserve $SU(3) \times U(1)^2$ symmetry.
It turns out that for the $2+2$ truncation there are only STU solutions in the anti-twist class and furthermore,
we have not found any hyperscalar solutions in this truncation and conjecture that they don't exist. On the other hand, for the $3+1$ truncation
 there are both
twist and anti-twist STU solutions as well as hyperscalar solutions, and so this truncation will be a focus in the sequel.

Returning to general STU solutions we can look\footnote{Similar comments, \emph{mutatis mutandis}, apply to solutions with $p^0=p^1$ and $p^2=p^3$.} 
for solutions with $p^1=p^2=p^3$. For coprime spindles,
$(n_N,n_S,p^I)$ uniquely specifies $m_{N,S}^I$ and correspondingly we have $m_N^1=m_N^2=m_N^3$ and 
$m_S^1 = m_S^2=m_S^3$ and so the solutions actually have $A^1=A^2=A^3$ and lie within the 3+1 truncation. However, for 
non-coprime solutions with $p^1 = p^2=p^3$, it is possible that $m_N^I, m_S^I$ will not satisfy this condition
and this corresponds to STU solutions that do not lie within the 3+1 truncation in which the $SU(3) \times U(1)^2$ symmetry (after uplift)
is broken in a subtle way, via the discrete orbibundle data $m_{N,S}^I$, similar to \cite{Arav:2025jee}; such a mechanism can only happen for non-coprime spindle solutions, and it would be interesting to investigate how this is manifested in the full operator spectrum of the dual SCFT.

\subsection{Solutions}\label{subsecstusols}
In the remainder of this section we discuss various examples of STU solutions\footnote{Additional STU examples will be presented in section \ref{hyperfluc} after discussing hyperscalar fluctuations.} both in the anti-twist class and the twist class.
For all the examples we have looked at, we find that the solutions for $x_{N,S}^I$ 
obtained by directly solving \eqref{eqn:STUcharges} and which 
also satisfy the necessary conditions \eqref{eqn:xIconstraint}, \eqref{eqn:BPS_Rsymmconstraint}
and \eqref{STUexistcond}, agree precisely with our general expression \eqref{STU_xI}.
Furthermore, we believe that whenever these necessary algebraic conditions for an STU solution to exist are satisfied, an analytic STU solution actually exists and we have checked this in some specific cases (this is in contrast to the hyperscalar solutions we discuss later). 

\subsubsection{Anti-twist STU solutions}\label{atwiststusolstext}
 Without loss of generality, we focus on the anti-twist solutions with\footnote{Solutions with $\kappa = +1$, with $(t_N, t_S) = (1,0)$, can be obtained by $n_N \leftrightarrow n_S$; $p^I \leftrightarrow - p^I$; $x_N^I \leftrightarrow - x_S^I$; $k \leftrightarrow - k$.} 
\begin{align}\label{atwistassumption}
\kappa = -1\,,\qquad  t_N=1\,, t_S=0\,.
\end{align}
To obtain solutions with the same $\kappa$ but with $(t_N,t_S) = (0,1)$, we can utilise 
the following symmetry: $t_N \leftrightarrow t_S$; $n_N\leftrightarrow n_S $; $x_N^I \leftrightarrow x_S^I$;
$k \leftrightarrow -k$.
In \cite{Couzens:2021cpk} the $(n_N,n_S)$ parameter space of STU solutions in the anti-twist class was argued to satisfy 
\begin{align}\label{anti-twist_bountiful}
	p^I < 0\  \ ( {\rm for \ all \ }I) \,,
\end{align}
which, in particular implies $p^1+p^2+p^3>n_S-n_N$ after using \eqref{eqn:BPS_Rsymmconstraint}.
A formal proof of this has not yet been obtained but strong numerical evidence was provided in \cite{Couzens:2021cpk} and our 
investigations offer further support (it can be proven for pairwise equal charges; see appendix \ref{2p2}).
We also recall from the previous section and appendix 
\ref{app:E}, that given \eqref{anti-twist_bountiful} we have proven \eqref{STU_xI}, \eqref{STU_k} and \eqref{eq:Couzens_BH} satisfy the algebraic constraints \eqref{eqn:xIconstraint}, \eqref{eqn:BPS_Rsymmconstraint} and \eqref{STUexistcond}.

We can illustrate some features of these anti-twist STU solutions with some examples; more will be presented later.
First consider a coprime case with $(n_N,n_S)=(11,3)$ and $p^I=-(4,1,1,2)$. We find:
\begin{align}\label{eqn:STU_ex_antitwist_11_3}
    k & %= 0.232480 
    = \frac{2}{33} \sqrt{\frac{103}{7}} \,,\nn
    x_N^I %& = (0.518621 \,,\, 0.183443 \,,\, 0.183443 \,,\, 0.332411 ) \nn
    & = \left( \frac{1}{2} + \frac{1}{2\sqrt{721}} \,,\, \frac{1}{2} - \frac{17}{2\sqrt{721}} \,,\, \frac{1}{2} - \frac{17}{2\sqrt{721}} \,,\, \frac{1}{2} - \frac{9}{2\sqrt{721}} \right) \,,\nn
    x_S^I 
    %& = ( -0.524154 \,,\, -0.0772502 \,,\,  -0.0772502 \,,\, -0.188976 ) \nn
    & = \left( \frac{1}{2} - \frac{55}{2\sqrt{721}} \,,\, \frac{1}{2} - \frac{31}{2\sqrt{721}} \,,\, \frac{1}{2} - \frac{31}{2\sqrt{721}} \,,\, \frac{1}{2} - \frac{37}{2\sqrt{721}} \right) \,,\nn
    S_{\textrm{BH}} & = \frac{8\pi}{99 \sqrt{27+\sqrt{721}}} N^{3/2}\sim  0.0345944 N^{3/2} \,.
\end{align}
By \eqref{eqn:fluxmIZcond} we have $m_N^I = (5,4,4,8)$, and $m_S^I = (1,1,1,2)$. It follows that $m_N^R = 21 = -1 \,\textrm{mod} \,n_N$, $m_S^R = 5 = -1 \,\textrm{mod} \,n_S$, from which we can verify that the regularity condition \eqref{eq:spinorregularity} is indeed satisfied.

We can also consider non-coprime solutions that are 
$h$-fold flux quotients 
of the form \eqref{eq:hfoldreduc} for $(\hat{n}_N,\hat{n}_S)=(11,3)$ and $\hat{p}^I=-(4,1,1,2)$.
As we saw in \eqref{hodd} we must have
$h$ odd to have smooth solutions with well-defined spinors and many solutions can be constructed.

Some additional anti-twist STU examples are presented in table \ref{table:STU_antitwist}. Note that in the last column we have indicated with $\rm{Y}$ or $\rm{N}$
whether or not, respectively, there are relevant deformations of the hyperscalar about the STU solution (to be discussed further
in the next section). 
\begin{table}[h]
\centering
\scalebox{0.85}{
\begin{tabular}{ |c|c|c|c| c|}
\hline
$(n_N,n_S)$ & $p^I$ & $\frac{1}{N^{3/2}}S_{\textrm{BH}}$ & $k$&$\delta>0$\\
\hline
$(5,1)$ & $-(1,1,1,1)$ & $\frac{4\sqrt{2}\pi}{15(\sqrt{13}+3)}$ & $\frac{\sqrt{13}}{5}$  &N\\ 
%$(9,2)$ & $-(2,3,1,1)$ & $0.068299$ & $0.350055$  \\ 
$(11,3)$ & $-(4,1,1,2)$ & $\frac{8\pi}{99\sqrt{27+\sqrt{721}}}$ & $\frac{2\sqrt{103}}{33\sqrt{7}}$&Y \\
$(11,3)$ & $-(5,1,1,1)$ & $\frac{2\pi}{99\sqrt{51-\sqrt{2581}}}$ & $\frac{\sqrt{2581}}{231}$&N\\
$(13,5)$ & $-(3,3,1,1)$ & $\frac{4\pi}{65\sqrt{87+9\sqrt{93}}}$ & $\frac{\sqrt{93}}{65}$& N\\
%
%$(5,1)$ & $-(1,1,1,1)$ & $0.17936$ & $0.721110$  \\ 
%%$(9,2)$ & $-(2,3,1,1)$ & $0.068299$ & $0.350055$  \\ 
%$(11,3)$ & $-(4,1,1,2)$ & $0.0345944$ & $0.232480$ \\
%$(11,3)$ & $-(5,1,1,1)$ & $0.0281306$ & $0.219929$ \\
%$(13,5)$ & $-(3,3,1,1)$ & $0.0146649$ & $0.148364$ \\
\hline
\end{tabular}
}
\caption{Some examples of coprime STU solutions in the anti-twist class.
}\label{table:STU_antitwist}
\end{table}

\subsubsection{Twist STU solutions}\label{subsection:STU_twist}
For twist class STU solutions, we focus on examples with\footnote{From these solutions, one can obtain solutions
with $\kappa=+1$ and $t_N=t_S=0$ by implementing 
$n_N \leftrightarrow n_S$, $p^I \leftrightarrow - p^I $, $x_N^I \leftrightarrow -x_S^I$, $k \leftrightarrow -k$, 
with $S_{\rm BH}$ left unchanged.}
\begin{align}
\kappa = -1\,,\qquad t_N=t_S=1\,,
\end{align}
and note that there are no solutions with $\kappa=-1$ and $t_N = t_S = 0$. It was argued based on numerical investigations
in \cite{Couzens:2021cpk}
that a necessary condition for these solutions
to exist is\footnote{For $AdS_2 \times S^2$ solutions that we discuss below, this has been shown in \cite{Benini:2015eyy}.}
\begin{align}\label{pconststwiststu}
\text{Three $p^I<0$ and one $p^I>0$}\,,
\end{align}
and in particular $p^0 p^1 p^2 p^3<0$.

A coprime example is given by $(n_N,n_S)=(17,3)$ and $p^I=(37,-5,-1,-11)$. We find:
\begin{align}
\label{eqn:STU_ex_twist_17_3}
    k & %= 1.72450 
    = \frac{\sqrt{379021}}{357} \,,\nn
    x_N^I %& = ( 1.20739\,,\, 0.184072\,,\, 0.0346358\,,\,0.505685) \nn
    & = \left( \frac{1}{2} + \frac{871}{2\sqrt{379021}} \,,\, \frac{1}{2} - \frac{389}{2\sqrt{379021}} \,,\, \frac{1}{2} - \frac{573}{2\sqrt{379021}} \,,\, \frac{1}{2} + \frac{7}{2\sqrt{379021}} \right) \,, \nn
    x_S^I %& = ( 2.04878\,,\, 0.0703706\,,\, 0.0118955\,,\, 0.255542 ) \nn
    & = \left( \frac{1}{2} + \frac{1907}{2\sqrt{379021}} \,,\, \frac{1}{2} - \frac{529}{2\sqrt{379021}} \,,\, \frac{1}{2} - \frac{601}{2\sqrt{379021}} \,,\, \frac{1}{2} - \frac{301}{2\sqrt{379021}} \right) \,, \nn
    S_{\textrm{BH}} & =  \frac{2\pi}{153} \sqrt{-609+\sqrt{379021}} N^{3/2}\sim 0.105875 N^{3/2}  \,.
\end{align}
By \eqref{eqn:fluxmIZcond} we have $m_N^I = (16,13,6,15)$, and $m_S^I = (5,2,1,2)$. 
Some additional twist STU examples are presented in table \ref{table:STU_twist}.

Non-coprime examples can also be obtained by taking $h$-fold flux quotients with $h$ odd from \eqref{hodd}.
Note that we will see in the
next section that there are never any relevant deformations of the hyperscalar about an STU solution
 in the twist class.
 
\begin{table}[h]
\centering
\scalebox{0.85}{
\begin{tabular}{ |c|c|c|c| }
\hline
$(n_N,n_S)$ & $p^I$ & $\frac{1}{N^{3/2}} S_{\textrm{BH}}$ & $k$ \\
\hline
%$(6,1)$ & $(11,-2,-1,-1)$ &  $0.343325$ & $3.06449$ \\ 
%$(12,1)$ & $(22,-3,-3,-3)$ &  $0.440866$ & $2.86940$ \\ 
%$(12,2)$ & $(22,-4,-2,-2)$ & $0.171663$ & $1.53225$ \\ 
$(3,1)$ & $(7,-1,-1,-1)$ & $\frac{2}{9} \sqrt{\sqrt{469}-21} \pi$ & $\frac{\sqrt{469}}{3}$  \\ 
$(7,1)$ & $(15,-1,-2,-4)$ & $\frac{2\pi}{21} \sqrt{2 \sqrt{2521}-98 }$ & $\frac{2\sqrt{2521}}{21}$ \\ 
$(13,1)$ & $(22,-3,-3,-2)$ & $\frac{4\pi}{13}\sqrt{\sqrt{23}-14/3}$ & $\frac{6\sqrt{23}}{13}$ \\ 
$(17,3)$ & $(23,-1,-1,-1)$ & $\frac{2\pi}{153}\sqrt{\sqrt{13781}-117}$ & $\frac{\sqrt{13781}}{357}$ \\
$(17,3)$ & $(37,-5,-1,-11)$ & $\frac{2\pi}{153} \sqrt{\sqrt{379021}-609}$ & $\frac{\sqrt{379021}}{357}$ \\
%
%
%$(13,1)$ & $(22,-3,-3,-2)$ & $0.347407$ & $2.21346$ \\ 
%$(17,3)$ & $(23,-1,-1,-1)$ & $0.0257283$ & $0.328831$ \\
%$(17,3)$ & $(37,-5,-1,-11)$ & $0.105875$ & $1.72450$ \\
\hline
\end{tabular}
}
\caption{Some examples of coprime STU solutions in the twist class.
}\label{table:STU_twist}
\end{table}

\subsection{\texorpdfstring{$AdS_2 \times S^2$}{AdS2 x S2} solutions}
We analyse the BPS equations for $AdS_2 \times S^2$ STU solutions in appendix \ref{app:S2}.
The scalars are all constant as is the warp factor $e^{V}$ and the spinor function
$\xi$ in \eqref{eqn:spindlespinor}. Hence the $x^I$ in \eqref{eqn:dressedscalar} are also constant.
These solutions only exist in the twist class with $t_N = t_S$.
They preserve supersymmetry with Killing spinors as in \eqref{eqn:spindlespinor}, with $\cos\xi = (-1)^t$ (so $\sin \xi = 0$), and have definite chirality. We recover\footnote{\label{scalarfootcompariosn}With $n_N = n_S =1$, if we identify $(p^0,p^1,p^2,p^3)$ here with $(\mathfrak{n}^4,\mathfrak{n}^1,\mathfrak{n}^2,\mathfrak{n}^3)$ there, then the neutral scalars $x^I/x^0$ for $I = 1,2,3$ agree with their $z_{1,2,3}$ in their (3.23), respectively.} the results for $AdS_2\times S^2$ solutions of \cite{Benini:2015eyy}
and, by taking $h$-fold flux quotients, also extend to orbifolds of the $S^2$.

One can obtain expressions for $x^I$ and the black hole entropy directly. One can also obtain these by considering
the results above for twist solutions and carefully taking the limit $1/k\to 0$. If one sets $n_N=n_S=1$ then we get the round $S^2$,
but 
for $n_N=n_S \ne 1$, we obtain an orbifold of the $S^2$. For example, one can write 
\eqref{eqn:BPS_flux} in the form $x_N^I = x_S^I - \frac{2}{k} \frac{p^I}{n_N n_S} $ to deduce that $x_N^I = x_S^I \equiv x^I$.
Furthermore, from \eqref{eqn:xIconstraint} we have the constraint 
\begin{align}
\sum_I x^I = -2 \kappa\,.
\end{align}
The flux condition \eqref{eqn:BPS_Rsymmconstraint} becomes
\begin{align}
p_R= \sum_Ip^I = 2(-1)^{t+1} n_N\,,
 \end{align}
which is the standard topological twist condition, upon setting $n_N = n_S = 1$.
 
The explicit values of the scalars can be obtained from substituting $\sigma=+1$ and $n_N=n_S$ into 
\eqref{STU_xI}. 
Expressions for the metric functions $f^2$ and $e^{2V}$ (not in conformal gauge) are given in \eqref{ads2f2e2v}.
Similarly, taking the limit in \eqref{BHentropy} we deduce that the entropy can be written (see \cite{BenettiGenolini:2024kyy})
\begin{align}
	S_{\rm BH} &= - (-1)^{t} \frac{\pi \sqrt{2}}{3} N^{3/2} \kappa \frac{p^I}{n_N^2 } \frac{\partial}{\partial x^I} i \cF(x^I) \,,\nn
 &= - (-1)^{t} \frac{\pi \sqrt{2}}{3} N^{3/2} \kappa \frac{\sqrt{x^0 x^1 x^2 x^3}}{n_N^2} \frac{p^I}{x^I} \,,\nn
 &= \frac{2\pi}{3n_N^2 } N^{3/2} \sqrt{\hat{P}^{(2)}+s - n_N^2} \,,
\end{align}
where the last line is in agreement with \eqref{eq:Couzens_BH}.

Notice the above formulae also incorporates non-coprime ``spheres" where $n_N = n_S \neq 1$, which are simply quotients of $S^2$.
Specifically, we consider $S^2/{\mathbb{Z}_h}$ where $(n_N,n_S,p^I) = h (1,1,\hat{p}^I)$ with the latter being associated with a topological twist on $S^2$. We then have $S_{\rm BH}(n_N,n_S,p^I) = \frac{1}{h} S_{\rm BH}(1,1,\hat{p}^I)$, as usual. 
From \eqref{hodd} we must have $h$ odd.
In Appendix \ref{app:S2}, we provide some additional details and also analyse hyperscalar fluctuations about the $AdS_2\times S^2$ solutions. 
In particular, we find no relevant deformations.

\section{Hyperscalar fluctuations for the STU solutions}\label{hyperfluc}
\subsection{Linearised analysis of hyperscalar fluctuation} 
In this section we
consider linearised perturbation of the hyperscalar $\rho e^{i \theta}$ about an STU solution of the form
\begin{equation}\label{hyperlinear}
    \rho = w(y) u^\delta \,,\quad \theta = \bar{\theta}z\,.
\end{equation}
Here  $u$ is the Poincar\'e radial coordinate on $AdS_2$, as in (\ref{eqn:frame}), and we have assumed the perturbation has fixed scaling dimension $\Delta = 1-\delta$. If $0<\Delta<1$, or equivalently $0<\delta<1$, then this mode is associated with having a source for a
relevant operator. If $\Delta>1$, or equivalently $\delta<0$, the mode is dual to an irrelevant operator. The perturbation breaks the 
$U(1)$ rotational symmetry of the spindle as well as the $U(1)_B$ symmetry, but preserves a diagonal subgroup. 
Recall that the gauge invariant one-form $D\theta$ is given by $D\theta=(\bar{\theta} + \zeta_I a^I) {d} z$ in the ansatz \eqref{eq:ansatz}.

We assume the perturbation preserves the same Poincar\'e supersymmetries of the STU solution (recall \eqref{eqn:proj}, 
\eqref{eqn:Poinproj}).
The superpotential (\ref{eqn:superpotential}) can be expanded $\mathcal{W} = \mathcal{W}_{\textrm{STU}} + \mathcal{O}(\rho^2)$,
 so the $\mathcal{O}(\rho)$ parts of the gravitino and gaugino variations in (\ref{eqn:4dKSEs}) are trivially satisfied, for the 
 linear perturbation. Next, using
\begin{equation}
    \partial_\rho \mathcal{W} = \frac{1}{2} \zeta_IX^I \rho + \mathcal{O}(\rho^2)\,,\quad \partial_\rho Q = -\frac{1}{2} \rho \,D\theta + \mathcal{O}(\rho^2)\,,
\end{equation}
we deduce that the linearised hyperino variation is given by
\begin{equation}
    \left[w \delta e^{-V} \Gamma^1 + \frac{w'}{f}\Gamma^2 + \zeta_IX^I w-i (D\theta)_z \frac{w}{h} \Gamma^3  \right] \epsilon = 0\,.
\end{equation}
Now using the conditions for Poincar\'e supersymmetries \eqref{eqn:proj}, \eqref{eqn:Poinproj}
it follows that
\begin{equation}
    \left[ \left(\frac{w'}{f\sin\xi} - \frac{w}{h\tan\xi} (D\theta)_z + \zeta_IX^I w\right) + i \left(\kappa w e^{-V} \delta + \frac{w}{h\sin\xi} (D\theta)_z - \frac{w'}{f\tan\xi} \right) \Gamma_{23} \right]\epsilon = 0 \,.
\end{equation}
With no further breaking of supersymmetry, $\epsilon$ and $\Gamma_{23}\epsilon$ are independent and so
\begin{align}
    \delta & = \kappa e^V \left( \frac{w'}{wf\tan\xi} -\frac{(D\theta)_z}{h\sin\xi} \right)\,,\nn
    \frac{w'}{w} & = \frac{f}{h}\cos\xi \,(D\theta)_z - \zeta_IX^I f\sin\xi \,.
\end{align}
Combining the two, replacing $h$ by (\ref{eqn:BPSint}), using 
conformal gauge $f=e^V$ \eqref{confgauge} and the definition of $\mathcal{I}^I$, 
given in (\ref{eqn:mathcalI}), we can write these as
\begin{align}
\label{eqn:ODE}
    \delta & = \frac{\kappa}{k} \left( \zeta_I\mathcal{I}^I -(D\theta)_z \right)\,,\nn
    \frac{w'}{w} & = \frac{1}{k} \left( \frac{(D\theta)_z}{\tan\xi} + \zeta_I \mathcal{I}^I \tan\xi \right)\,.
\end{align}

We next assume that as we approach the poles in the two patches, the hyperscalar behaves as
\begin{equation}
	w \sim (y-y_N)^{r_N} \,, \qquad w \sim (y_S - y)^{r_S} \,,
\end{equation}
with $r_{N,S}\ge 0$. Recalling (\ref{eqn:sincosBC}) and how the metric is degenerating at the poles, for
the hyperscalar to be smooth we demand that $r_{N,S}\in\mathbb{Z}_{\ge 0}$. Then from 
\eqref{eqn:sincosBC}, \eqref{eqn:ODE}
we deduce that
\begin{equation}
\label{eqn:wDthetaatpoles}
    \left.(D\theta)_z\right\vert_N = (-1)^{t_N}\frac{r_N}{n_N}\,,\qquad \left.(D\theta)_z\right\vert_S = -(-1)^{t_S}\frac{r_S}{n_S}\,.
\end{equation}
Next, we evaluate $\delta$ at the poles. From (\ref{eqn:BPSFequiv}) we have $\frac{p^I}{n_N n_S} = \frac{1}{2}\left.\mathcal{I}^I\right\vert^S_N$. Using (\ref{eqn:wDthetaatpoles}), this leads to the following constraint on the broken flux
\begin{equation}
\label{eq:brokenflux}
    p_B = (-1)^{t_N+1}r_Nn_S + (-1)^{t_S+1}r_Sn_N \,.
\end{equation}
Thus, given an STU solution, specified by $n_{N,S}$, $t_{N,S}$ and $p^I$, the condition \eqref{eq:brokenflux} imposes
a constraint on $r_N$ and $r_S$ to have a smooth hyperscalar perturbation. In particular, if $r_N=r_S=0$, so that the hyperscalar is 
non-vanishing at both poles, as studied in \cite{Suh:2022pkg,BenettiGenolini:2024kyy}, we must have $p_B=0$.
All of these results are consistent with the general comments about smooth hyperscalar bundles
that we discussed in section \ref{hyperscalar}. As discussed there, and comparing
\eqref{eq:Dthetahyper} with \eqref{eqn:wDthetaatpoles}, smoothness of the hyperscalar implies
that we must have
\begin{align}
\label{eq:rNrSconst}
    (-1)^{t_N} r_N = n_N \bar{\theta}_N + 2 \zeta_I m_N^I\,,\nn
    -(-1)^{t_S}r_S = n_S \bar{\theta}_S + 2 \zeta_I m_S^I \,,
\end{align}
with $\bar{\theta}_{N,S}\in\mathbb{Z}$ which, in particular, is consistent with
$r_{N,S}\in \mathbb{Z}_{\ge 0}$.

Using the explicit result \eqref{STU_xI}, we can express the scaling dimension $\Delta$ of the operator dual to the perturbation as $\Delta = 1 - \delta$ with
\begin{align}\label{deltafinal}
	\delta  =  \frac{1}{2} - \frac{n_N n_S}{2s} &+ \frac{\sigma}{12s} \left( p_R^2 - 2 p_B p_R - 2 p_B^2 + 2 (p_{F_1}^2 + p_{F_1} p_{F_2} + p_{F_2}^2) \right) \nn
	&- \frac{1}{4s}\left({n_N n_S (r_N + r_S) - \sigma (r_N n_S^2 + r_S n_N^2)}\right)  \,.
\end{align}
Notice that the value of $\delta$ will explicitly depend on $m_N^I$ and $m_S^I$ ($p_B$ can be expressed as in \eqref{eq:brokenflux}, where $r_N, r_S$ are given by \eqref{eq:rNrSconst}). This implies that STU solutions with the same $n_{N,S}$, $t_{N,S}$, and $p^I$, but different $m_N^I$ and $m_S^I$ (in the non-coprime case), will have the same Bekenstein-Hawking entropy given by \eqref{eq:Couzens_BH}, but in general different spectra for the hyperscalar.
It is convenient to introduce a ``Kaluza-Klein" integer $n_{\rm KK}$ for the hyperscalar modes \eqref{hyperlinear}:
\begin{equation}
	n_{\rm KK} \equiv (-1)^{t_N} \bar{\theta}_N \,.
\end{equation}
Then we can write
\begin{align}\label{rNrSnKK}
	\frac{r_N}{n_N} & = n_{\rm KK} + (-1)^{t_N} \frac{2\zeta_I m_N^I}{n_N} \,, \nn
	\frac{r_S}{n_S} & = (-1)^{t_N + t_S + 1} \frac{r_N}{n_N} - (-1)^{t_S} \frac{p_B}{n_N n_S} \,,
\end{align}
where in the second equation we used \eqref{eq:brokenflux}. 

These expressions allow us to see that the spectrum of hyperscalar fluctuations in the twist and anti-twist class are substantially different.
For the twist class, we have $t_N = t_S$, which implies that
\begin{equation}\label{constratinrstwist}
	0 \leq \frac{r_N}{n_N} \leq (-1)^{t_S+1} \frac{p_B}{n_N n_S} \,.
\end{equation}
It follows that there are only a finite number of fluctuations with $r_{N,S} \geq 0$ which are allowed, labelled by different values of $n_{\rm KK}$. In particular, if $(-1)^{t_S+1} p_B < 0$ we don't obtain any solutions.
However, for the anti-twist class, we have $t_N + t_S = 1$, so we must have 
\begin{equation}
	r_N \geq (-1)^{t_S} \frac{p_B}{n_S} \,,
\end{equation}
where $r_N \geq 0$. This leads to an infinite number of solutions for $n_{\rm KK}$.

Using \eqref{eqn:ODE}, \eqref{eqn:mathcalI}, \eqref{eqn:wDthetaatpoles} and \eqref{eqn:xIconstraint}
we can also show that at both poles we can write $\delta$ in the form
\begin{align}\label{rzeroantitwist}
-2\kappa (-1)^{t+1}\delta=2r\kappa(-1)^{t+1}&+(r-1)(-1)^{t+1}x^0+
(r+1)(-1)^{t+1}x^1\nn
&+(r+1)(-1)^{t+1}x^2+
(r+1)(-1)^{t+1}x^3\,.
\end{align}
Since $(-1)^{t+1}x^I>0$ at the poles, 
when $r\ge 1$ the last four terms on the right hand side are all positive. 
Thus, for the anti-twist class, setting
$\kappa=-1$ for convenience,
we can conclude the following: 
when $t_N=1$ and $t_S=0$ then $\delta>0$ requires $r_S=0$
and 
when $t_N=0$ and $t_S=1$ then $\delta>0$ requires $r_N=0$.

\begin{figure}[h!]	
 \centering
	\includegraphics[scale=0.45]{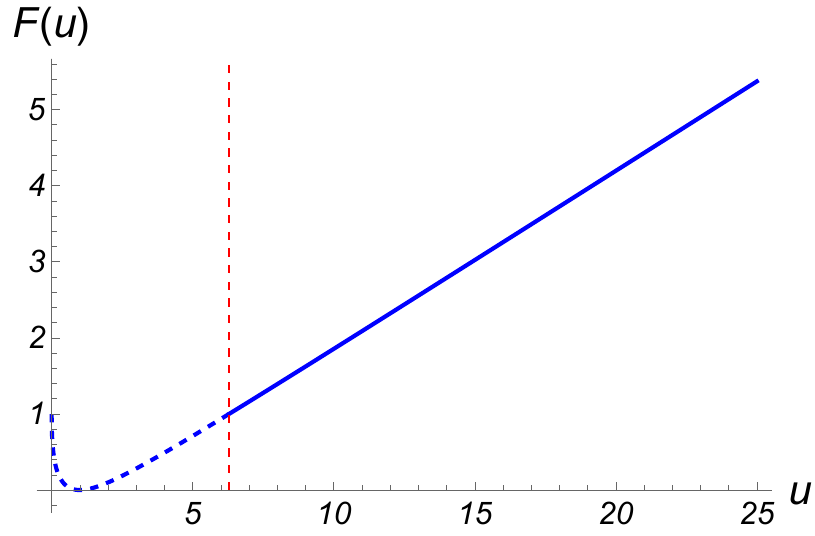}~
	\caption{Plot of $F(u)$, with $r_N\le F(u)$ and $u > 1+2\sqrt{7}$ (dashed line) required in order that $\delta>0$, for $r_N\ne 0$ 
	for the anti-twist STU solutions with $p^1=p^2=p^3$.  
	}
	\label{fig:Fplot}
\end{figure}

We can make some further observations regarding STU solutions that admit relevant modes with $\delta>0$.
Similar to \cite{Arav:2025jee}, it is convenient to introduce the rescaled variables
\begin{equation}
	u = \frac{n_N}{n_S} \,,\qquad v_1 = \frac{p_{F_1}}{n_S} \,,\qquad v_2 = \frac{p_{F_2}}{n_S} \,.
\end{equation}
For anti-twist solutions, focussing for definiteness on the class
\begin{equation}\label{atwistclassconds}
	\kappa = -1 \,,\qquad t_N =1 \,, t_S = 0 \qquad  \Rightarrow \qquad r_S = 0 \,,
\end{equation}
the fluxes (constrained by \eqref{eqn:BPS_Rsymmconstraint}, \eqref{eq:brokenflux}) can be written as
\begin{align}
	p^0 & = \frac{n_S}{2} ( r_N - u + 1) \,,\qquad\qquad \qquad \quad p^1  = - \frac{n_S}{6} ( r_N + u - 1 - 4 v_1 - 2 v_2 )\,, \nn
	%p^1 & = - \frac{n_S}{6} ( r_N + u - 1 - 4 v_1 - 2 v_2 ) \,,\nn
	p^2 & = - \frac{n_S}{6} ( r_N + u - 1 + 2 v_1 - 2 v_2 ) \,,\quad p^3  = - \frac{n_S}{6} ( r_N + u- 1 + 2 v_1 + 4 v_2 )\,.
	%p^3 & = - \frac{n_S}{6} ( r_N + u- 1 + 2 v_1 + 4 v_2 ) \,.
\end{align}
We also have the conditions \eqref{STUexistcond} and \eqref{anti-twist_bountiful}.
In particular, with $p^0<0$, from \eqref{anti-twist_bountiful}, we have
\begin{equation}\label{defofu}
	u > 1 + r_N \,.
\end{equation}
For the particular case of $r_N = r_S = 0$ (so $p_B=0$), we find that demanding  
$\delta > 0$ does not provide any additional constraints over and above \eqref{STUexistcond}, \eqref{anti-twist_bountiful}
and so these STU anti-twist solutions will have relevant hyperscalar modes.
Next, for the case of pairwise equal fluxes, we always have $r_N>0$ (see \eqref{2p2Bsymm}), and furthermore
one can show that there is never a relevant hyperscalar mode for these anti-twist solutions.
This also rules out relevant hyperscalar modes for anti-twist STU solutions with $p^0 = p^1 = p^2 = p^3$, some of which are solutions of minimal gauged supergravity (the ones with $m^I$ all equal at each pole; see appendix \ref{minimal}).
Turning to the case of three equal fluxes, $p^1 = p^2 = p^3$, for $r_N>0$, 
a necessary condition for $\delta>0$ is given by
\begin{equation}\label{ubound}
	u > 1 + 2 \sqrt{7} \sim 6.3\,.
\end{equation}
The positive integer $r_N$ is then bounded from above as a function of $u$,\footnote{$F(u)$, for $u\ge 0$, is the middle root of the cubic: $2 - 3 u + u^3 - ( 2u^2 + 14 u + 2 ) x - (2+10 u) x^2 + 2x^3 = 0$.}
\begin{align}\label{rNbound}
	r_N \leq F(u) \,,
\end{align}
and a plot of $ F(u)$, for which $F(u = 1 + 2 \sqrt{7})=1$ is given in figure \ref{fig:Fplot}.

Finally, we turn to twist class STU solutions. Setting $\kappa=-1$, there are only solutions with $t_N=t_S=1$.
We have been able to show that there are no relevant hyperscalar modes for $AdS_2 \times S^2$ solutions (see appendix \ref{app:S2}) 
and also for STU twist solutions in the 3+1 truncation (see appendix \ref{3p1}).
For unequal fluxes, in an extensive numerical search we always find $\delta < 0$. Therefore, we conjecture that relevant hyperscalar deformations do not exist for any twist class STU solutions.

For the special case of $AdS_2 \times S^2$ twist solutions, and their $h$-fold flux quotients, we set $n_N = n_S$. 
The expression for $\delta$ in \eqref{deltafinal}
only depends on $r_{N,S}$ via their sum, since from \eqref{eq:brokenflux} we have
$p_B = n_N (r_N + r_S )$. This leads to a degeneracy of the scaling dimensions for the modes. 
When $n_N = n_S = 1$, associated with the round $S^2$, this is to be expected due to the enhanced $SO(3)$ isometry group. The degeneracy can also persist for $h$-fold flux quotients of the $S^2$, sometimes with reduced degeneracy,
as we illustrate in appendix \ref{app:S2}.

In the remainder of this section, we present some more examples of STU solutions for both anti-twist and twist class, and
summarise the spectrum of hyperscalar modes.

\subsection{Hyperscalar modes for anti-twist STU solutions, 3 $p^I$ equal}\label{stu3pequal}

\begin{figure}[h!]	
 \centering
	\includegraphics[scale=.8]{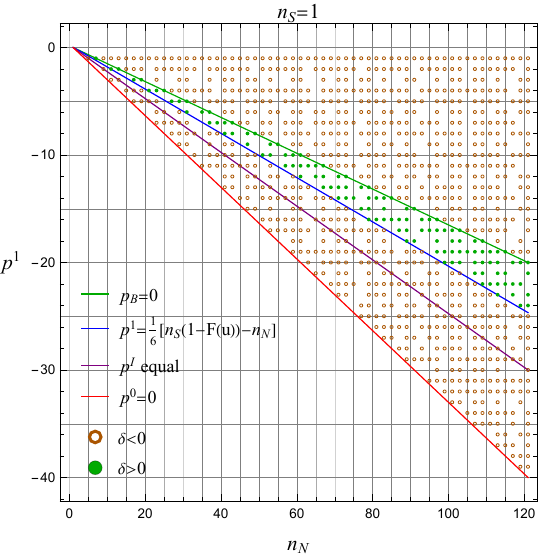}~
	\includegraphics[scale=.8]{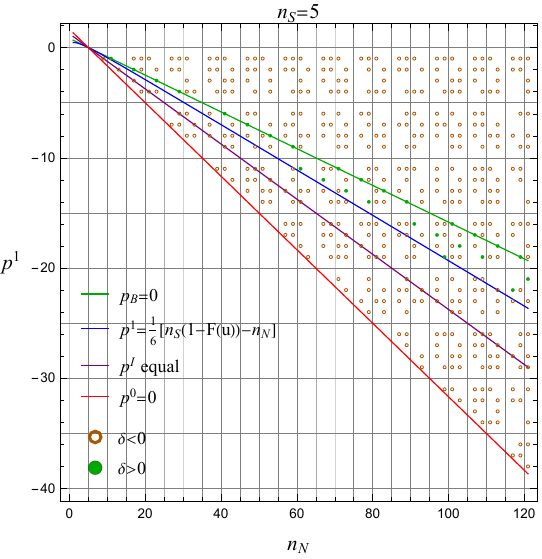}~
	\caption{Examples of anti-twist $STU$ solutions with $n_N, n_S$ coprime and 3 equal fluxes, $p^1=p^2=p^3$. Left panel $n_S=1$, right panel $n_S=5$. Brown circles denote smooth, supersymmetric
		STU solutions. Green dots denote smooth, supersymmetric solutions that also have a relevant hyperscalar mode. We have also marked
		 the line associated with solutions that have 4 equal fluxes, which never have relevant modes.}
	\label{fig:dotplot}
\end{figure}

In Figure \ref{fig:dotplot} we have made some plots of the landscape of smooth anti-twist STU solutions \eqref{atwistclassconds}
with three equal fluxes
$p^1=p^2=p^3$. When in addition $m^1=m^2=m^3$ at both poles (which is automatically true for the coprime spindles), the solutions
are in the $3+1$ truncation and
the uplifted solutions preserve an enhanced $SU(3)$ flavour symmetry (i.e. $SU(3)\times U(1)\times U(1)_R$ symmetry).
In addition, the figure  also indicates whether or not an individual solution has a relevant hyperscalar mode. 
The solutions are as in \eqref{atwistclassconds} and all solutions have $n_N, n_S$ coprime.

The plots have $n_S=1$ and $n_S=5$, for various values of $n_N$.  
We highlight some general features of the plots.
 For smooth spindles we necessarily have $n_N$ is odd from \eqref{thensareodd}.
 For coprime solutions we also require
$p^0$, $p^1$ to be coprime to both $n_N$ and $n_S$ (from \eqref{pcoprconds}) and for anti-twist solutions $p^0+3p^1=n_S-n_N$ (from \eqref{eqn:BPS_Rsymmconstraint}). For relevant hyperscalar modes we require $r_S=0$ and so
 $p_B=r_N n_S\ge 0$, with $r_N$ an integer $\ge 0$ (from
\eqref{eq:brokenflux}). Recall that the STU solutions have $p^I<0$, as in \eqref{anti-twist_bountiful} and so $n_N>n_S$. Furthermore,
we have $3p^1=n_S-n_N-p^0$ and we only find solutions for $p^1$ taking values
greater than the $p^0=0$ line (marked on the figure),
and also for $p^1<0$.
The actual solutions satisfying the smoothness conditions are marked with brown circles in this region.
We also have $p_B=n_S-n_N-6p^1$ and since $p_B=r_N n_S$ and $r_N\ge 0$, the relevant hyperscalar modes
must lie on or below the line $p_B=0$, which is also marked on the figure.
From \eqref{rNbound}, relevant hyperscalar modes also require $p_B\le n_S F(u)$, or equivalently
$p^1\ge p^1_\text{crit}\equiv \frac{1}{6}[n_S(1-F(u))-n_N]$,
with $F(u)$ given in figure \ref{fig:Fplot}. The solutions with relevant modes are marked with green dots
in the figure. Note that once one has found all of the brown circles in the wedge where relevant modes
can exist (i.e. the values of $p^0$ and hence $p^1$ and $p_B$), the condition that $p_B=r_N n_S$ needs to be satisfied
for integer $r_N\ge 0$. For $n_S=1$ this is trivially satisfied, but not for other values of $n_S$: this explains
the fact that in the figure all of the allowed solutions in the wedge for $n_S=1$ also have relevant hyperscalar
modes, but not for $n_S=5$.

We highlight another general result for anti-twist solutions with $p^1=p^2=p^3$ regarding
the existence of relevant hyperscalar deformations when $n_S$ is a multiple of 3. 
First recall that below \eqref{rzeroantitwist} we showed that $\delta>0$ implies $r_S=0$.
Next if $m^1_S=m^2_S=m^3_S$, then for smooth supersymmetric solutions
from \eqref{eq:rNrSconst} we have $m^0_S-3m_S^1=0$ mod $n_S$ and from
\eqref{eq:spinorregularity} we also have $m^0_S+3m^1_S=-1$ mod $n_S$. But when $n_S$ is a multiple of 3 both of these
conditions cannot be satisfied. Thus, for coprime spindles with $p^1=p^2=p^3$, we automatically have
$m^1=m^2=m^3$ at both poles and hence we cannot have any relevant hyperscalar deformations.
In the non-coprime case with $p^1=p^2=p^3$ it is possible for $m^1_S$, $m^2_S$ and $m^3_S$ not to be all equal,
in which case they break an $SU(3)$ flavour symmetry, and it is only in such cases that we can have a relevant hyperscalar
deformation; we shall see some examples later.

\subsection{Hyperscalar modes for anti-twist STU solutions, 4 $p^I$ equal }

We now consider STU solutions with $p^0 = p^1 = p^2 =p^3\equiv p$, 
which only exist in the anti-twist class. These are, of course, a subset of the solutions considered in the previous
subsection.
When in addition $m^I_{N,S}$ are all equal the uplifted solutions preserve $SU(4)$ 
flavour symmetry (i.e. $SU(4)\times U(1)_R$ symmetry) and are solutions of D = 4 minimal gauged supergravity, as discussed in
appendix \ref{appminimalallequal}. For coprime spindles
these are the solutions of \cite{Ferrero:2020twa}; here we also comment on the case of non-coprime spindles. 

Solutions with all $p^I$ equal have $p=(n_S-n_N)/4$ and so $(n_S-n_N)$ must be divisible by 4, 
with $n_N$ and $n_S$ both odd (from \eqref{thensareodd}) then there is a unique smooth, supersymmetric solution with
all $m_N^I$ equal and all $m_S^I$ equal, which preserves $SU(4)$ flavour symmetry. 

In the coprime case, since the $m^I_{N,S}$ are uniquely specified by the fluxes we preserve the $SU(4)$ flavour symmetry.
Some examples, with the spectrum of hyperscalar modes are presented in table \ref{AT_STU_4pequal}.
These solutions never have any relevant hyperscalar modes, 
as noted below \eqref{defofu}, and
as one can see in figure \ref{fig:dotplot}.

We can have non-coprime cases, with $h$ odd from \eqref{hodd},
where the $SU(4)$ flavour symmetry is broken: in table \ref{minimalnoncoprime}
we have summarised an $h=3$-fold flux reduction of an example in table \ref{AT_STU_4pequal} and
the entropy of these are reduced by $1/h$ as in \eqref{entropyfluxred}. Notice that there are solutions
that preserve the $SU(4)$ flavour symmetry, but there are also solutions that break this 
to an $SU(3)\times U(1)$ flavour symmetry
(which are related by a permutation of the fluxes).

\begin{table}[h]
\centering
\scalebox{0.85}{
\begin{tabular}{ |c|c|c|c|c|c| }
\hline
$(n_N,n_S)$ & $p^I$ & $r_N$ & $r_S$ & $\delta$ & $\frac{1}{N^{3/2}} S_{\rm BH}$ \\
\hline
$(5,1)$ & $-(1,1,1,1)$ & $2+5 n_{\rm KK} $ & $ n_{\rm KK}$ & $-0.193 -\frac{5}{\sqrt{13}} \nkk$ & $\frac{4\sqrt{2}\pi}{15(\sqrt{13}+3)}$  \\ 
$(7,3)$ & $-(1,1,1,1)$ & $3+7\nkk' $ & $ 1+3\nkk'$ & $-1.45-\frac{21}{\sqrt{29}} \nkk'$ & $\frac{4 \sqrt{2} \pi}{63(\sqrt{29}+5)}$  \\ 
$(9,5)$ & $-(1,1,1,1)$ & $4+9\nkk $ & $2+ 5\nkk$ & $-2.59-\frac{45}{\sqrt{53}}\nkk$ & $\frac{4 \sqrt{2} \pi}{135(\sqrt{53}+7)}$  \\ 
$(9,1)$ & $-(2,2,2,2)$ & $4+9\nkk$ & $\nkk $ & $-0.203- \frac{9}{\sqrt{41}}\nkk$ & $\frac{16 \sqrt{2}\pi}{27(\sqrt{41}+5)}$  \\ 
$(11,3)$ & $-(2,2,2,2)$ & $5+11 \nkk'$ & $1+3\nkk' $ & $-1.55 - \frac{33}{\sqrt{65}} \nkk'$ & $\frac{16  \sqrt{2} \pi}{99 (\sqrt{65}+7)}$  \\ 
\hline
\end{tabular}
}
\caption{
Examples of smooth, coprime anti-twist STU solutions with $p^0 = p^1 = p^2 = p^3$, and the spectrum of hyperscalar modes.
We denote $\nkk'=\nkk+1$: whichever quantity $n_{KK}$ or  $\nkk'$ that appears is an integer that is $ \ge 0$.}\label{AT_STU_4pequal}
\end{table}

\begin{table}[h!]
 \begin{center}
 $(n_N,n_S)=3(7,3),~p^I = -3(1,1,1,1)$\\ \smallskip
 \scalebox{0.85}{
\begin{tabular}{ |c|c|c|c|c| } 
 \hline
 $m_N^I$ &  $m_S^I$ & $r_N$&$r_S$& $\delta$\\ 
 \hline 
 $(5,5,5,5)$ & $(2,2,2,2)$ & $10+21 \nkk$ & $4+9 \nkk $ & $\delta[3 \nkk+1]$ \\ 
\hline
\hline
$(5,19,19,19)$ & $(2,8,8,8)$ & $10+21 \nkk''$ & $4+9 \nkk''$ & $\delta[3 \nkk''+1]$\\ 
$(19,5,19,19)$ & $(8,2,8,8)$ & $3+21 \nkk'$ & $1+9 \nkk'$ & $\delta[3 \nkk']$\\ 
$m^1\leftrightarrow m^2\leftrightarrow m^3$ & $m^1\leftrightarrow m^2\leftrightarrow m^3$ & $\dots$ & $\dots$ & $\dots$\\ 
%$(19,19,19,5)$ & $(8,8,8,2)$ & $3+21 \nkk'$ & $1+9 \nkk'$ & $\delta[3 \nkk']$\\ 
 \hline
\end{tabular}
}
\caption{
Examples of non-coprime, anti-twist STU solutions with $p^0 = p^1 = p^2 = p^3$, and the spectrum of hyperscalar modes. 
The solutions are an $h=3$-fold flux quotient of the $n_N=7$, $n_S=3$ solution in table \ref{AT_STU_4pequal} and correspondingly we
have defined $\delta[n]\equiv\frac{1}{2}-\frac{21 + 42 n}{2\sqrt{29}}$.
We denote $\nkk'=\nkk+1$, $\nkk''=\nkk+2$: whichever quantity $n_{KK}$, $\nkk'$, $n_{KK}''$ that appears is an integer that is $ \ge 0$.
There are two more solutions obtained from the third, by permuting $m^1, m^2, m^3$, with all other entries identical.
\label{minimalnoncoprime}}
\end{center}
 \end{table}

\subsection{Hyperscalar modes for twist STU solutions}

We now consider the spectrum of hyperscalar modes about twist class STU solutions. We present some coprime
spindle examples in
table \ref{twist_STU_modes} and we have taken 
\begin{align}
\kappa=-1,\qquad t_N=t_S=1\,.
\end{align}
As the modes must have $r_N, r_S\ge0$, recalling
\eqref{constratinrstwist}, we only have a finite number of hyperscalar modes and this number is denoted by 
$\# (\Delta)$ in table \ref{twist_STU_modes}.
In the last example with $(n_N,n_S)=(9 , 7)$, from \eqref{rNrSnKK} we find $r_N= 9 \nkk + 7$ and $r_S = -3 - 7 \nkk$
and so there is no value of $\nkk$ that gives rise to $r_{N,S}\ge 0$.
Hence there are no hyperscalar modes at all for this example.
As noted above, we believe that, in general, there are no STU twist-class solutions with relevant hyperscalar modes.
\begin{table}[h]
\centering
\scalebox{0.85}{
\begin{tabular}{ |c|c|c|c|c|c| }
\hline
$(n_N,n_S)$ & $p^I$ & $r_N$ & $r_S$ & $\delta$ & $\# (\Delta)$ \\
\hline
$(3,1)$ & $(7,-1,-1,-1)$ & $3 n_{\rm KK} +1$ & $3 - n_{\rm KK}$ & $\frac{1}{2} -  \frac{39+6 \nkk}{2\sqrt{469}} $ & 4   \\ 
%$(5,1)$ & $(9,-1,-1,-1)$ & $5 n_{\rm KK} +2$ & $2 - n_{\rm KK}$ & $\frac{1}{2} + \frac{55 + 20 \nkk}{2\sqrt{877}}$ & 3   \\ 
$(5,3)$ & $(11,-1,-1,-1)$ & $5 \nkk+3$ & $1-3 \nkk$ & $\frac{1}{2} - \frac{111 + 30 \nkk}{2\sqrt{2069}}$ & 1\\
$(7,5)$ & $(18,-2,-2,-2)$ & $7\nkk''+2$ & $2-5 \nkk''$ & $\frac{1}{2} - \frac{295 + 70 \nkk''}{2\sqrt{17737}}$ & 1 \\
$(9,7)$ & $(19,-1,-1,-1)$ & $-$ & $-$ & $-$ & 0  \\
%$(7,1)$ & $(15,-1,-2,-4)$ & $7 n_{\rm KK} +1$ & $3 - n_{\rm KK}$ & $-0.266786 - \frac{21}{2\sqrt{2521}} n_{\rm KK}$ & 4  \\ 
%$(13,1)$ & $(22,-3,-3,-2)$ & $13 n_{\rm KK} +4$ & $2 - n_{\rm KK}$ & $-0.334058 - \frac{13}{6\sqrt{23}} n_{\rm KK}$ & 3  \\ 
%$(17,3)$ & $(23,-1,-1,-1)$ & $17 n_{\rm KK} +3$ & $1 - 3 n_{\rm KK}$ & $-0.892764 - \frac{357}{\sqrt{13781}} n_{\rm KK}$ & 1  \\
%$(17,3)$ & $(37,-5,-1,-11)$ & $17 n_{\rm KK} + 18$ & $ -3 n_{\rm KK}$ & $-0.855485 - \frac{357}{\sqrt{379021}} n_{\rm KK}$ & 1  \\
%$(?,?)$ & $(?,?,?,?)$ & $-$ & $-$ & $-$ & 0  \\
\hline
\end{tabular}
}
\caption{Some examples of smooth, coprime STU solutions in the twist class and their spectrum of BPS hyperscalar modes. As before $\nkk'' =\nkk+2$.
For all allowed values of $n_{\rm KK}$ we have $\delta < 0$ and so the BPS hyperscalar modes are always irrelevant. }\label{twist_STU_modes}
\end{table}

We discuss some further examples of $AdS_2\times S^2$ solutions, as well as their $h$-fold quotients
in appendix \ref{app:S2}.

\section{Hyperscalar solutions}\label{sec5}
We now consider solutions with non-vanishing hyperscalar. While those with the hyperscalar non-vanishing at both poles were constructed in \cite{Suh:2022pkg} and necessarily have $p_B=0$, here we allow the hyperscalar to vanish at the poles allowing for $p_B\ne 0$. We continue using the conformal gauge $f=e^V$.
In this section we analyse the BPS equations and make some general observations. In the next section we summarise
some solutions constructed numerically.

The superpotential (\ref{eqn:superpotential}) can be written as
\begin{equation}
\label{eqn:superpotentialx}
    \cos\xi e^V \mathcal{W} = \frac{1}{2}\sum_{I=0}^3 x^I - \zeta_Ix^I \sinh^2\frac{\rho}{2}\,.
\end{equation}
Then
\begin{equation}
    \cos\xi e^V\partial_\rho \mathcal{W} = -\frac{1}{2}\zeta_Ix^I \sinh\rho\,.
\end{equation}
We assume that near the poles the hyperscalar behaves as
\begin{equation}\label{hsexppoles}
    \rho = C_N (y-y_N)^{r_N} + \mathcal{O}(y-y_N)^{r_N+1}\,, \qquad \rho = C_S (y_S-y)^{r_S} + \mathcal{O}(y_S-y)^{r_S+1}\,,
\end{equation}
where $r_{N,S} \in \Z_{\geq 0}$ as discussed in subsection \ref{hyperscalar}, with the constants $C_{N,S}>0$.
Using the BPS equation for $\rho$ (\ref{eqn:BPSeq}), near the pole we find
\begin{equation}
\label{eqn:zetaxNS}
    \zeta_Ix^I_N = \frac{(-1)^{t_N}r_N}{kn_N}\,,\qquad \zeta_Ix^I_S = \frac{(-1)^{t_S+1}r_S}{kn_S}\,.
\end{equation}
Similarly, the second constraint in \eqref{eqn:BPSconstraints}, together with (\ref{eqn:BPSint}), can be written as
\begin{equation}
    \zeta_Ix^I = \frac{1}{k} (D\theta)_z\,.
\end{equation}
Hence, at the poles we deduce (c.f. \eqref{Dthetasigma})
\begin{equation}
    \left.(D\theta)_z\right\vert_N = (-1)^{t_N} \frac{r_N}{n_N}\,,\qquad \left.(D\theta)_z\right\vert_S = -(-1)^{t_S}\frac{r_S}{n_S}\,.
\end{equation}

Next we evaluate (\ref{eqn:superpotentialx}) at the poles. If $r_{N,S}>0$ then $\rho$ vanishes at that pole, while if $r_{N,S}=0$ then (\ref{eqn:zetaxNS}) implies $\zeta_Ix^I=0$. Thus, in both cases
\begin{equation}
    \left.e^V\mathcal{W}\right\vert_{N,S} = \frac{1}{2} \sum_{I=0}^3 (-1)^{t_{N,S}} x^I_{N,S}\,.
\end{equation}
Using (\ref{eqn:bdycond2}), we then find
\begin{equation}
\label{eqn:xIconstrainthyper}
    \sum_I x_N^I = -2\kappa + (-1)^{t_N} \frac{2}{kn_N}\,,\qquad \sum_I x_S^I = -2\kappa - (-1)^{t_S} \frac{2}{kn_S}\,.
\end{equation}
Using (\ref{eqn:zetaxNS}) and (\ref{eqn:xIconstrainthyper}) in the expression for the fluxes (\ref{eqn:BPS_flux}), we have (c.f. \eqref{eq:Rsymmconst}, \eqref{eq:brokenflux_v2})
\begin{align}\label{pRpBhyper}
    \frac{p_R}{n_Nn_S} & = \frac{(-1)^{t_N+1}}{n_N} + \frac{(-1)^{t_S+1}}{n_S}\,,\nn
        \frac{p_B}{n_Nn_S} & = (-1)^{t_N+1} \frac{r_N}{n_N} + (-1)^{t_S+1} \frac{r_S}{n_S}\,.
\end{align}
The two flavour fluxes are given by
\begin{align}\label{flavourhyper}
    \frac{p_{F_1}}{n_Nn_S} = \frac{k}{2} \left( (x_S^1-x_S^2) - (x_N^1-x_N^2) \right)\,,\nn
    \frac{p_{F_2}}{n_Nn_S} = \frac{k}{2} \left( (x_S^2-x_S^3) - (x_N^2-x_N^3) \right)\,.
\end{align}
We also have three conserved charges (\ref{eqn:fluxconstraint3}):
\begin{align}\label{conservedhyper}
    \frac{1}{\kappa}\mathcal{E}_R & = \frac{i}{2} (-1)^{t_{N,S}+1} \mathcal{F}(x_{N,S}^I) \left(\frac{3}{x_{N,S}^0} + \frac{1}{x_{N,S}^1} + \frac{1}{x_{N,S}^2} + \frac{1}{x_{N,S}^3} + 6\kappa \right)\,,\nn
    \frac{1}{\kappa}\mathcal{E}_{F_1} & = \frac{i}{2} (-1)^{t_{N,S}+1} \mathcal{F}(x_{N,S}^I)\left( \frac{1}{x_{N,S}^1} - \frac{1}{x_{N,S}^2} \right)\,,\nn
    \frac{1}{\kappa}\mathcal{E}_{F_2} & = \frac{i}{2} (-1)^{t_{N,S}+1} \mathcal{F}(x_{N,S}^I) \left( \frac{1}{x_{N,S}^2} - \frac{1}{x_{N,S}^3} \right)\,.
\end{align}

Therefore, for given spindle data $n_{N,S}, (-1)^{t_{N,S}}, r_{N,S}$ and freely specified flavour fluxes $p_{F_1}, p_{F_2}$, we have nine algebraic equations:
two from \eqref{eqn:zetaxNS}, two from \eqref{eqn:xIconstrainthyper}, two from \eqref{flavourhyper}, and three from \eqref{conservedhyper}. These can be used to solve for $x^I_{N,S}$ and $k$ and hence obtain the Bekenstein-Hawking entropy \eqref{BHentropy}, without solving the BPS equations, just assuming they exist.

We are only interested in such solutions that satisfy the necessary conditions
\begin{equation}\label{necconds}
	S_{\rm BH} > 0 \,,\qquad \left. (-1)^{t+1} x^I \right\vert_{N,S} > 0 \,.
\end{equation}
We find that closed-form expressions analogous to \eqref{STU_xI} in general involve roots of quartic polynomials and are algebraically complicated. However, many individual solutions to these conditions can be found, both for the anti-twist and the twist classes.
Interestingly, we have found that these necessary conditions are, in general, not sufficient for the existence of fully back-reacted hyperscalar solutions, as we discuss further below, and in particular we only find hyperscalar solutions, by solving
the BPS equations numerically, in the anti-twist class.

Before doing that we obtain some further information from these necessary conditions.
Using \eqref{eqn:zetaxNS} and \eqref{eqn:xIconstrainthyper}, we first observe the following for anti-twist solutions at either pole:
\begin{align}\label{antitwistconstraint}
    -\kappa(-1)^{t+1} = \frac{1}{2r} \Big[ & (r-1) (-1)^{t+1} x^0 + (r+1) (-1)^{t+1} x^1 \nn
    & +(r+1) (-1)^{t+1} x^2+(r+1) (-1)^{t+1} x^3 \Big] \,.
\end{align}
Now $t_{N,S}$ is either $0$ or $1$, and $r_{N,S}\in\mathbb{Z}_{\geq 0}$. Focussing on $\kappa = -1$, since $(-1)^{t+1}x^I>0$, if $r\geq1$ at a pole we have $t=1$. In other words, anti-twist solutions with $t_N = 0, t_S=1$ necessarily have $r_N=0$, while those with $t_N = 1, t_S=0$ necessarily have $r_S=0$. Notice that this exactly mirrors the conclusions below
\eqref{rzeroantitwist} regarding the existence of hyperscalar fluctuations about the STU solutions.

For solutions in the anti-twist class, without loss of generality, for definiteness we again set
$\kappa = -1$ and focus on solutions with $t_N = 1$,  $t_S = 0$ which implies $r_S = 0$
from \eqref{antitwistconstraint}. 
In addition we can also argue that $r_N$ is necessarily even. To see this, we first add the two conditions in
\eqref{pRpBhyper}, and with $r_S=0$ we deduce that $2p^0=(1+r_N) n_S-n_N$. Thus, if $r_N$ were odd then $n_N$ would
be even which contradicts the fact that
$n_N$ and $n_S$ are both odd, as in \eqref{thensareodd} .

To solve the BPS equations for hyperscalar solutions, for coprime spindles, we specify the spindle data $n_N,n_S$, $t_N, t_S$, $r_N, r_S$, with the fluxes $p^I$ constrained by \eqref{pRpBhyper} (so two fluxes are undetermined) which is sufficient to 
determine the boundary conditions at both poles of all fields except the hyperscalar.
We then solve the ODEs numerically, by performing a search over the values of the leading coefficient of the 
hyperscalar at one of the poles as in \eqref{hsexppoles}. In the non-coprime case one should also specify $m_{N,S}^I$, satisfying the necessary smoothness conditions \eqref{eqn:smoothcond}. In addition, the $m_{N,S}^I$ must also satisfy \eqref{eq:spinorregularity} to have well-defined spinors, as well as \eqref{eq:Dthetahyper}, \eqref{Dthetasigma}, such that the hyperscalar is a section of the associated line bundle.

We noted above that solving the algebraic conditions satisfying the necessary conditions \eqref{necconds} is not, in general
sufficient to find an actual hyperscalar solution. However, in all cases we have found that if we impose the further condition
that the STU solution with the same fluxes $p^I$ has a relevant mode with $\delta>0$, then this appears to be sufficient
for the existence of a hyperscalar solution. 
This is highly suggestive that whenever a hyperscalar solution actually exists,
it can be reached via a holographic RG flow from an STU solution with the same $p^I$, perturbed by the relevant hyperscalar mode.
We shall call this the \emph{RG scenario}.
It would be very interesting to construct such RG flow solutions, but this goes well beyond the scope of this paper.
Since the STU solutions in the twist class appear to never have relevant hyperscalar modes, the RG scenario
implies that hyperscalar solutions only exist in the anti-twist class, which are in fact the only ones we have found numerically.
We also emphasise that when the hyperscalar solution exists
it always has a smaller entropy than that of the STU solution:
\begin{align}\label{comparestuhbhentropy}
S^H_{\rm BH}<S^{STU}_{\rm BH}\,.
\end{align}
Thus, assuming the RG scenario, the entropy would always decrease coming from the UV to the IR. 

It is interesting to point out that there are some values of $n_N,n_S$ and fluxes $p^I$, where there
are multiple solutions to the algebraic conditions that satisfy \eqref{necconds}. Furthermore, we find that in
comparing with the analogous STU solution with the same fluxes, one can obtain 
both $S^H_{\rm BH}<S^{STU}_{\rm BH}$ and $S^H_{\rm BH}>S^{STU}_{\rm BH}$. However,
in such cases, we find that when in addition the STU solution has a relevant mode, the only
hyperscalar solution that actually exists is the one with $S^H_{\rm BH}<S^{STU}_{\rm BH}$.
A concrete example is given by
$(n_N, n_S) = (11,1)$ and $p^I = - (4,2,2,2)$,
for which $S_{\rm BH}^{STU} \sim 0.223833 N^{3/2}$ and there is a relevant hyperscalar mode with
$\delta= \frac{1}{2} - \frac{45}{2\sqrt{2081}} \sim 0.00677$.
One finds two solutions to the algebraic conditions for the $x^I$'s, the first of which is
\begin{align}\label{funnycase1}
	k & = 0.693757\,,\nn
    x_N^0 & = 0.606883\,,\quad x_N^1=x_N^2=  x_N^3 =0.377013\,,\nn
    x_S^0 & =-0.441427\,,\quad x_S^1=x_S^2=  x_S^3=-0.147142\,,\nn
    S_{\textrm{BH}}^H & = 0.223813 N^{3/2}\,,
\end{align}
whereas the second solution is given by
\begin{align}\label{funnycase2}
	k & = 0.273792\,,\nn
    x_N^0 & = 0.00388779\,,\quad x_N^1=x_N^2=  x_N^3 =0.444012\,,\nn
    x_S^0 & =-2.65241\,,\quad x_S^1=x_S^2=  x_S^3= -0.884137 \,,\nn
    S_{\textrm{BH}}^H & = 0.556469 N^{3/2}\,.
\end{align}
Notice that both examples are consistent with \eqref{necconds}.
Only the first is consistent with $S^H_{\rm BH}<S^{STU}_{\rm BH}$, and is the one that is associated with an actual hyperscalar solution obtained by solving the BPS equations numerically.

Many of the above aspects regarding the existence of hyperscalar solutions are in direct analogy
with what was seen for the $D=5$ $AdS_3$ solutions in
\cite{Arav:2025jee}. There are a couple of differences. Firstly, in the $D=5$ case one finds that
 there are never examples analogous to \eqref{funnycase1}, \eqref{funnycase2} where solving the necessary algebraic conditions
 gives rise to multiple solutions.
  Secondly, in
 the $D=5$ case, when the hyperscalar solution exists, the fact that the central charge 
 is smaller than the STU solution is consistent with there being an RG flow from the STU solution
 to the hyperscalar solution, due to the monotonicity of the central charge under the RG flow.
 By contrast for the $AdS_2$ solutions we are discussing here, we don't know of an analogous
reason why the entropy should decrease under an ``RG flow" of the dual supersymmetric quantum mechanical theories
(if indeed the flow exists).

\section{Examples of anti-twist hyperscalar solutions}\label{sec:examplesofhssols}
By solving the BPS equations numerically, we have not found any hyperscalar solutions in the twist class and we conjecture that they do not exist.
Furthermore, we only found irrelevant hyperscalar modes in the fluctuations about the STU solutions in the twist class.
Together, these results are consistent with the RG scenario mentioned at the end of the last section.

We therefore focus on anti-twist hyperscalar solutions. We set 
\begin{equation}\label{signchoicehyper}
	\kappa = -1 \,,\quad t_N = 1\,,\quad t_S = 0 \qquad \Rightarrow \qquad r_S=0,\quad  \text{$r_N$ even},\quad  \text{$n_S$ odd}\,.
\end{equation}
with the implications following from the discussion below \eqref{antitwistconstraint}.

We will illustrate with some examples that have three equal fluxes: $p^1=p^2=p^3$.
Recall from the end of section
\ref{stu3pequal} that, in the coprime case,  
the associated STU solutions don't have relevant modes if $n_S$ 
is a multiple of 3 and, consistent with the RG scenario, we never find hyperscalar solutions with
$p^1=p^2=p^3$ and $n_S$ a multiple of 3.
For solutions with $p^1=p^2=p^3$, if we also have $m^1=m^2=m^3$ at both poles then
the uplifted solutions preserve an $SU(3)$ flavour symmetry\footnote{i.e. $SU(3)\times U(1)\times U(1)_R$ symmetry for the STU solutions and $SU(3)\times U(1)_R$ symmetry for the hyperscalar solutions}; this is automatically the
case for coprime spindles but for non-coprime spindles the $SU(3)$ flavour symmetry
can be broken, and indeed must be broken when $n_S$ is a multiple of 3 (which can be deduced using a similar argument to that of section
\ref{stu3pequal}). 

We have summarised many hyperscalar solutions in tables \ref{pbzerotablecoprime}-\ref{noncoprime_pBneq0} below.
These were generated by demanding that all of the necessary algebraic conditions (discussed in section \ref{sec5})
for smooth, supersymmetric hyperscalar solutions are satisfied.
We have verified that many of these solutions exist (and in fact many other solutions, including those without three equal fluxes)
by solving the BPS equations numerically; some further discussion
and representative examples are presented in appendix \ref{plotssols}.
The tables also include the associated STU solutions with the same fluxes, as well as
the spectrum of their hyperscalar modes.
 All examples
are consistent with the RG scenario i.e. for each hyperscalar solution the associated STU solution has a relevant
hyperscalar mode and furthermore $S^H_{BH}<S^{STU}_{BH}$.

\subsection{Hyperscalar non-vanishing at both poles, \texorpdfstring{$p_B = 0$}{pB = 0}}
The solutions with $p_B=0$ necessarily have $r_N = r_S = 0$ from \eqref{pRpBhyper}.
The sub-class of solutions with $p^1=p^2=p^3$ then have $p^0=3p^1$. 
Hyperscalar solutions
that also have $A^1=A^2=A^3$ and $A^0=3A^1$ (which automatically follows in the coprime case)
actually have constant scalars fixed at the mABJM $AdS_4$ vacuum values;
in fact they are solutions
of the minimal gauged supergravity associated with the mABJM vacuum, as
discussed in appendix \ref{appminimalmabjm}. In the coprime case, these
are included in the solutions constructed in \cite{Suh:2022pkg}.

\subsubsection{Coprime spindles, $\hcf(n_N, n_S)=1$}

Various examples are summarised in\footnote{Note that we have included some coprime examples of hyperscalar solutions 
whose existence was suggested by the extremization carried out in section 6.2 of \cite{BenettiGenolini:2024kyy} with $p_B=0$.} table \ref{pbzerotablecoprime}. 
For example, consider the case of $(n_N,n_S)=(11,3)$, $p^I=-(4,1,1,2)$, and $r_N = r_S = 0$.
This example can be compared with an anti-twist STU solution (cf. \eqref{eqn:STU_ex_antitwist_11_3}); specifically
we find that the entropy of the hyperscalar solutions is lower than that of the STU solution and furthermore
the minimal value of $\delta$, when $n'_{KK}=0$ is $\delta = \frac{1}{2} - \frac{11}{\sqrt{721}} \sim 0.0903$, corresponding
to a relevant mode. This is consistent with the RG scenario outlined above. In all cases
\eqref{necconds}, \eqref{comparestuhbhentropy} are satisfied. 
As noted above the solutions with $p^1=p^2=p^3$ in this table are actually solutions of minimal gauged supergravity associated with the mABJM vacuum, with the hyperscalar taking its constant vacuum value, as in appendix \ref{appminimalmabjm}.

\begin{table}[h]
\centering
\scalebox{0.85}{
\begin{tabular}{ |c|c|c|c|c|c|c| }
\hline
$(n_N,n_S)$ & $p^I$ & $r_N$ & $r_S$ & $\delta$ & $\frac{1}{N^{3/2}} S_{\rm BH}^{STU}$ & $\frac{1}{N^{3/2}} S_{\rm BH}^H$ \\
\hline
$(7,1)$ & $-(3,1,1,1)$ & $7 n_{\rm KK}$ & $n_{\rm KK}$& $0.152 -\frac{28}{\sqrt{349}} n_{\rm KK}$ & $0.169$ & $0.163$  \\ 
$(9,1)$ & $-(4,1,1,2)$ & $9 n_{\rm KK}$ & $n_{\rm KK}$ & $ 0.161 -\frac{45}{2\sqrt{217}} n_{\rm KK}$ & $0.171$ & $0.165$  \\ 
$(11,1)$ & $-(5,1,1,3)$ &$11\nkk$&$\nkk$& $ 0.157 -\frac{66}{\sqrt{1789}} \nkk$& $0.160$ & $0.155$   \\ 
$(11,1)$ & $-(5,2,2,1)$ &$11\nkk$&$\nkk$& $ 0.175 -\frac{33}{4\sqrt{29}} \nkk$& $0.182$ & $0.176$   \\ 
$(11,3)$ & $-(4,1,1,2)$ & $11 n_{\rm KK}'$ & $3 n_{\rm KK}'$& $0.0903-\frac{33\sqrt{7}}{2\sqrt{103}} n_{\rm KK}'$ & $0.0346$ & $0.0339$   \\ 
$(13,1)$ & $-(6,1,1,4)$ &$13\nkk$&$\nkk$ & $ 0.150  -\frac{91}{2\sqrt{817}} \nkk$& $0.147$ & $0.144$  \\ 
$(13,1)$ & $-(6,1,2,3)$ &$13\nkk$&$\nkk$  & $0.177  -\frac{91}{24\sqrt{6}} \nkk$ & $0.177$ & $0.171$  \\ 
$(13,1)$ & $-(6,2,2,2)$ &$13\nkk$&$\nkk$ & $0.189  -\frac{91}{\sqrt{3529}} \nkk$ & $0.203$ & $0.195$   \\ 
$(13,5)$ & $-(4,1,1,2)$ & $13 n_{\rm KK}'$ & $5 n_{\rm KK}'$& $0.0572 - \frac{585}{2\sqrt{1841}} n_{\rm KK}'$ & $0.0139$ & $0.0137$  \\ 
$(23,5)$ & $-(9,3,3,3)$ &$23 \nkk''$&$5\nkk''$ & $ 0.117 - \frac{1610}{\sqrt{48757}} \nkk''$& $0.0270$ & $0.0262$   \\ 
\hline
\end{tabular}
}
\caption{Examples of coprime, hyperscalar solutions with $p_B=0$ all in the anti-twist class. 
The KK spectrum of the BPS hyperscalar perturbation of the associated STU solution is given, as well
as the entropies of the STU solution and the hyperscalar solution.
As before, $n_{\rm KK}' \equiv n_{\rm KK}+1$ and $\nkk'' = \nkk +2$;
whichever quantity $n_{KK}$, $\nkk'$, $n_{KK}''$ that appears is an integer that is $ \ge 0$. The solutions with $p^1=p^2=p^3$ are solutions
of minimal gauged supergravity as in appendix \ref{appminimalmabjm}.}
\label{pbzerotablecoprime}
\end{table}

\subsubsection{Non-coprime spindles}

Next, we consider some non-coprime examples obtained as
 $h$-fold flux quotients, with $h$ odd from \eqref{hodd}. 
Some interesting solutions with $h=3$ are presented in table \ref{table339}.
The first three entries preserve an $SU(2)$ flavour symmetry: it is only the first of these that has an STU solution
with a relevant hyperscalar mode and correspondingly it is only
for this case that a hyperscalar solution exists, with smaller entropy. The bottom two entries in the table break
the $SU(2)$ flavour symmetry and are related by exchanging the values of $m^1$ and $m^2$, with
all other values remaining the same. In this case the associated STU solutions have a relevant mode
and there is a hyperscalar solution with smaller entropy.
\begin{table}[h!]
\centering
 $(n_N,n_S)=3(11,3),~p^I = -3(4,1,1,2)$\\ \smallskip
\scalebox{0.85}{
\begin{tabular}{ |c|c|c|c|c|c|c| } 
 \hline
 $m_N^I$ &  $m_S^I$ & $r_N$&$r_S$& $\delta$& $\frac{1}{N^{3/2}} S_{\rm BH}^{STU}$&$\frac{1}{N^{3/2}} S_{\rm BH}^H$\\ 
 \hline 
 $(16,4,4,8)$ & $(4,1,1,2)$ & $33 n_{KK}$&$9 n_{KK}$&  $ 0.0903%386
  - \frac{693}{2\sqrt{721}} n_{KK}$&$0.0115%315
  $& $0.0113%047
  $\\ 
$(5,4,4,19)$ & $(1,1,1,5)$ & $22 + 33 n_{KK}$&$6 + 9 n_{KK}$& $-8.51%255
 - \frac{693}{2\sqrt{721}} n_{KK}$&$0.0115%315
 $& $-$\\ 
$(5,26,26,8)$ & $(1,7,7,2)$ & $22 + 33 n_{KK}'$&$6+25n_{KK}'$& $-8.51%255
- \frac{693}{2\sqrt{721}} n_{KK}'$&$0.0115%315
$& $-$\\ 
\hline
\hline
$(16,26,4,19)$ & $(4,7,1,5)$ & $33n_{KK}'$&$9 n_{KK}'$& $0.0903%386
 - \frac{693}{2\sqrt{721}} n_{KK}'$& $0.0115%315
 $& $0.0113%047
 $\\ 
$m^1 \leftrightarrow m^2$& $m^1 \leftrightarrow m^2$ & $\dots$&$\dots$&$\dots$& $\dots$& $\dots$\\ 
 \hline
\end{tabular}%
}
\caption{
Examples of non-coprime hyperscalar solutions with $p_B=0$.
 As before
$n_{KK}'\equiv n_{KK}+1$.
The top 3 STU solutions preserve an $SU(2)$ flavour symmetry, while the bottom 2 solutions break the $SU(2)$ flavour symmetry:
only one solution given, the other is obtained by exchanging $m^1 \leftrightarrow m^2$ with all other entries identical.
\label{table339}}
 \end{table}

We highlight that some examples of non-coprime cases with $p_B=0$ were discussed in section 6 of \cite{BenettiGenolini:2024kyy}, 
based on extremizing an entropy function (as in section \ref{section:equivlocal}). However, it turns out that none of those
specific examples satisfy the conditions for the spindle solution to be a smooth supersymmetric solution (which weren't known for non-coprime spindles
at the time of \cite{BenettiGenolini:2024kyy}). For example, we can consider the case given in (6.25) of \cite{BenettiGenolini:2024kyy}
which had $(n_N, n_S) = (15,3)$ and $p^I = - (6,3,2,1)$. Since $h = \hcf{(15,3)}=3$, for a smooth solution from
\eqref{eq:hfoldreduc} we would require, in particular, that all $p^I$ are divisible by $3$ which is not true. For similar reasons,
none of the non-coprime cases $(n_N,n_S) = (8,2) ,\, (10,2) ,\, (9,3)$ spindles presented in Table 1 of \cite{BenettiGenolini:2024kyy}
correspond to smooth solutions.

Finally, we summarise some features of other non-coprime solutions in table \ref{noncoprime_pBeq0sec612}. Specifically, we consider $h$-fold flux quotients of
the case $(n_N,n_S)=(23,5)$ 
and $p^I = -(9,3,3,3)$ of table \ref{pbzerotablecoprime}, for various $h$ (that is necessarily odd from \eqref{hodd}).
We list the number of STU solutions that preserve an $SU(3)$ flavour symmetry
and those that preserve an $SU(2)$ flavour symmetry (but not also an $SU(3)$ flavour symmetry). We also list similar numbers for
hyperscalar solutions, when they exist.
\begin{table}[h]
\centering
\scalebox{0.85}{
\begin{tabular}{ |c|c|c|c||c|c|c|}
\hline
$(n_N,n_S)$ & STU& $SU(2)$ inv't.  & $SU(3)$ inv't  & Hyper & $SU(2)$ & $ SU(3)$ \\
\hline
$\mathrm{even}(23,5)$ & $0$ & $-$ & $- $& $0$ & $-$ & $- $  \\ 
$\mathrm{3}(23,5)$ & $5$ & $3$ & $2$& $3$ & $3$ & $0 $  \\ 
$\mathrm{5}(23,5)$ & $51$ & $30$ & $3$& $13$ & $6$ & $1 $  \\ 
$\mathrm{7}(23,5)$ & $89$ & $36$ & $5$& $19$ & $6$ & $1 $  \\ 
$\mathrm{9}(23,5)$ & $135$ & $63$ & $6$& $27$ & $9$ & $0 $  \\ 
%$3(11,1)$ & $1$ & $0$ & $$  \\ 
%$5(11,1)$ & $17$ & $7$ & $$  \\ 
%$7(11,1)$ & $$ & $$ & $$  \\ 
%$9(11,1)$ & $$ & $$ & $$  \\ 
%$11(11,1)$ & $$ & $$ & $$  \\ 
\hline
\end{tabular}
}
\caption{Examples of non-coprime, anti-twist, hyperscalar solutions with $p_B =0$, obtained by an $h$-fold flux quotient of $(n_N,n_S)=(23,5)$, $p^I = -(9,3,3,3)$ of table \ref{pbzerotablecoprime}.}\label{noncoprime_pBeq0sec612}
\end{table}

Notice that these are $h$-fold flux quotients of a coprime case with $p^1=p^2=p^3$.
Recall from the end of section
\ref{stu3pequal} we showed that when $n_S$ is a multiple of 3, so in the table $h=3$ or $9$,
the STU solutions do not have any relevant hyperscalar deformations that preserve the $SU(3)$ flavour symmetry
(after uplifting)
and, consistent with the RG scenario, we see that there are no associated hyperscalar solutions preserving $SU(3)$.
However, for $h=3,9$ there are STU solutions with relevant modes that break the $SU(3)$ to $SU(2)\times U(1)$ or $U(1)^2$ and
correspondingly these are associated with the hyperscalar solutions in the table.

\subsection{Hyperscalar vanishing at one of the poles, {$p_B \neq 0$}}

We now consider examples of hyperscalar solutions with $p_B\ne 0$.
Recalling \eqref{signchoicehyper}, we necessarily have $r_S=0$, so the hyperscalar is non-zero at the $S$ pole,
and at the $N$ pole where the hyperscalar vanishes, $r_N>0$ is even.

\subsubsection{Coprime spindles, $\hcf(n_N, n_S)=1$}
We first consider the coprime case. Various examples are summarised in table \ref{tablecoprimepbnonzero}.
The bottom entries have $p^1=p^2=p^3$.
\begin{table}[h]
\centering
\scalebox{0.85}{
\begin{tabular}{ |c|c|c|c|c|c|c| }
\hline
$(n_N,n_S)$ & $p^I$ & $r_N$ & $r_S$ & $\delta$ & $\frac{1}{N^{3/2}} S_{\rm BH}^{STU}$ & $\frac{1}{N^{3/2}} S_{\rm BH}^H$ \\
\hline
$(29,1)$ & $-(13,5,6,4)$ & $2+29n_{\rm KK}$ & $n_{\rm KK}$& $0.138 - \frac{435}{2 \sqrt{20641}} n_{\rm KK}$ & $0.2358$ & $0.2304$  \\ 
$(31,1)$ & $-(13,5,7,5)$ & $4 + 31 n_{\rm KK}$ & $n_{\rm KK}$& $0.0789 - \frac{496}{\sqrt{111309}} n_{\rm KK}$ & $0.2470$ & $0.2450$  \\ 
$(47,1)$ & $-(22,4,7,13)$ & $2 + 47 n_{\rm KK} $ & $n_{\rm KK}$& $0.156 - \frac{188}{\sqrt{14569}} n_{\rm KK}$ & $0.2080$ & $0.2029$  \\ 
$(53,3)$ & $-(22,2,7,19)$ & $2 + 53 \nkk'$ & $3 \nkk'$ & $0.0210 - \frac{1113}{\sqrt{56137}} n_{\rm KK}'$  & $0.04614$ & $0.04612$\\ 
\hline
\hline$(11,1)$ & $-(4,2,2,2)$ & $2+11 \nkk $ & $  n_{\rm KK}$ & $0.00677 - \frac{66}{\sqrt{2081}} \nkk$& $0.22383$ & $0.22381$  \\ 
$(17,1)$ & $-(7,3,3,3)$ & $2+17 \nkk $ & $  n_{\rm KK}$ & $0.079 -\frac{9}{\sqrt{37}} \nkk$& $0.2335$ & $0.2314$  \\ 
$(21,1)$ & $-(8,4,4,4)$ & $4+21 \nkk$ & $  n_{\rm KK}$& $0.0179 -\frac{231}{\sqrt{25177}} \nkk$ & $0.2509$ & $0.2507$  \\ 
$(61,5)$ & $-(23,11,11,11)$ & $2+61 \nkk''$ & $  5n_{\rm KK}''$ & $0.0262 - \frac{3355}{\sqrt{212653}} \nkk''$& $0.04533$ & $0.04527$   \\ 
$(67,5)$ & $-(26,12,12,12)$ & $2+67 \nkk''$ & $  5n_{\rm KK}''$ & $0.043 - \frac{12060}{\sqrt{2720497}} \nkk''$& $0.04578$ & $0.04564$   \\ 
\hline
\end{tabular}
}
\caption{
Some coprime examples of hyperscalar solutions with $p_B \neq 0$ in the anti-twist class. 
As before, $n_{\rm KK}' \equiv n_{\rm KK}+1,~ \nkk'' = \nkk+2$.}
\label{tablecoprimepbnonzero}
\end{table}

\subsubsection{Non-coprime spindles}

We now turn to non-coprime spindles with $p_B \neq 0$ and $h$ odd from \eqref{hodd}.
In table \ref{noncoprime_pBneq0} we illustrate with various $h$-fold quotients of the $(n_N,n_S) = (11,1)$ 
and $(21,1)$ examples in table \ref{tablecoprimepbnonzero} (with three equal $p^I$). In particular we present the number of solutions that
are invariant under an $SU(3)$ flavour symmetry and also the
number that are invariant under an $SU(2)$ flavour symmetry (but not also an $SU(3)$).
Interestingly, the number of solutions is not monotonic with increasing $h$. 

These are again $h$-fold flux quotients for coprime spindles with 3 equal $p^I$. 
Recalling the discussion at the end of section \ref{stu3pequal}, it is possible 
to have non-coprime STU solutions with relevant hyperscalar modes and hence, via the RG scenario,
hyperscalar solutions with $h$ a multiple of 3 but only if the $SU(3)$ flavour symmetry is broken
as we see in the bottom of table \ref{noncoprime_pBneq0}.
\begin{table}[h]
\centering
\scalebox{0.85}{
\begin{tabular}{ |c|c|c|c||c|c|c|}
\hline
$(n_N,n_S)$ & STU& $SU(2)$ inv't.  & $SU(3)$ inv't  & Hyper & $SU(2)$ & $ SU(3)$ \\
\hline
$\mathrm{even}(11,1)$ & $0$ & $-$ & $- $& $0$ & $-$ & $- $  \\ 
$\mathrm{3}(11,1)$ & $1$ & $0$ & $0$& $0$ & $-$ & $- $  \\ 
$\mathrm{5}(11,1)$ & $17$ & $9$ & $2$& $7$ & $0$ & $1 $  \\ 
$\mathrm{7}(11,1)$ & $91$ & $39$ & $4$& $16$ & $9$ & $1 $  \\
$\mathrm{9}(11,1)$ & $27$ & $18$ & $3$& $0$ & $-$ & $- $  \\  
\hline
\hline
$\mathrm{even}(21,1)$ & $0$ & $-$ & $-$& $0$ & $-$ & $- $  \\  
$\mathrm{3}(21,1)$ & $5$ & $3$ & $2$& $3$ & $3$ & $0 $  \\  
$\mathrm{5}(21,1)$ & $16$ & $15$ & $1$& $0$ & $-$ & $- $  \\  
$\mathrm{7}(21,1)$ & $185$ & $78$ & $5$& $31$ & $12$ & $1 $  \\  
$\mathrm{9}(21,1)$ & $135$ & $63$ & $6$& $27$ & $9$ & $0 $  \\  
%$3(11,1)$ & $1$ & $0$ & $$  \\ 
%$5(11,1)$ & $17$ & $7$ & $$  \\ 
%$7(11,1)$ & $$ & $$ & $$  \\ 
%$9(11,1)$ & $$ & $$ & $$  \\ 
%$11(11,1)$ & $$ & $$ & $$  \\ 
\hline
\end{tabular}
}
\caption{Examples of non-coprime, anti-twist, hyperscalar solutions with $p_B\neq 0$, obtained by an $h$-fold flux quotient of the
$(n_N,n_S)=(11,1)$, $p^I = -(4,2,2,2)$ and $(n_N,n_S)=(21,1)$, $p^I = -(8,4,4,4)$ solutions of table \ref{tablecoprimepbnonzero}.
}\label{noncoprime_pBneq0}
\end{table}

\section{Equivariant Localization}
\label{section:equivlocal}

In this section we obtain the Bekenstein-Hawking entropy of spindle black holes with non-trivial hyperscalars. 
While the analysis is parallel to that outlined in subsections \ref{subsection:BPSanalysis}-\ref{sec:smooth}, we shall see how an extremization principle can be utilised without solving the BPS equations, just assuming the solutions exist.
This is a direct generalisation of the $D=4$ analysis in \cite{BenettiGenolini:2024kyy}, which considered the case when the hyperscalar was non-vanishing at both poles; here we now allow for the hyperscalar to vanish at the poles.

Our starting point is the metric ansatz in the conformal gauge \eqref{eqn:AdS2ansatz}, which we rewrite as
\begin{equation}
\label{eqn:AdS2Sig2ansatz}
    {\rd} s_4^2 = e^{2V} \left[{\rd} s^2(AdS_2) + {\rd} s^2(\Sigma) \right] \, , \qquad {\rd} s^2(\Sigma) = {\rd} y^2+g(y)^2 {\rd} z^2\,.
\end{equation}
We assume the $D=4$ Killing spinor has the form
\begin{equation}
    \epsilon = \vartheta \otimes e^{V/2} \zeta \,,
\end{equation}
where $\zeta$ is a spinor on $\Sigma$ as in \eqref{eqn:spindlespinor}. The spinor $\vartheta$ on $AdS_2$ satisfies \eqref{eqn:AdS2spinor} as before, where we fix
\begin{equation}
\label{eqn:kappachoice}
    \kappa = -1 \,,
\end{equation}
following the choice in \cite{BenettiGenolini:2024kyy}.
Imposing the metric and Killing spinor ans\"atze on the $D=4$ action and Killing spinor equations (\ref{eqn:4daction}), (\ref{eqn:4dKSEs}) gives a $D=2$ action on $\Sigma$ and a set of $D=2$ Killing spinor equations that $\zeta$ must satisfy.
Further imposing the equation of motion for $V$ on the $D=2$ effective action, we obtain the following ``partially off-shell action" \cite{BenettiGenolini:2024kyy}:
\begin{equation}
\label{eqn:POS}
    S_2|_{\textrm{POS}} = \int_{\Sigma} \Big[ e^{4V} \textrm{vol}_2 - \frac{1}{2} \sum_{I=0}^3 (X^I)^{-2} F_{12}^I F^I \Big]\,,
\end{equation}
where $F^I \equiv F^I_{12} \textrm{vol}$, with $F^I_{12}$ a function on $\Sigma$. If (\ref{eqn:AdS2Sig2ansatz}) is the near-horizon limit of an extremal black hole, (\ref{eqn:POS}) can be used to define an off-shell entropy function
\begin{equation}
    S_{\textrm{BH}} = -\frac{1}{8G_4} S_2|_{\textrm{POS}} \,,
\end{equation}
which, on-shell, is the Bekenstein-Hawking entropy of the black hole.

Recall we have the following real bilinears in $\zeta$:
\begin{equation}
\label{eqn:bilinears}
    S = \zeta^\dagger \zeta = 1\,,\qquad P = \zeta^\dagger \gamma_3 \zeta = - \cos \xi \,,\qquad \xi^\mu \partial_\mu = - i \zeta^\dagger \gamma^\mu\gamma_3\zeta \partial_\mu = \frac{1}{k} \partial_z \,.
\end{equation}
These satisfy a set of algebraic and differential conditions implied by the $D=2$ Killing spinor equations, which can be found in Appendix B of \cite{BenettiGenolini:2024kyy}.
In particular, one finds that $\xi^\mu\partial_\mu$ is precisely the azimuthal Killing vector that rotates $\Sigma$, and that the one-form $\xi^\flat$ metrically dual to the Killing vector satisfies
\begin{equation}
\label{eqn:dxiflat}
    {\rd} \xi^\flat = 2 \left(1+e^V \mathcal{W}PS^{-1} \right) P\,\textrm{vol}_2\,.
\end{equation}
The differential conditions satisfied by the bilinears imply that the bilinear $S$ is constant, and without loss of generality we have normalised it to be 1. At fixed points it can be shown that $P^2=S^2=1$, so $P=\pm1$. Notice from the definition of $P$ this sign is associated with the chirality of the spinor $\zeta$ at a fixed point. We leave $P_S$ and $\sigma \equiv P_S/P_N$ arbitrary for now.\footnote{To compare with the language of section \ref{subsection:BCs}, we have $P_{N,S} = (-1)^{t_{N,S}+1}$.}

\subsection{Localization}
Constructed as a spinor bilinear, the Killing vector $\xi$ is globally defined \cite{Ferrero:2021etw}. To make connections with \cite{BenettiGenolini:2024kyy}, we write
\begin{equation}
    \xi = b_0 \partial_z\,,
\end{equation}
where $b_0 = 1/k$ is a constant. As before, $\Delta z=2\pi$, and we take $y\in[y_N,y_S]$, writing $y_{N,S}$ as the location of the poles of the spindle. We may locally model the neighbourhood of the poles as $\mathbb{R}^2/\mathbb{Z}_{n_{N,S}}$, where $n_N,n_S$ are positive integers. In doing so, we assume that 
$g(y)\rightarrow \frac{1}{n_N}(y-y_N)$ and $g(y) \rightarrow \frac{1}{n_S} (y_S-y)$ as one approaches $y_N$ and $y_S$, respectively.

The BVAB theorem allows us to evaluate the integral of the two-form component of equivariantly closed forms, in terms of the zero-form component evaluated on the fixed points of $\xi$. That is, given a polyform $\Phi\equiv \Phi_{(2)}+\Phi_{(0)}$ that satisfies
\begin{equation}
    {\rd} _\xi \Phi = -\xi\hook \Phi_{(2)} + {\rd} \Phi_{(0)} = 0\,,
\end{equation}
we have
\begin{equation}
    \int_\Sigma \Phi = \sum_{\textrm{fixed pt}} \frac{1}{n_{N,S}}\frac{2\pi}{\epsilon_{N,S}} \left.\Phi_{(0)}\right\vert_{N,S}\,,
\end{equation}
where $\epsilon_{N,S}$ are the weights of the action of $\xi$ at the two poles. Recalling that ${\rd} \xi^\flat|_{N,S} = 2\epsilon_{N,S} \textrm{vol}_2$, we find that the weights are given as $\epsilon_{N,S}= \pm b_0/n_{N,S}$.

Several equivariantly closed forms were constructed in \cite{BenettiGenolini:2024kyy}, using the bilinears above and the supergravity fields. Associated with action (\ref{eqn:POS}), the polyform
\begin{equation}
\label{eqn:PhiS}
    \Phi^S \equiv e^{4V}\textrm{vol}_2 - \frac{1}{2} \sum_{I=0}^3 (X^I)^{-2}F_{12}^I F^I + e^{3V}\mathcal{W}S + \frac{1}{2} e^V P \sum_{I=0}^3 (X^I)^{-1} F^I_{12}\,,
\end{equation}
is equivariantly closed. Associated with the gauge fields,
\begin{equation}
\label{eqn:PhiF}
    \Phi^I \equiv F^I - X^I e^V P \,,
\end{equation}
are equivariantly closed, where
\begin{equation}
\label{eqn:xIdef}
    x^I \equiv X^I e^V P\,,
    \end{equation}
is precisely the ``dressed scalar" introduced in \eqref{eqn:dressedscalar}, with $P=-\cos\xi$.

By applying the BVAB formula to the magnetic fluxes through the spindle, we find
\begin{equation}
\label{eqn:BVAB_F}
    \frac{p^I}{n_Nn_S} \equiv \frac{1}{4\pi}\int_\Sigma F^I = \frac{1}{2b_0} (x_S^I - x_N^I) \,.
\end{equation}
Applying the BVAB formula to (\ref{eqn:PhiS}), one finds \cite{BenettiGenolini:2024kyy}:
\begin{equation}
\label{eqn:BVAB_S}
    S_{\textrm{BH}} = -P_S \frac{\pi\sqrt{2}}{3} N^{3/2} \frac{1}{b_0} \left(\sqrt{x^0_Sx^1_Sx^2_Sx^3_S} - \sigma \sqrt{x^0_Nx^1_Nx^2_Nx^3_N}  \right)\,.
\end{equation}
Finally, the following linear combination of the $\Phi^I$, corresponding to the $U(1)$ the hyperscalar transforms under, satisfies
\begin{equation}
\label{eqn:equivhyper}
    \zeta_I \Phi^I = {\rd}_{\xi} D\theta \,.
\end{equation}
If $D\theta$ is globally defined, $\zeta_I\Phi^I$ is equivariantly exact. That is, if we assume the hyperscalar to be non-vanishing
at both poles, the broken flux $p_B \equiv 2\zeta_I p^I$ through the spindle $\Sigma$ is necessarily zero. In what follows, we will be lifting this assumption, allowing for $\rho$ to smoothly go to zero at one or both of the poles.

The value of the superpotential $\mathcal{W}$ (\ref{eqn:superpotential}) at the poles only depends on the real scalars $\varphi^i$ in the vector multiplets:
\begin{equation}
\label{eqn:Watpoles}
    \mathcal{W}|_{N,S} = -\frac{1}{2} \sum_{I=0}^3 X^I|_{N,S} \,.
\end{equation}
This is obvious from (\ref{eqn:superpotential}) when $\rho=0$ at the poles, but it is also true when $\rho$ is non-zero at the poles. In that case, regularity implies $D\theta|_{N,S} = 0$, and from the zero-form component of (\ref{eqn:equivhyper}) we see that $\zeta_I X^I|_{N,S} = 0$.

Next, using (\ref{eqn:Watpoles}), we can evaluate (\ref{eqn:dxiflat}) near the poles. This gives
\begin{equation}
\label{eqn:Rsymmconstraint}
    \sum_{I=0}^3 x_S^I = 2+P_S\frac{2b_0}{n_S}\,,\qquad \sum_{I=0}^3x_N^I = 2-\sigma P_S\frac{2b_0}{n_N}\,.
\end{equation}
Using (\ref{eqn:BVAB_F}), we obtain the following constraint on the R-symmetry flux:
\begin{equation}
\label{eqn:constraint_R}
    p_R \equiv \sum_{I=0}^3 p^I = P_S (n_N+\sigma n_S)\,.
\end{equation}

As mentioned below \eqref{eqn:equivhyper}, if we assume that the hyperscalar is non-vanishing at the poles, the broken flux $p_B$ is necessarily zero. These were the considerations in e.g. \cite{Arav:2022lzo,Suh:2022pkg,BenettiGenolini:2024kyy}. Instead, we are interested in cases where the hyperscalars smoothly go to zero, in the orbifold sense, at one or both of the poles. Recalling discussions in subsection \ref{sec:smooth}, from \eqref{Dthetasigma} we have
\begin{equation}
    D\theta|_S = P_S \frac{r_S}{n_S} {\rd} z\,,\qquad D\theta|_N = -P_N\frac{r_N}{n_N} {\rd} z\,,
\end{equation}
with $r_{N,S} \in \mathbb{Z}_{\geq 0}$ describing how the hyperscalar vanishes at the poles. The zeroth component of (\ref{eqn:equivhyper}) then gives the constraints
\begin{equation}
\label{eqn:brokensymmconstraint}
    \zeta_Ix^I_S = b_0 P_S \frac{r_S}{n_S}\,,\qquad \zeta_I x^I_N = - b_0 P_N \frac{r_N}{n_N}\,.
\end{equation}
From (\ref{eqn:BVAB_F}), we then see that the broken flux is given by
\begin{equation}
\label{eqn:constraint_B}
    p_B = P_S r_Sn_N + P_Nr_Nn_S \,.
\end{equation}
Observe that by setting $r_{N,S}=0$, we recover the conditions used in \cite{Arav:2022lzo,BenettiGenolini:2024kyy} when the complex scalar is non-vanishing at both poles of the spindle. Note the conditions (\ref{eqn:Rsymmconstraint}), (\ref{eqn:constraint_R}), (\ref{eqn:brokensymmconstraint}), (\ref{eqn:constraint_B}), as well as (\ref{eqn:BVAB_F}), (\ref{eqn:BVAB_S}) are invariant under
\begin{equation}
\label{eqn:flippsymm}
    x_N^I \leftrightarrow x_S^I\,,\quad P_N \leftrightarrow P_S\,,\quad n_N \leftrightarrow n_S \,,\quad r_N\leftrightarrow r_S\,,\quad b_0 \leftrightarrow -b_0\,,
\end{equation}
with the fluxes left unchanged. This symmetry can be utilised to relate anti-twist class solutions with $P_N=-P_S=1$ and $P_N=-P_S=-1$.

Given the spindle data $n_{N,S}$, $P_{N,S}$, and $r_{N,S}$, the R-symmetry flux and the broken flux are determined by (\ref{eqn:constraint_R}) and (\ref{eqn:constraint_B}), respectively. This leaves two independent fluxes $p_{F_1}$ and $p_{F_2}$, associated with (\ref{eqn:flavoursymm}). The action (\ref{eqn:BVAB_S}) depends on 9 variables, $b_0$ and $x^I_{N,S}$, where we can eliminate $x_N^I$ using (\ref{eqn:BVAB_F}). The constraints (\ref{eqn:Rsymmconstraint}) and (\ref{eqn:brokensymmconstraint}) determine two of the $x_S^I$. Recalling that the action (\ref{eqn:BVAB_S}) is only partially on-shell, we now extremize over the remaining three variables ($b_0$ and the remaining two of the $x_S^I$) to get the on-shell action and hence the Bekenstein-Hawking entropy. Extremal values of $x^I_{N,S}$ then determine the values at the poles of the warp factor and the vector multiplet scalars, from $X^I|_{N,S}=Pe^{-V}x^I|_{N,S}$ and $e^{4V}|_{N,S} = x^0x^1x^2x^3|_{N,S}$, where the second comes from the constraint (\ref{eqn:sectionsconstraint}).\footnote{For the STU model, with zero hyperscalar, there is an analogous extremization problem \cite{BenettiGenolini:2024kyy}. In this case there is no analogue of the constraint \eqref{eqn:brokensymmconstraint} and so one extremizes
over 4 independent variables, which we may take to be $b_0$ and three of the independent $x_N^I$; we have checked for explicit examples that the result agrees with \eqref{STU_xI}.}

\subsection{Extremization}
We now explicitly study the extremization problem constructed in the previous subsection.
There are immediate positivity constraints we should take into account, given they are necessary conditions for a solution to exist. These include
\begin{equation}\label{posconstraints}
    S_{\textrm{BH}} > 0\,,\qquad P_{N,S} x^I_{N,S} > 0\,,
\end{equation}
where the latter follows from (\ref{eqn:xIdef}). In addition, for given $P_{N,S}$, from (\ref{eqn:Rsymmconstraint}) we have
\begin{equation}
\label{eqn:twistpositivity}
    P_N=P_S= +1:\quad b_0<n_N\,,\quad P_N=P_S=-1:\quad b_0<-n_N\,,
\end{equation}
for the twist class ($\sigma=+1$), where only the former is allowed, given that we set $\kappa = -1$.
For the anti-twist class ($\sigma=-1$), we have
\begin{equation}
\label{eqn:antitwistpositivity}
    P_N=-P_S=+1:\quad n_S<b_0<n_N\,, \quad P_N=-P_S = -1: \quad -n_S<b_0<-n_N\,.
\end{equation}
The two anti-twist cases can be related by \eqref{eqn:flippsymm}. The last case corresponds to the one studied in section 6.2 of \cite{BenettiGenolini:2024kyy}, where we again stress that there the hyperscalar was assumed to be non-vanishing at the poles. Here we are assuming that the hyperscalar vanishes at the poles, and we need to input how they vanish via the data $r_{N,S}$.

The fluxes $(p^0,p^3)$ are determined from $(p^1,p^2)$ via
\begin{align}\label{hyperconstraints}
    p^0 & = \frac{1}{2}(P_Sn_N+P_Nn_S)+\frac{1}{2}(P_Sr_Sn_N+P_Nr_Nn_S) \,,\nn
    p^3 & = p^0-p^1-p^2 - (P_Sr_Sn_N+P_Nr_Nn_S)\,,
\end{align}
from (\ref{eqn:constraint_R}) and (\ref{eqn:constraint_B}).

We have the following general result for the anti-twist class (analogous to what was observed in \cite{Arav:2025jee}), following
from (\ref{eqn:Rsymmconstraint}) and (\ref{eqn:brokensymmconstraint}):
\begin{equation}
\label{eqn:antitwistconstraint}
    P_{N,S} = \left. \frac{1}{2r} \left[ (r-1)Px^0 + (r+1)Px^1+(r+1)Px^2+(r+1)Px^3 \right]\right\vert_{N,S} \,,
\end{equation}
at either pole. Recall that $P_{N,S}$ is either $+1$ or $-1$, and $r_{N,S}\in\mathbb{Z}_{>0}$. Since $Px^I>0$, if $r\geq1$ at a pole we have $P=+1$. In other words, anti-twist solutions with $P_N=-P_S=-1$ necessarily have $r_N=0$, while those with $P_N=-P_S=+1$ necessarily have $r_S=0$.

The necessary conditions \eqref{posconstraints} are not sufficient for the existence of solutions. 
For example, in the anti-twist class for $(n_N, n_S) = (11,1)$ and $p^I = - (4,2,2,2)$ 
one can obtain multiple solutions from extremisation that we already saw in
 \eqref{funnycase1},
\eqref{funnycase2}. As discussed earlier, only
\eqref{funnycase1} is associated with an actual hyperscalar solution. 
For an example
 in the twist class, consider
\begin{equation}
	\kappa = -1 \,,\qquad P_N = P_S = +1 \,,
\end{equation}
Taking $(n_N,n_S)=(17,3)$, $p^I=(37,-5,-1,-11)$ (cf. \eqref{eqn:STU_ex_twist_17_3}), and $r_N = 1,\, r_S = 3$, 
from extremization
we get
\begin{align}\label{eqn:hyper_extr_ex_twist_17_3}
    k & = 0.901832\,,\nn
    x_N^0 & = 0.869547\,,\quad x_N^1=0.275624\,,\quad x_N^2= 0.0532762\,,\quad x_N^3 = 0.671100\,,\nn
    x_S^0 & = 2.47847\,,\quad x_S^1=0.0582015\,,\quad x_S^2= 0.00979179\,,\quad x_S^3=0.192771\,,\nn
    S_{\textrm{BH}} & = 0.101595 N^{3/2}\,.
\end{align}
The $x^I$ and $S_{\textrm{BH}}$ are all positive, i.e. they satisfy \eqref{posconstraints}. This example can be compared with the STU twist solution \eqref{eqn:STU_ex_twist_17_3}; notice that $S_{\rm BH}^{STU} > S_{\rm BH}^H$.
However, we have not found any hyperscalar solutions in the twist class.
As discussed in section \ref{sec5}, what seems to be a sufficient condition for the existence of a hyperscalar solution
is that the necessary conditions \eqref{posconstraints} are satisfied and in addition, the associated STU
solution with the same charges has a relevant hyperscalar mode. Here,
the associated STU solution only has irrelevant hyperscalar modes (the mode with the smallest value of $\Delta=1-\delta$ has
$\delta = -0.275607$). This is again consistent with the RG scenario. 

Finally, we highlight that in the localization approach of this section, we did not utilise the conserved charges 
\eqref{eqn:fluxconstraint3} for the hyperscalar solutions (or \eqref{eqn:fluxconstraint4} for the STU solutions).
This is the reason that one still has an extremization to carry out.

\section{Final comments}\label{sec:finalcomments}

We have constructed rich new classes of $AdS_2\times \Sigma$ solutions where $\Sigma$ is a spindle.
The solutions have been constructed in the $D=4$ STU gauged supergravity model coupled to a charged hyperscalar.
We have focussed on solutions that uplift to supersymmetric  solutions of $D=11$ supergravity that are regular with
globally defined spinors.

We have constructed both coprime and non-coprime spindle solutions. 
We have also constructed hyperscalar 
solutions allowing the hyperscalar to vanish at the poles. In practice, we found that the hyperscalar solutions
only exist in the anti-twist class and furthermore, we have also found that the hyperscalar must be non-vanishing at one of the poles.
The non-coprime solutions with the same magnetic fluxes $p^I$ but different values of the discrete orbibundle data
$m^I_{N,S}$ have the same entropy in the large $N$ limit that we consider 
and uplift to different solutions; for the STU solutions they have different BPS spectra for the hyperscalar modes, in general.

We have seen that the existence of the hyperscalar solutions is in alignment with what we refer to as the ``RG scenario". 
In particular, this says that
a hyperscalar solution exists if and only if there is an STU solution (with vanishing hyperscalar) with the same magnetic fluxes and,
in addition, with a hyperscalar mode that is relevant. 
The algebraic conditions for the existence of a relevant mode for an STU solution 
that we derived in section \ref{hyperfluc} are closely related
to the necessary algebraic conditions satisfied by BPS hyperscalar solutions discussed in section \ref{sec5}. Heuristically,
it might then seem reasonable that the existence of a hyperscalar solution implies the existence of  
a relevant mode in a ``nearby" STU solution. However, the existence of the hyperscalar solutions requires solving
the BPS equations with different boundary conditions than the STU solution, so we do not consider this to be a good explanation
of the RG scenario.
Furthermore, we have no \emph{a priori} field theory arguments that indicate that there is any relationship between
the different gravitational solutions.
It is therefore striking that our numerical investigations have uncovered very strong evidence for the RG scenario. If true, this then
strongly suggests that there exist new holographic RG flows from
the STU $AdS_2\times \Sigma$ solutions in the UV to the hyperscalar $AdS_2\times \Sigma$ solutions in the IR. 
Such RG flow solutions will have a dependence both on the spindle coordinate and
a holographic radial coordinate ($y$ in \eqref{eq:ansatz} and $u$ in \eqref{eqn:frame}), implying that one will need to solve
partial differential equations. 

There are many similarities between the $AdS_2\times \Sigma$ solutions here and the $AdS_3\times \Sigma$ solutions of \cite{Arav:2025jee}.
We can summarise some of the differences that we observed.
Firstly, for non-coprime spindles we saw in \eqref{thensareodd} that $n_N$ and $n_S$ are both
odd integers in general, and hence for non-coprime spindles $h$ is necessarily odd). This result essentially follows from
the fact that we have four gauge fields and was
not the case for the solutions of \cite{Arav:2025jee} which had three gauge fields. 

Second, for hyperscalar solutions we saw (e.g. \eqref{signchoicehyper}) that the hyperscalar is necessarily  
non-vanishing at one pole and vanishes at the other pole with an even power of the locally defined radial coordinate
on the spindle i.e. $r_N$ is even; this is in contrast to \cite{Arav:2025jee} where $r_N$ could be both even or odd.

Third, the RG scenario is broadly the same as for the $AdS_3\times \Sigma$ solutions in \cite{Arav:2025jee}.
For $AdS_3\times \Sigma$ solutions, strong numerical evidence for a one-to-one correspondence between the existence of 
hyperscalar solutions and STU solutions with relevant hyperscalar modes was found, and this suggests that there
are corresponding RG flows between them. It was also found that the central charge of the STU solution is always greater than  
that of the hyperscalar solution. Since it is known that the central charge should be monotonic under RG flow 
this provides additional support for the existence of the RG flows. For the $AdS_2\times \Sigma$ solutions
we observed analogous behaviour of the entropy $S_{BH}$. A substantial difference is that we are unaware of any argument
as to why $S_{BH}$ should decrease, if indeed the RG flows exist. From the gravitational side, by analysing the
PDEs for the RG flows it may be possible to argue that $S_{BH}$ must always decrease and this would be an interesting
new result\footnote{Recall that simple $c$-theorems for RG flows are based on solving ODEs e.g. \cite{Freedman:1999gp}.}.
On the field theory side, we currently lack a foundational understanding of the RG flows of
the dual quantum mechanical theories. Assuming progress can be made in that direction, it may be possible to 
establish a field-theoretic explanation for the observed monotonicity.

Fourth, another novelty is that in 
solving the algebraic conditions for BPS solutions one could find multiple potential hyperscalar
solutions (as in \eqref{funnycase1}, \eqref{funnycase2}) also with an STU solution with a relevant mode; only the one with 
an entropy smaller than the STU solution actually exists.
This is precisely the solution associated with the RG scenario.
From the localization approach of section
\ref{section:equivlocal} the existence of multiple potential solutions arises after extremisation (which is equivalent to imposing all of
the algebraic conditions for BPS solutions). Of course the localisation approach is based on the premise that if a solution exists
then one can extract a supersymmetric observable, in this case the black hole entropy. Here the RG scenario provides
a sharp conjecture for when the solutions actually exist.  In \cite{Arav:2025jee} such multiple potential solutions were not possible for the analogous $AdS_3\times \Sigma$ solutions.

Fifth, for the class of solutions with 3 equal $p^I$ (which is roughly analogous to the solutions in \cite{Arav:2025jee} with
two equal $p^I$), we saw at the end of section \ref{stu3pequal} that for coprime solutions, which necessarily preserve an $SU(3)$ flavour symmetry, there were no relevant hyperscalar deformations of STU solutions and hence no coprime hyperscalar solutions if $n_S$ is 
a multiple of 3. However, switching to the non-coprime case, it is possible to have 
relevant hyperscalar deformations of STU solutions and correspondingly non-coprime hyperscalar solutions with $h$ a multiple of 3 
as we see in table \ref{noncoprime_pBneq0}, with, necessarily, the $SU(3)$ flavour symmetry broken.

For STU solutions there is an understanding of how the $AdS_2\times \Sigma$ solutions
arise as the near horizon limit of accelerating black holes: interestingly this is best understood by embedding them
in a larger family of supersymmetric solutions that are also rotating and carry electric charge, as shown in the context of minimal
gauged supergravity in \cite{Ferrero:2020twa}. It would be interesting to construct similar families of black hole solutions 
with the hyperscalar non-vanishing.

To date all of the hyperscalar solutions we have found, both in the $AdS_2\times \Sigma$ 
constructions here and in \cite{Suh:2022pkg,BenettiGenolini:2024kyy}, as well as
the $AdS_3\times \Sigma$ constructions in \cite{Arav:2022lzo,Arav:2025jee}, arise in the anti-twist class.
That anti-twist solutions are in some sense preferred in the gravitational solutions has been seen before. For
example, in the original
constructions of \cite{Ferrero:2020laf,Ferrero:2020twa} in minimal gauged supergravity, there were only anti-twist solutions, but
after the inclusion of extra vector multiplets to get the STU model, solutions
in the twist class were found \cite{Ferrero:2021ovq,Ferrero:2021etw}. It would be very interesting to know whether there are other
constructions which can lead to $AdS_2\times \Sigma$ and $AdS_3\times \Sigma$ hyperscalar solutions in the twist class. For the
case of $AdS_2\times \Sigma$ there is the specific possibility of adding electric fluxes and this will be explored elsewhere.

\section*{Acknowledgements}
\noindent
This work was supported in part by STFC grant ST/X000575/1; 
in the framework of H.F.R.I. call “Basic research Financing
(Horizontal support of all Sciences)” under the National Recovery and
Resilience Plan “Greece 2.0” funded by the European Union –
NextGenerationEU (H.F.R.I. Project Number: 15384);
by FWO projects G094523N and G003523N, as well as by KU Leuven grant C16/25/010.
JP is supported by a Dean's PhD studentship at Imperial College.

\appendix

\section{\texorpdfstring{$AdS_2 \times S^2$}{AdS2 x S2} solutions}\label{app:S2}
In this appendix we consider $AdS_2\times S^2$ solutions of the STU model,
setting $n_N = n_S$,
with a round metric on $\Z_{n_N}$ quotients of $S^2$ and supersymmetry preserved with a topological twist. These solutions lie within the ansatz (\ref{eq:ansatz}); 
we don't use conformal gauge \eqref{confgauge} here, but instead take
$h=\frac{1}{n_N}f\sin y$, $a^I=-\frac{p^I}{n_N^2}\cos y$
with $y \in [0,\pi]$,
and $V,f$ and the scalars are all constants. They preserve supersymmetry with Killing spinors as in (\ref{eqn:spindlespinor}) that are now (anti-)chiral everywhere, with $\cos\xi=(-1)^t$ and $\sin\xi=0$, i.e.
\begin{equation}
    i \Gamma_{23}\epsilon = \cos\xi\,\epsilon\,.
\end{equation}
Restricting to the STU model, the BPS equations
\eqref{eqn:BPSeq} and \eqref{eq:fieldstrength} imply
\begin{align}\label{S2BPS}
   p_R= \sum_Ip^I = 2 (-1)^{t+1}n_N \,,\qquad \sum_Ix^I = -2\kappa\,,\nn
    p^I = n_N (-1)^{t+1} f^2 e^{-2V} \left[\kappa x^I + (x^I)^2 \right]\,,
\end{align}
where we replaced $X^I$ with $x^I = (-1)^{t+1} e^V X^I$.
Since $x^0 x^1 x^2 x^3 = e^{4V}$, these can be solved in terms of the $p^I$.
Setting
\begin{equation}\label{kapconagain}
	\kappa = -1 \,,\qquad t = 1 \,,
\end{equation}
we find
\begin{align}\label{ads2f2e2v}
	f^2 &= \frac{1}{\sqrt{2}n_N}\left( \hat{P}^{(2)} +s - n_N^2\right)^{1/2} \,,\nn
e^{2V} &=\frac{1}{ \sqrt{2} s^2}\left(n_N^3 -  \hat{P}^{(2)} n_N +   \hat{P}^{(3)} \right)\left( \hat{P}^{(2)} +s - n_N^2\right)^{1/2} \,,
\end{align}
where $s = [(\hat{P}^{(2)} - n_N^2)^2 - 4 \hat{P}^{(4)}]^{1/2}$
and $x^I$ given by \eqref{STU_xI}, with $\sigma = +1$ and $n_N = n_S$. 
From \eqref{expressionforsbh}, the entropy is given by
\begin{align}
S_{BH}=\frac{2\pi\sqrt{2}}{3}\frac{f^2}{n_N} N^{3/2}\,.
\end{align}
Setting $n_N=1$ these expressions agree with equation (3.22) of \cite{Benini:2015eyy}, after noting that
we can also write
\begin{align}
 p_R^3 - 4 \hat{P}^{(2)} p_R + 8  \hat{P}^{(3)}=(p^0+p^1-p^2-p^3)(p^0-p^1+p^2-p^3)(p^0-p^1-p^2+p^3)\,,
\end{align}
and we also find agreement with the scalars, as in footnote \ref{scalarfootcompariosn}.
Note in particular there are no solutions in the case of pairwise equal fluxes.

We can analyse linear hyperscalar perturbations about the STU solutions as in Section \ref{hyperfluc}. With \eqref{kapconagain} we find
\begin{equation}\label{deltaS2}
	\delta = -\frac{1}{2} (x^0 - x^1 - x^2 - x^3) \,,
\end{equation}
which is purely a function of the fluxes.
As reviewed in subsection \ref{subsection:STU_twist}, with \eqref{kapconagain}, twist class solutions 
are believed to exist when there are three 
negative and one positive magnetic charges \cite{Couzens:2021cpk}. Noting that the fluxes are constrained via \eqref{S2BPS}, as well as (cf. \eqref{eq:brokenflux_v2}) 
\begin{equation}
	p_B = (r_N + r_S) n_N \,,
\end{equation}
one can show that there are no solutions for which $\delta > 0$. This is consistent with the RG scenario, as there are no $AdS_2 \times S^2$ solutions with non-vanishing hyperscalar \cite{Bobev:2018uxk}.

Notice from \eqref{deltaS2} that the scaling dimension does not depend on $m^I_{N,S}$. Hyperscalar fluctuations around $AdS_2 \times S^2$ solutions
exhibit a degeneracy of modes, attributed to the $SO(3)$ isometry of the round $S^2$.
Indeed, from \eqref{deltafinal} it is possible to see that, with $n_N = n_S$, $\delta$ only depends on $r_N, r_S$ via their sum, since $p_B = (r_N + r_S) n_N$.
For example, with $n_N = n_S= 1$ and $p^I = (5,-1,-1,-1)$ we have 7 hyperscalar modes, all of which are irrelevant with the same value of $\delta = \frac{1}{2} - \frac{3\sqrt{21}}{14} < 0$.
In table \ref{noncoprimeS2}, we present some examples of non-coprime ``sphere" solutions, i.e. $\mathbb{Z}_{n_N}$ quotients of the $S^2$, and recall from \eqref{thensareodd} that $n_N$ must be odd.
Notice that the degeneracy of modes can persist in the non-coprime case even though the $SO(3)$ isometry is broken, however the number of degenerate modes can change.

\begin{table}[h!]
 \begin{center}
 $(n_N,n_S)=3(1,1),~p^I = 3(5,-1,-1,-1)$\\ \smallskip
 \scalebox{0.85}{
\begin{tabular}{ |c|c|c|c|c|c|} 
 \hline
 $m_N^I$ &  $m_S^I$ & $r_N$&$r_S$& $\# (\Delta)$& $\frac{1}{N^{3/2}} S_{\rm BH}^{STU}$ \\ 
 \hline 
 $(2,2,2,2)$ & $(1,1,1,1)$ & $3 n_{KK} + 4 $&$4 - 3 n_{KK}$&  $ 2$ & $ \frac{2\pi}{9} \sqrt{3\sqrt{21}-13} $ \\ 
 \hline
\end{tabular}
}
\bigskip

 $(n_N,n_S)=5(1,1),~p^I = 5(5,-1,-1,-1)$\\ \smallskip
 \scalebox{0.85}{
\begin{tabular}{ |c|c|c|c|c|c|} 
 \hline
 $m_N^I$ &  $m_S^I$ & $r_N$&$r_S$& $\# (\Delta)$& $\frac{1}{N^{3/2}} S_{\rm BH}^{STU}$ \\ 
 \hline 
 $(2,4,4,4)$ & $(2,3,3,3)$ & $5 n_{KK}' + 5$ & $3 - 5 n_{KK}'$ & $1$ & $\frac{2\pi}{15} \sqrt{3\sqrt{21}-13}$ \\ 
$(3,2,2,2)$ & $(3,1,1,1)$ & $5 n_{KK} + 3$ & $5 - 5 n_{KK}$ &$2$ & $\frac{2\pi}{15} \sqrt{3\sqrt{21}-13}$ \\ 
\hline
\hline
$(1,2,2,4)$ & $(1,1,1,4)$ & $5n_{KK}'+2$&$6 - 5 n_{KK}'$& $2$& $\frac{2\pi}{15} \sqrt{3\sqrt{21}-13}$ \\ 
$(1,2,3,3)$ & $(1,1,2,2)$ & $5n_{KK}'+2$&$6 - 5 n_{KK}'$& $2$& $\frac{2\pi}{15} \sqrt{3\sqrt{21}-13}$ \\ 
$(2,2,2,3)$ & $(2,1,1,2)$ & $5n_{KK}'$&$8 - 5 n_{KK}'$& $2$& $\frac{2\pi}{15} \sqrt{3\sqrt{21}-13}$ \\ 
$(3,3,4,4)$ & $(3,2,3,3)$ & $5n_{KK}'+3$&$5 - 5 n_{KK}'$& $2$& $\frac{2\pi}{15} \sqrt{3\sqrt{21}-13}$ \\ 
$(4,2,4,4)$ & $(4,1,3,3)$ & $5n_{KK}'+1$&$7 - 5 n_{KK}'$& $2$& $\frac{2\pi}{15} \sqrt{3\sqrt{21}-13}$ \\ 
$(4,3,3,4)$ & $(4,2,2,3)$ & $5n_{KK}'+1$&$7 - 5 n_{KK}'$& $2$& $\frac{2\pi}{15} \sqrt{3\sqrt{21}-13}$ \\ 
$ m^1 \leftrightarrow m^2 $ & $\dots$ & $\dots$&$\dots$&$\dots$& $\dots$ \\ 
$ m^2 \leftrightarrow m^3 $ & $\dots$ & $\dots$&$\dots$&$\dots$& $\dots$ \\ 
 \hline
\end{tabular}
}
\caption{
Examples of non-coprime $AdS_2 \times S^2$ solutions.
The number of hyperscalar modes are denoted as $\# (\Delta)$.
Top table: $(n_N,n_S)=3(1,1)$ and $p^I = 3(5,-1,-1,-1)$.
Bottom table: $(n_N,n_S)=5(1,1)$ and $p^I = 5(5,-1,-1,-1)$.
The top 2 STU solutions preserve an $SU(3)$ flavour symmetry, while the bottom 18 solutions break this flavour symmetry to $SU(2)$
(only 6 solutions given, the other 12 are obtained by exchanging $m^1 \leftrightarrow m^2$ or $m^2 \leftrightarrow m^3$). 
 $n_{KK}$ and $n_{KK}'$ are as before.
\label{noncoprimeS2}}
\end{center}
 \end{table}

\section{Minimal gauged supergravity}\label{minimal}
\subsection{Truncation via ABJM}\label{appminimalallequal}
Minimal gauged supergravity can be obtained via a consistent truncation by
setting the scalars to zero, i.e. $\rho = \varphi^i=0$, as well as $A^0 = A^1 = A^2 = A^3$. The scalar potential is then $\mathcal{V} = -6$. The bosonic Lagrangian can be written as
\begin{equation}
	\cL = \frac{1}{16\pi G_4} \left[ R + 6 - F^0_{\mu\nu}F^{0\mu\nu} \right] \,.
\end{equation}
Similar to section \ref{hyperfluc}, we may consider a linearised perturbation of the hyperscalar. As we show in appendix \ref{2p2}, there are no relevant hyperscalar deformations when the fluxes are pairwise equal, and hence there are also none
for this minimal gauged supergravity truncation.

Amongst the general class of STU solutions discussed in section \ref{section:STU}, those with 
\begin{align}\label{minimalcase}
p^0 &= p^1 = p^2 = p^3 \equiv p\,,\nn
m_{N,S}^0&=m_{N,S}^1=m_{N,S}^2=m_{N,S}^3 \equiv m_{N,S}\,,
\end{align}
are solutions of minimal gauged supergravity (with the latter always following from the former in the coprime case). As shown in \cite{Ferrero:2021etw,Couzens:2021cpk}, solutions in this class necessarily preserve supersymmetry by an anti-twist. We now comment on the uplift of these solutions to $D=11$ on $S^7$, which extends the discussion in \cite{Ferrero:2020twa} to non-coprime spindles. With small changes we can also consider the uplift on other seven-dimensional Sasaki-Einstein manifolds.

For definiteness and without loss of generality, we again set
\begin{equation}
	\kappa = -1 \,,\qquad t_N = 1\,,\qquad t_S = 0\,.
\end{equation}
The R-symmetry constraint \eqref{eq:Rsymmconst} reads
\begin{equation}
	p = \frac{1}{\beta} (n_S - n_N) \,,
\end{equation}
with\footnote{This value of $\beta$ is associated with uplifting the solutions to $D=11$ on $S^7$, as we are discussing. We can also uplift
solutions on other Sasaki-Einstein spaces if we choose different values of $\beta$; see e.g. table 1 of \cite{Ferrero:2020twa}.} 
$\beta = 4$; thus $n_S - n_N$ must be divisible by 4.
These STU solutions have (in alignment with
\eqref{STU_xI}, \eqref{STU_k}, \eqref{eq:Couzens_BH})
\begin{align}\label{ming_xI_4p1AT}
	k & = \frac{\sqrt{n_N^2 + n_S^2}}{\sqrt{2} n_N n_S} \,, \nn
x_N^0 &=  x_N^1 = x_N^2 = x_N^3 = \frac{1}{2} \left( 1 - \frac{\sqrt{2} n_S}{\sqrt{n_N^2 + n_S^2}} \right) \,, \nn
	x_S^0 & = x_S^1 = x_S^2 = x_S^3 = \frac{1}{2} \left( 1 - \frac{\sqrt{2} n_N}{\sqrt{n_N^2 + n_S^2}} \right) \,, \nn
	S_{\rm BH}^H & = \frac{ \pi}{3 \sqrt{2}} \frac{ \sqrt{2(n_N^2 + n_S^2)} - (n_N + n_S) }{n_N n_S} N^{3/2} \,.
\end{align}

Recall that the magnetic flux through the spindle is specified by the discrete data of the bundle, \eqref{eqn:fluxmIZcond}. For the R-symmetry bundle we have
\begin{equation}\label{p_minimal}
	p = n_N m_S - n_S m_N \,,
\end{equation}
where for smoothness of the solutions we require $\hcf(m_N, n_N) = 1$, $\hcf(m_S, n_S) = 1$.
For well-defined spinors we need \eqref{eq:spinorregularity}, which given \eqref{minimalcase}
reduces to
\begin{equation}
	\beta m_N = -1 \mod n_N \,,\qquad \beta m_S = -1 \mod n_S \,.
\end{equation}
As pointed out below \eqref{eq:spinorregularity}, this condition is automatically satisfied for coprime spindles.
More generally,  this condition implies that for smooth, supersymmetric ($SU(4)$ invariant) solutions we require
$\hcf(n_N,\beta) = 1$ and  $\hcf(n_S,\beta) = 1$, or since $\beta=4$, we must have both $n_N$ and $n_S$ odd;
however, recalling \eqref{thensareodd}, this is not an extra condition.
Thus, for $n_N$, $n_S$ not necessarily coprime, there is a unique $SU(4)$ invariant supersymmetric 
uplifted solution with
\eqref{minimalcase} and $4p=n_S-n_N<0$, if and only if  $n_N$, $n_S$ are both odd and $(n_S-n_N)$ is divisible by 4.

Following an argument in appendix C of \cite{Arav:2025jee} we also have a converse result and
so for $n_N$, $n_S$, not necessarily coprime there is a unique supersymmetric smooth uplift with all
$p^I$ equal and all $m_N^I$ equal and all $m_S^I$ equal (i.e. $SU(4)$ invariant)
if and only if $n_S$, $n_N$ are odd and $n_S - n_N$ is divisible by 4.

As noticed in \cite{Arav:2025jee}, if we write $\beta m_S = -1 + l n_S$ for some $l \in \Z$, then $\beta m_N = -1 + l n_N$, for the same $l$. By theorem 4.4 of \cite{geiges2017seifertfibrationslensspaces}, the Lens space $L(p,q)$ associated with the R-symmetry orbibundle can be systematically identified in terms of the Seifert invariants $((n_N, m_N),(n_S, m_S))$. As laid out in appendix B of \cite{Arav:2025jee}, we have $p$ as in \eqref{p_minimal}, and $q = n_N b - m_N a$, for some integers $a, b$ satisfying $n_S b - m_S a = 1$. For (non-coprime) spindle solutions of minimal gauged supergravity we have $1 = l n_S - \beta m_S = n_N l - m_N \beta$, which shows that we have $q=1$. Therefore, after uplifting on $S^7$, we obtain $AdS_2 \times Y_9$ solutions of M-theory with $Y_9$ an $L(p,1)$ bundle over $\mathbb{CP}^3$.
In the coprime case, this is in agreement with appendix A of \cite{Ferrero:2020twa}. We have $n_- = n_S$, $n_+ = n_N$, $I^{\rm there} = \beta = 4$ and $k^{\rm there}=1$ (to uplift on $S^7$), $\texttt{p}^{\rm there} = n_N$, $\texttt{q}^{\rm there} = p = \frac{1}{4}(n_S - n_N)$.
The above arguments show that, with $n_N$, $n_S$ not divisible by 2, and $4 p = n_S - n_N$ divisible by 4, we have a supersymmetric uplift on $S^7$.
Notice that we can write $n_S = \texttt{q}^{\rm there} \beta + \texttt{p}^{\rm there}$. Given $n_N = \texttt{p}^{\rm there}$, in the coprime case $\hcf(n_N, n_S) = 1$ implies $\hcf(\texttt{p}^{\rm there},\texttt{q}^{\rm there}) = 1$. Thus, as in \cite{Ferrero:2020twa}, the total space $Y_9$ is simply connected in the coprime case. In the non-coprime case, $Y_9$ is no longer simply connected, with the topology being a $\Z_h$ quotient of the covering space, with $h = \hcf{(\texttt{p}^{\rm there}, \texttt{q}^{\rm there})}$. Note this is still an $L(p,1)$ bundle over $\mathbb{CP}^3$.

\subsection{Truncation via mABJM}\label{appminimalmabjm}
Minimal gauged supergravity can also be obtained as a consistent truncation by
setting the scalars to 
the values in the other supersymmetric vacuum,
$e^{\frac{1}{2}\varphi^i} = 3^{1/4}$, $\tanh\frac{\rho}{2} = \frac{1}{\sqrt{3}}$,
as well as setting $\frac{1}{3}A^0 = A^1 = A^2 = A^3$, leading to the Lagrangian
\begin{equation}
	\cL = \frac{1}{16\pi G_4} \Big[ R + \frac{9 \sqrt{3}}{2} - \sqrt{3}F^1_{\mu\nu}F^{1\mu\nu} \Big] \,.
\end{equation}

This allows us to obtain many anti-twist hyperscalar solutions with $\frac{1}{3}p^0=p^1=p^2=p^3$ (and correspondingly for
the $m_{N,S}^I$) and all scalars
being constant everywhere: $e^{\frac{1}{2}\varphi^i} = 3^{1/4}$, $\tanh\frac{\rho}{2} = \frac{1}{\sqrt{3}}$.
Indeed one just needs to suitably embed the minimal gauged supergravity spindle solutions of \cite{Ferrero:2020twa}.
In the language of this paper, setting $\kappa = -1$, and $(t_N,t_S) = (1,0)$ (as usual for anti-twist), we have
\begin{align}\label{peesminhyper}
p^0=(n_S-n_N)/2\,,\qquad p^1=p^2=p^3=(n_S-n_N)/6\,,
\end{align}
and 
\begin{align}\label{hyper_xI_3p1AT}
	k & = \frac{\sqrt{n_N^2 + n_S^2}}{\sqrt{2} n_N n_S} \,, \nn
	x_N^0 & = 1 - \frac{\sqrt{2} n_S}{\sqrt{n_N^2 + n_S^2}} \,,\qquad x_N^1 = x_N^2 = x_N^3 = \frac{1}{3} \left( 1 - \frac{\sqrt{2} n_S}{\sqrt{n_N^2 + n_S^2}} \right) \,, \nn
	x_S^0 & = 1 - \frac{\sqrt{2} n_N}{\sqrt{n_N^2 + n_S^2}} \,,\qquad x_S^1 = x_S^2 = x_S^3 = \frac{1}{3} \left( 1 - \frac{\sqrt{2} n_N}{\sqrt{n_N^2 + n_S^2}} \right) \,, \nn
	S_{\rm BH}^H & = \frac{2\sqrt{2} \pi}{9 \sqrt{3}} \frac{ \sqrt{2(n_N^2 + n_S^2)} - (n_N + n_S) }{n_N n_S} N^{3/2} \,.
\end{align}
The ratio of the entropy for these solutions compared with \eqref{ming_xI_4p1AT} is explained by
\eqref{eqn:G4N32} and \eqref{freemABJM}.

We can illustrate how these solutions relate to the RG scenario with a specific example. Consider
the anti-twist STU solution with $(n_N,n_S)=(13,1)$ and $p^I = -(6,2,2,2)$. From table \ref{pbzerotablecoprime}
we have $S_{\textrm{BH}}   = \frac{2\pi}{39} \sqrt{61 - \sqrt{3529} } N^{3/2}\sim  0.203443 N^{3/2}$ and
we have a hyperscalar mode with $r_N=r_S=0$ and $\delta = \frac{1}{2} - \frac{37}{2\sqrt{3529}}\sim  0.188580$.
Also as in table \ref{pbzerotablecoprime} we have a hyperscalar anti-twist solution as in \eqref{hyper_xI_3p1AT}, with
$S^H_{\textrm{BH}} = \frac{4\sqrt{2} \pi}{117\sqrt{3}} (-7 + \sqrt{85}) N^{3/2}\sim  0.194645 N^{3/2} $.

A general point is that for this truncation the hyperscalar is a non-zero constant everywhere and so $r_S=r_N=0$.
Using the argument at the end of section \ref{stu3pequal}, for smooth, supersymmetric solutions of this truncation (with $\frac{1}{3}m^0=m^1=m^2=m^3$ at both poles), both $n_N$ and $n_S$ must not be divisible by 3, as well as both being odd from \eqref{thensareodd}. Furthermore, \eqref{peesminhyper} implies the extra condition that $n_S-n_N$ is divisible by 6.
One can see this on the green dots on the green line in the two plots of figure \ref{fig:dotplot} (recalling also the coprime condition
\eqref{pcoprconds}). 

\section{2+2 truncation}\label{2p2}
Here we discuss a truncation of the model \eqref{eqn:4daction}-\eqref{eqn:Q}, considering pairwise equal gauge fields, keeping the charged hyperscalar and a single neutral scalar:
\begin{align}
	& \varphi^1 \equiv \varphi \,, \qquad \varphi^2 = \varphi^3 = 0 \,, \nn
	& A^0 = A^1  \,,\qquad A^2 = A^3  \,,
\end{align}
which implies $X^0=X^1$ and $X^2=X^3$.
If one further sets the hyperscalar to be zero, the resulting supergravity model is a truncation of the STU model considered in \cite{Ferrero:2021ovq}\footnote{With the axion $\chi^{\rm there} = 0$; in writing \eqref{eqn:4daction} we already restrict to solutions with $F^I \wedge F^J = 0$.} (see also \cite{Couzens:2021rlk}), with prepotential $\cF(Y^I) = -2i X^0 X^2$, where
\begin{equation}
	X^0 = e^{\frac{1}{2} \varphi} \,,\qquad X^2 = e^{-\frac{1}{2} \varphi} \,.
\end{equation}
The bosonic Lagrangian, with non-vanishing hyperscalar, can be written as
\begin{align}
	\cL = \frac{1}{16 \pi G_4} \Big[ R - \frac{1}{2} (\partial \varphi)^2 - \frac{1}{2} \left( e^{-\varphi} F^0_{\mu\nu} F^{0\mu\nu} + e^{\varphi} F^2_{\mu\nu} F^{2\mu\nu} \right) - \cV \nn
	- \frac{1}{2} (\partial \rho)^2 - \frac{1}{2} \sinh^2 \rho (D\theta)^2 \Big] \,,
\end{align}
 and the scalar potential $\cV$ is given by 
\begin{equation}
	\cV = -2 (2+ \cosh \varphi) +2\sinh^2\frac{\rho}{2}(\cosh \varphi \sinh^2\frac{\rho}{2}-2 )\,.
\end{equation}
The real superpotential is
\begin{align}
	\cW & = - (X^0 + X^2) - X^2 \sinh^2 \frac{\rho}{2} \nn
	& = - 2 \cosh \frac{\varphi}{2} - e^{-\frac{1}{2} \varphi} \sinh^2 \frac{\rho}{2} \,.
\end{align}
This model has the ABJM $AdS_4$ vacuum solution and also the 
non-supersymmetric and unstable $SU(4)^-$
$AdS_4$ vacua with $\varphi=0$, $e^{\frac{\rho}{2}}=\sqrt{2}\pm 1$.

We first consider solutions with zero hyperscalar.
From (\ref{eqn:bdycond2}) the dressed scalars are constrained by
\begin{align}\label{2p2constraint}
	x_N^0 + x_N^2 & = - \kappa + (-1)^{t_N} \frac{1}{k n_N} \,, \nn
	x_S^0 + x_S^2 & = - \kappa - (-1)^{t_S} \frac{1}{k n_S} \,.
\end{align}
Defining the fluxes (with $a = 0,2$)
\begin{equation}\label{2p2flux}
	\frac{p^a}{n_N n_S} \equiv \frac{1}{4\pi} \int_\Sigma F^a = \frac{k}{2} (x_S^a - x_N^a) \,, \qquad p^a \in \Z \,,
\end{equation}
the R-symmetry flux $p_R = 2(p^0 + p^2)$ is constrained by
\begin{equation}\label{2p2Rsymm}
	\frac{p_R}{n_N n_S} = \frac{(-1)^{t_N+1}}{n_N} + \frac{(-1)^{t_S+1}}{n_S} \,.
\end{equation}
Following the analysis of subsection \ref{subsection:BCs}, the Bekenstein-Hawking entropy can be written as
\begin{equation}\label{2p2SBH}
	S_{\rm BH} = - (-1)^{t_S} \frac{\pi \sqrt{2}}{3} N^{3/2} k \kappa \left[ x_S^0 x_S^2 - \sigma x_N^0 x_N^2 \right] \,,
\end{equation}
where $\sigma \equiv (-1)^{t_N-t_S}$.

For given $\kappa$, the spindle data within the truncation are specified by $n_{N,S}, t_{N,S}$, and the fluxes $p^a$, constrained by \eqref{2p2Rsymm}.
For non-coprime spindles, additional data is specified by $m_{N,S}^0, m_{N,S}^2$. However, solutions with the same $n_{N,S}, t_{N,S}, p^a$ will have the same Bekenstein-Hawking entropy, as discussed in the main text.
For given independent flux $p^0$, we have a system of 5 variables: $x_{N,S}^0, x_{N,S}^2, k$. We may eliminate $x_S^a$ using \eqref{2p2flux}, and $k$ using the constraint \eqref{2p2constraint}.
We can then aim to use the two equations for the conserved charges \eqref{eqn:STUcharges}, to solve for $x_{N,S}^a$.
We treat the twist and anti-twist class separately.

For the twist class the solution to the algebraic conditions for $x_{N,S}^a$ is simply the relation $x_N^0 + x_N^2  = - \kappa$ and hence from \eqref{2p2constraint} we see that it implies $k=\infty$. This would be a sphere case, but
as noted in appendix \ref{app:S2} such solutions do not exist for pairwise equal charges.
Thus, twist class solutions ($\sigma = +1$) do not exist in the 2+2 truncation, as already shown in \cite{Ferrero:2021etw,Couzens:2021cpk} and also consistent with \eqref{pconststwiststu}. 

We therefore focus on the anti-twist case below and set
\begin{align}
	\sigma=-1\,,
\end{align}
in the remainder of this appendix. We now find two branches of solutions that solve \eqref{eqn:STUcharges},
\begin{align}\label{2p2branch12}
	x_N^0 & = - \frac{\kappa}{2} \mp \frac{n_N+n_S}{4} \frac{(-1)^{t_N} (n_N + n_S) + 4 \sigma p^0 }{2s} \,, \nn
	x_N^2 & = - \frac{\kappa}{2} \pm \frac{n_N+n_S}{4} \frac{(-1)^{t_N} n_N + (-1)^{t_S} n_S - ((-1)^{t_N} - (-1)^{t_S}) n_S + 4 \sigma p^0 }{2s} \,,
\end{align}
with
\begin{equation}\label{s22exexpappc}
	s = \frac{n_N + n_S}{4} \sqrt{(n_N + n_S)^2 + 8 p^0 ((-1)^{t_N} n_N + (-1)^{t_S} n_S) + 16 \sigma (p^0)^2} \,.
\end{equation}
Recall that necessary conditions for an STU solution to exist are given by \eqref{STUexistcond}. Setting $\kappa = -1$ for definiteness, \eqref{2p2branch12} comprises 4 solutions in total: lower/upper branch for $(t_N,t_S) = (1,0)$, and likewise 2 branches of solutions for $(t_N,t_S) = (0,1)$. For the conditions \eqref{STUexistcond} to be satisfied we
find it is the lower branch for $(t_N,t_S) = (1,0)$ (and the upper branch for  $(t_N,t_S) = (0,1)$) with, in particular, a positive entropy given by \eqref{2p2SBH}.
Setting $t_N = 1$, $t_S = 0$, and $\kappa = -1$, for definiteness, 
we now continue with the lower branch.
This solution can be written as:
\begin{align}\label{2p2xIs}
	x_N^a & = \frac{1}{2} - \frac{n_N+n_S}{4s} \left( \frac{n_N + n_S}{2} + 2 p^a \right) \,.
\end{align}
Note in the 2+2 truncation the quantity $s$ in \eqref{STU_defn_s} reduces to
\begin{equation}
	s = \sqrt{\Big( \sigma n_N n_S - (p^0)^2 - (p^2)^2 - 4 p^0 p^2 \Big)^2 - 4 (p^0)^2 (p^2)^2} \,,
\end{equation}
agreeing with \eqref{s22exexpappc} for the anti-twist case $\sigma=-1$ (which we are assuming).
It is then straightforward to see that \eqref{2p2xIs} agrees with \eqref{STU_xI}.
Furthermore, \eqref{2p2SBH} then gives
\begin{equation}\label{2p2trunc_Sbh}
	S_{\rm BH} = - \frac{\pi \sqrt{2}}{3} N^{3/2} \left( \frac{n_N + n_S}{2 n_N n_S} - k \right) \,,
\end{equation}
where
\begin{equation}\label{2p2_k}
	k = - \frac{2 s}{n_N n_S ((-1)^{t_N} n_N - (-1)^{t_S} n_S)}  \,.
\end{equation}
Note from \eqref{2p2Rsymm} we have the following identity (with $\sigma = -1$)
\begin{align}
\hat{P}^{(2)} - \sigma n_N n_S =\frac{(n_N+n_S)^2}{4} + 2p^0 p^2 \,,
\end{align}
which allows us to equate \eqref{2p2trunc_Sbh} with \eqref{eq:Couzens_BH}.
Furthermore, imposing the conditions \eqref{STUexistcond} we find that solutions can only exist with
 $p^0,p^2<0$, and this proves the condition \eqref{anti-twist_bountiful} for this class.

The above results show that \eqref{STU_xI} is \emph{not} a general solution to the constraint equations \eqref{eqn:STUcharges} for general $t_N,t_S$. Rather, it is a general expression that reduces to the correct solution satisfying \eqref{STUexistcond}, whenever such a solution exists. The following observation makes this clear: in the 2+2 truncation, if we use \eqref{STU_xI} for the twist class also, one finds that we still solve 
\eqref{2p2constraint} and \eqref{2p2flux}.
However, in this truncation the twist class is obstructed, as pointed out above, and indeed instead of \eqref{eqn:STUcharges} one finds
\begin{equation}
x^0_N x^2_N \left(\frac{1}{x_N^a} + \kappa \right) =- x^0_S x^2_S \left(\frac{1}{x_S^a} + \kappa \right) \,,
\end{equation}
which is only the right equation for the anti-twist class. In the anti-twist class, we saw that there are two branches of solutions that solve \eqref{2p2constraint}, \eqref{2p2flux}, and \eqref{eqn:STUcharges}. However, only one of the two branches satisfies \eqref{STUexistcond}, in particular a positive entropy, and it is precisely this branch that coincides with \eqref{STU_xI}.

Following the discussion in section \ref{hyperfluc}, we may consider a linearised perturbation of the hyperscalar about the STU background. The truncation allows us to write an analytic expression for $\delta$ from \eqref{eqn:ODE}, which takes the same constant value at both poles, as ensured by the regularity condition
\begin{equation}\label{2p2Bsymm}
	p_B = -2 p^2 = (-1)^{t_N+1} r_N n_S + (-1)^{t_S+1} r_S n_N \,.
\end{equation}
For definiteness let us again focus on solutions in the anti-twist class where
\begin{equation}
	\kappa = -1 \,,\qquad t_N = 1 \,,\qquad t_S = 0 \,.
\end{equation}
Focussing on the possibility of relevant hyperscalar modes,
following the discussion around \eqref{rzeroantitwist}, we now set $r_S=0$. It follows that $p_R = n_S - n_N$ and $p_B = r_N n_S$.
Together these imply
\begin{equation}
	p^0 = \frac{1}{2} (n_S -n_N + n_S r_N) \,,\qquad p^2 = -\frac{r_N n_S}{2} \,.
\end{equation}
Using the expressions \eqref{2p2xIs}, \eqref{2p2_k}, as well as \eqref{eqn:wDthetaatpoles}, $\delta$ is then uniquely determined in terms of $(n_N, n_S)$ and $r_N$:
\begin{equation}
	\delta_{2+2} = \frac{1}{2} - \frac{n_N+n_S+ 2 n_S r_N}{2\sqrt{  (n_N+n_S)^2 + 4(n_N - n_S) n_S r_N - 4 n_S^2 r_N^2  }} \,.
\end{equation}
Recall from above that we have $p^0,p^2<0$ (i.e. \eqref{anti-twist_bountiful} is satisfied)
which implies
\begin{equation}
	r_N < \frac{n_N}{n_S} - 1 \,,
\end{equation}
and we find that under these conditions there are no solutions for which $\delta_{2+2} > 0$. Thus, there are no relevant hyperscalar modes when the fluxes are pairwise equal. This also implies
that there are no relevant modes when all four $p^I$ are equal.

\section{3+1 truncation}\label{3p1}
We now consider a 3+1 truncation; this theory admits both twist and anti-twist solutions, as we shall see below. Starting from the model \eqref{eqn:4daction}-\eqref{eqn:Q}, we keep the charged hyperscalar and a single neutral scalar, with a 3+1 split for the gauge fields:
\begin{align}
	& \varphi^1 =\varphi^2=\varphi^3\equiv \varphi \,,  \nn
	& A^0 \equiv C^0 \,, \qquad A^1 = A^2 = A^3 \equiv C^1 \,.
\end{align}
Notice that the identification of the scalars implies that $X^1=X^2=X^3$.
Defining $H^a = {\rd}C^a$, $a=0,1$, the bosonic Lagrangian can be written as
\begin{align}
	\cL = \frac{1}{16 \pi G_4} \Big[ R - \frac{3}{2} (\partial \varphi)^2 - \frac{1}{4} 
	e^{-3\varphi} H^0_{\mu\nu} H^{0\mu\nu} -\frac{3}{4} e^{\varphi} H^1_{\mu\nu} H^{1\mu\nu} 
	 - \cV \nn
	- \frac{1}{2} (\partial \rho)^2 - \frac{1}{2} \sinh^2 \rho (D\theta)^2 \Big] \,,
\end{align}
where the scalar potential $\cV$ is given by
\begin{align}
	\cV = - 6\cosh\varphi+\frac{1}{4}\sinh^2\frac{\rho}{2}e^{-\varphi}(-9-12e^{2\varphi}+e^{4\varphi}+(3+e^{4\varphi})\cosh\rho    )\,,
\end{align}
and the real superpotential is
\begin{equation}
	\cW = -\frac{1}{2}(e^{3\varphi/2}+3e^{-\varphi/2})+\frac{1}{2}\sinh^2\frac{\rho}{2}(e^{3\varphi/2}-3e^{-\varphi/2})\,.
\end{equation}
This model has the ABJM vacuum, with $\varphi=\rho=0$,
the $SU(3)\times U(1)$ invariant vacua with $e^{2\varphi}=3, e^\rho=2\pm \sqrt{3}$ 
and the 
(non-susy and unstable \cite{Pope:1984bd,Bobev:2010ib}) $SU(4)^-$ invariant 
vacua with $\varphi=0$, $e^{\frac{\rho}{2}}=\sqrt{2}\pm 1$.

We first consider solutions with zero hyperscalar.
From (\ref{eqn:bdycond2}) the dressed scalars are constrained by
\begin{align}\label{3p1constraint}
	x_N^0 + 3 x_N^1 & = - 2 \kappa + (-1)^{t_N} \frac{2}{k n_N} \,, \nn
	x_S^0 + 3 x_S^1 & = - 2 \kappa - (-1)^{t_S} \frac{2}{k n_S} \,.
\end{align}
Defining the fluxes
\begin{equation}\label{3p1flux}
	\frac{p^a}{n_N n_S} \equiv \frac{1}{4\pi} \int_{\Sigma} H^a = \frac{k}{2} (x_S^a - x_N^a) \,,\qquad p^a \in \Z \,,
\end{equation}
the R-symmetry flux $p_R = p^0 + 3 p^1$ is constrained by
\begin{equation}\label{3p1Rsymm}
	\frac{p_R}{n_N n_S} = \frac{(-1)^{t_N+1}}{n_N} + \frac{(-1)^{t_S+1}}{n_S} \,.
\end{equation}
Following the analysis of subsection \ref{subsection:BCs}, the Bekenstein-Hawking entropy can be written as
\begin{equation}
	S_{\rm BH} = \frac{\pi \sqrt{2}}{3} N^{3/2} k \kappa \left( x_S^1 \sqrt{x_S^0 x_S^1} -  x_N^1 \sqrt{x_N^0 x_N^1}  \right) \,,
\end{equation}

For given $\kappa$, the spindle data within the truncation are specified by $n_{N,S}, t_{N,S}$, and the fluxes $p^a$, constrained by \eqref{3p1Rsymm}.\footnote{For non-coprime spindles, additional data is specified by $m_{N,S}^0, m_{N,S}^1$. However, solutions with the same $n_{N,S}, t_{N,S}, p^a$ will have the same Bekenstein-Hawking entropy, as discussed in the main text.}
We have a system of 5 variables: $x_{N,S}^0, x_{N,S}^1, k$. We may eliminate $x_S^0, x_S^1$ using \eqref{3p1flux}, and also relate $k$ with $x_N^0, x_N^1$ using the constraint \eqref{3p1constraint}.
By utilising the expression for the conserved charges \eqref{eqn:STUcharges}, we can solve for $x_{N,S}^0, x_{N,S}^1$ and $k$. 
Within the 3+1 truncation, assuming \eqref{STUexistcond},
\eqref{eqn:STUcharges} reduces to
\begin{equation}
x^1_N \sqrt{x^0_N x^1_N} \left(\frac{1}{x_N^a} + \kappa \right) = x^1_S \sqrt{x^0_S x^1_S} \left(\frac{1}{x_S^a} + \kappa \right) \,,
\end{equation}
for both twist and anti-twist class solutions.
Taking \eqref{STU_xI}, we have checked numerically that this is indeed satisfied for various examples,
and in appendix \ref{app:E} we have proven it for anti-twist solutions, assuming $\kappa p^I>0$.
Some examples for the anti-twist class are given in table \ref{table:STU_antitwist} and also figure \ref{fig:dotplot}.

Following the discussion in section \ref{hyperfluc}, we may consider a linearised perturbation of the hyperscalar about the STU background. The truncation allows us to write an analytic expression for $\delta$ from \eqref{eqn:ODE}, which takes the same constant value at both poles, as ensured by the regularity condition
\begin{equation}\label{3p1Bsymm}
	p_B = p^0 - 3 p^1 = (-1)^{t_N+1} r_N n_S + (-1)^{t_S+1} r_S n_N \,.
\end{equation}

Let us first consider the twist class. Setting
\begin{equation}
	\kappa = -1 \,,\qquad t_N = 1\,,\qquad t_S = 1 \,,
\end{equation}
the fluxes are constrained by
\begin{equation}\label{3p1twistflux}
	p^0 = \frac{1}{2} ( n_N(1+r_S) + n_S(1 + r_N)) \,,\qquad p^1 = \frac{1}{6} (n_N(1-r_S) + n_S(1 - r_N) ) \,,
\end{equation}
and $\delta$ is uniquely determined in terms of $(n_N, n_S, r_N, r_S)$.
For example, in the special case when $r_N = r_S = 0$, $\delta$ can be written as
\begin{equation}
	\delta_{3+1} = \frac{1}{2} + \frac{\sqrt{3}(n_N^2 - 4 n_N n_S + n_S^2)}{2 \sqrt{108 n_N^2 n_S^2 - 72 n_N n_S (n_N + n_S)^2 + 11(n_N + n_S)^4}} \,.
\end{equation}
The expression for $\delta_{3+1}$ for general values of  $(n_N, n_S, r_N, r_S)$ is rather unwieldy.
Nevertheless, it is possible to show that $\delta_{3+1}$ is necessarily negative, provided the fluxes in \eqref{3p1twistflux} are quantised, which imposes additional constraints on $(n_N, n_S, r_N, r_S)$. Therefore, we conclude that there are no relevant hyperscalar modes in the twist class in the 3+1 truncation.

Next, we consider the anti-twist class.
Setting
\begin{equation}
	\kappa = -1 \,,\qquad t_N = 1\,,\qquad t_S = 0 \,,
\end{equation}
the fluxes are constrained by
\begin{equation}\label{3p1antitwistflux}
	p^0 = \frac{1}{2} ( n_S(1+r_N) - n_N(1+r_S) ) \,,\qquad p^1 = \frac{1}{6} ( n_S(1-r_N) + n_N(-1 +r_S )) \,.
\end{equation}
Again, the general closed form expression for $\delta_{3+1}$ is quite involved. 
Recall from \eqref{atwistclassconds} that in order to have a relevant mode with $\delta>0$ we must have $r_S=0$.
For $r_N = r_S = 0$, it reduces to
\begin{equation}
	\delta_{3+1} = \frac{1}{2} - \frac{\sqrt{3}(n_N^2 + 4 n_N n_S + n_S^2)}{2 \sqrt{11 n_N^4 + 28 n_N^3 n_S + 30 n_N^2 n_S^2 + 28 n_N n_S^3 + 11 n_S^4}} \,,
\end{equation}
and one can see that (with $n_N\ne n_S$) we have $\delta_{3+1} >0$
(examples can be found in table \ref{pbzerotablecoprime}).
There are also many solutions with $r_N \neq 0$ with $\delta_{3+1} > 0$, some of which can be found in table \ref{tablecoprimepbnonzero}. These necessarily satisfy \eqref{ubound}-\eqref{rNbound}.

\section{Analysis of algebraic conditions for anti-twist STU solutions}\label{app:E}

In this appendix we consider spindle solutions in the anti-twist class
 ($\sigma=-1$).
 Without loss of generality, in this appendix (in contrast to the main text) we take
\begin{align}\label{othercaseconvs}
\kappa=1\,,\qquad t_N=0\,,\qquad t_S=1\,.
\end{align} 
We will consider STU solutions with spindle data $n_N,n_S>0$ and $p^I$ satisfying
\begin{align}\label{prconsandpsign}
p_R = p^0 + p^1 + p^2 + p^3 = n_N - n_S\,,\qquad p^I>0\,,
\end{align}
 which implies $n_N > n_S $, and  recall that the second condition is
expected to be true for all explicit anti-twist STU solutions but has not rigorously been proven.
We want to show that the expressions for $x_N^I$, $x_S^I$ and $k$ in  \eqref{STU_xI} and \eqref{STU_k}
satisfy the following conditions (with general $\kappa$, $t_N$, $t_S$ still included below).
First, the BPS constraints:
\begin{equation}
\label{eqn:xIconstraintapp}
    \sum_I x_N^I = -2\kappa + (-1)^{t_N} \frac{2}{kn_N}\,,\qquad \sum_I x_S^I = -2\kappa - (-1)^{t_S} \frac{2}{kn_S}\,.
\end{equation}
Second, the flux constraints:
\begin{equation}
\label{eqn:BPS_fluxapp}
    \frac{p^I}{n_Nn_S} \equiv \frac{1}{4 \pi} \int_{\Sigma} F^I = \frac{k}{2}(x^I_S - x^I_N)\,.
\end{equation}
Third, the conserved charge constraints:
\begin{equation}
\label{eqn:STUchargesapp}
    \rho=0\,:\quad\frac{1}{\kappa}\mathcal{E}^I = \frac{1}{2} (-1)^{t_N+1} i \mathcal{F}(x^I_N) \left(\frac{1}{x_N^I} + \kappa \right) = \frac{1}{2}(-1)^{t_S+1} i \mathcal{F}(x^I_S) \left(\frac{1}{x_S^I} + \kappa \right)\,,
\end{equation}
where the prepotential is given in \eqref{prepotstu}.
Fourth, the inequalities:
\begin{equation}\label{STUexistcondapp}
    S_{\textrm{BH}} > 0\,,\quad \left.(-1)^{t+1} x^I\right\vert_{N,S} > 0\,,
\end{equation}
where $S_{\textrm{BH}}$ is given by
\begin{equation}\label{BHentropyapp}
    S_{\textrm{BH}} = -(-1)^{t_S} \frac{\pi\sqrt{2}}{3} N^{3/2} \frac{k}{2}\kappa \left[ i \mathcal{F}(x_S^I) - \sigma i \mathcal{F}(x_N^I) \right]\,.
\end{equation}
We also want to show that, given the above conditions, the expression for $S_{\textrm{BH}}$ reduces to
\begin{equation}
\label{eq:Couzens_BHapp}
    S_{\textrm{BH}} = \frac{2\pi}{3n_N n_S} N^{3/2} \sqrt{\hat{P}^{(2)}- \sigma n_N n_S  + \sigma s} \,.
\end{equation}
Now \eqref{eqn:xIconstraintapp} and \eqref{eqn:BPS_fluxapp} can be simply checked to be satisfied, so the goal is to show
\eqref{eqn:STUchargesapp}-\eqref{eq:Couzens_BHapp} are satisfied, given \eqref{prconsandpsign}.

Our approach will make extensive use of elementary symmetric polynomials. 
Before continuing, we pause to make two comments. First, we chose \eqref{othercaseconvs} so that the 
fluxes for the anti-twist solutions are all positive $p^I>0$ and hence all the symmetric polynomials
$\hat P^{(k)}$, $k=1,2,3,4$, defined in the next section are all positive $\hat P^{(k)}>0$, making the analysis
clearer. 
Using the discrete symmetries mentioned in section \ref{atwiststusolstext},
we can switch to $\kappa=-1$, $t_N=1$, $t_S=0$, as used in the main text, by
taking $p^I\leftrightarrow -p^I$, $x^I_{N,S}\leftrightarrow -x^I_{N,S}$, keeping $n_{N,S}$ and $k$ unchanged,
and then
making suitable changes to the proof below. Second, for the STU solutions in the twist class, 
the fluxes do not all have a fixed sign (recall \eqref{pconststwiststu}), and so the approach below using symmetric polynomials becomes more involved.

\subsection{Elementary symmetric polynomials}\label{symmpolys}
Let $\hat{Y}^{(k)}$ be the elementary symmetric polynomial of degree $k$ in the variables $Y^I (I=0,1,2,3)$:
\begin{align}
\hat{Y}^{(1)} &= \sum_{0\leq I\leq 3} Y^I \,, \qquad\qquad\qquad
\hat{Y}^{(2)} = \sum_{0\leq I<J\leq 3} Y^I Y^J \,, \nn
\hat{Y}^{(3)} &= \sum_{0\leq I<J<K\leq 3} Y^I Y^J Y^K \,, \qquad
\hat{Y}^{(4)} = Y^0 Y^1 Y^2 Y^3 \,.
\end{align} 
These satisfy the following properties:
\begin{enumerate}
\item Any symmetric polynomial in $Y^I$ can be written as a polynomial of the elementary symmetric polynomials above.
\item The elementary symmetric polynomials are positive if and only if each of their variables is positive, i.e.\ $ \hat{Y}^{(k)} >0 $ for $ k=1,2,3,4$ if and only if $Y^I>0$ for $I=0,1,2,3$. This follows from Descartes' rule of signs applied to the polynomial $p(t) = \prod_{0\leq I \leq 3}(t-Y^I)$.
\item When all variables $Y^I$ are positive ($Y^I>0$), the elementary symmetric polynomials satisfy
Newton's and Maclaurin's inequalities, given by
\begin{align}\label{newtmcl}
& \frac{\tilde{Y}^{(k+1)}}{\tilde{Y}^{(k)}} \leq \frac{\tilde{Y}^{(k)}}{\tilde{Y}^{(k-1)}} \,,\qquad\qquad 
 (\tilde{Y}^{(k+1)})^{1/(k+1)} \leq (\tilde{Y}^{(k)})^{1/k} \,,
\end{align}
where $\tilde{Y}^{(k)}$ are defined by
\begin{equation}
\tilde{Y}^{(k)} \equiv \frac{\hat{Y}^{(k)}}{\binom{4}{k}} \,.
\end{equation}
\end{enumerate}

\subsection{Rewriting the constraints in terms of permutation invariant quantities}

To show \eqref{eqn:STUchargesapp} and \eqref{STUexistcondapp} are satisfied, we start by rewriting them in terms of quantities which are invariant under $S_4$ permutations of the $p^I$. 
It is helpful to first view $x^I_N(s,n_N,n_S,p^I)$ and $x^I_S(s,n_N,n_S,p^I)$ in \eqref{STU_xI} as polynomials in $p^I$ with $s$ kept independent.
Then any symmetric polynomial of $x^I_N$ or $x^I_S$ can be viewed as a symmetric polynomial of the fluxes,
which in turn can be written in terms of $\hat P^{(k)}$, $k=1,2,3,4$.
We can then use the fact that $\hat{P}^{(1)}=p_R=n_N-n_S$ and $\hat{P}^{(4)}=\frac{1}{4}\left( (\hat{P}^{(2)} - \sigma n_N n_S)^2 - s^2 \right)$ to end up with an expression that depends only on $n_N$, $n_S$, $s$, $\hat{P}^{(2)}$ and $\hat{P}^{(3)}$. Going through this process for the elementary symmetric polynomials of the $x^I_N$ and $x^I_S$, we end up with the following expressions:
\begin{align}\label{eq:XIElemSymmPol}
s \hat{X}^{(1)}_N &= -2s + n_S(n_N+n_S) \,,\nn
s \hat{X}^{(1)}_S &= -2s + n_N(n_N+n_S) \,,\nn
2s^2 \hat{X}^{(2)}_N &= n_S(n_N+3n_S)(\hat{P}^{(2)}+n_N n_S) + 2(n_N + 2n_S) \hat{P}^{(3)} - 3n_S(n_N+n_S)s + 2s^2 \,,\nn
2s^2 \hat{X}^{(2)}_S &= n_N(3n_N+n_S)(\hat{P}^{(2)}+n_N n_S) -2(2n_N+n_S) \hat{P}^{(3)} - 3n_N(n_N+n_S)s + 2s^2 \,,\nn
4s^3 \hat{X}^{(3)}_N &= -A_1 \left( 2 \hat{P}^{(3)} + (n_N+3n_S)(\hat{P}^{(2)}+n_N n_S -s) \right) \,,\nn
4s^3 \hat{X}^{(3)}_S &= A_2 \left( 2\hat{P}^{(3)}-(3n_N+n_S)(\hat{P}^{(2)}+n_N n_S -s) \right) \,,\nn
8s^4 \hat{X}^{(4)}_N &= A_1^2 (\hat{P}^{(2)}+n_N n_S -s) \,,\nn
8s^4 \hat{X}^{(4)}_S &= A_2^2 (\hat{P}^{(2)}+n_N n_S -s) \,,
\end{align}
where
\begin{align}
A_1 &\equiv (n_N-n_S)(\hat{P}^{(2)}+n_N n_S) - 2\hat{P}^{(3)} + (n_N+n_S) s \,,\nn
A_2 &\equiv -(n_N-n_S)(\hat{P}^{(2)}+n_N n_S) + 2\hat{P}^{(3)} + (n_N+n_S) s \,.
\end{align}
For future use, observe that $A_1+A_2=2(n_N+n_S) s$. Also, notice that we immediately obtain \eqref{eqn:xIconstraintapp} from the first two lines of \eqref{eq:XIElemSymmPol}.

Applying property number 2 in section \ref{symmpolys}  to $(-1)^{t+1} x^I$, we see that the inequalities 
$ \left.(-1)^{t+1} x^I\right\vert_{N,S} > 0 $ in \eqref{STUexistcond} are equivalent to
\begin{equation}\label{eq:XXIneqConstraints}
(-1)^{k} \hat{X}^{(k)}_N > 0,
\qquad
\hat{X}^{(k)}_S > 0 \quad (k=1,2,3,4) \,.
\end{equation}
We next consider the conserved charge constraints \eqref{eqn:STUchargesapp}. To express them in terms of permutation invariant quantities, we can look at the elementary symmetric polynomials of the conserved charges expressed either in terms of the north or south pole quantities $\hat{\mathcal{E}}^{(k)}_N$, $\hat{\mathcal{E}}^{(k)}_S$. The conservation constraints \eqref{eqn:STUchargesapp} imply $\hat{\mathcal{E}}^{(k)}_N = \hat{\mathcal{E}}^{(k)}_S \, (k=1,2,3,4)$. 
The converse is also true: if  $\hat{\mathcal{E}}^{(k)}_N = \hat{\mathcal{E}}^{(k)}_S $, then we have
the values of $\mathcal{E}^I_N$ are equal to $\mathcal{E}^I_S$ up to a permutation in $I=0,1,2,3$. However, the latter
ambiguity can be resolved
since we can show that $(\mathcal{E}^I_N)^2$ are equal to $(\mathcal{E}^I_S)^2$
as in \eqref{eqn:STUcharges2}.

However, note that $\hat{\mathcal{E}}^{(k)}_N$, $\hat{\mathcal{E}}^{(k)}_S$ are not polynomials of the $x^I$ (and therefore of the fluxes). In order to apply the above process, we can instead consider the products $ \left( \hat{X}^{(4)}_N \right)^{(2-k)/2} \hat{\mathcal{E}}^{(k)}_N $. One can easily verify that these are indeed symmetric polynomials in $x^I_N$, and we can therefore apply the process to them and finally divide again by $\left( \hat{X}^{(4)}_N \right)^{(2-k)/2}$ using the last two lines of \eqref{eq:XIElemSymmPol}. A similar procedure can be done for the south pole quantities, and this leads to the following results:
\begin{equation}\label{eq:ConsChargeSymmPolRatios}
\frac{ \hat{\mathcal{E}}^{(1)}_N}{\hat{\mathcal{E}}^{(1)}_S} = \frac{\operatorname{sign}(A_1)}{\operatorname{sign}(A_2)} \,,
\quad
\frac{ \hat{\mathcal{E}}^{(2)}_N}{\hat{\mathcal{E}}^{(2)}_S} = 1 \,,
\quad
\frac{ \hat{\mathcal{E}}^{(3)}_N}{\hat{\mathcal{E}}^{(3)}_S} = \frac{\operatorname{sign}(A_2)}{\operatorname{sign}(A_1)} \,,
\quad
\frac{ \hat{\mathcal{E}}^{(4)}_N}{\hat{\mathcal{E}}^{(4)}_S} = 1 \,.
\end{equation}
Finally, by plugging the expressions for $\hat X^{(4)}_N$,$\hat X^{(4)}_S$ into equation \eqref{BHentropyapp} we obtain the following expression for the black hole entropy:
\begin{equation}\label{eq:SBHSymmPolExpr}
S_{\textrm{BH}} = \frac{\pi}{3n_N n_S (n_N+n_S) s} N^{3/2} \sqrt{\hat{P}^{(2)}+n_N n_S - s} \, \left( |A_1| + |A_2| \right) \,.
\end{equation}

\subsection{Proving the constraints are satisfied}

We are now ready to prove the various constraints are satisfied given \eqref{prconsandpsign}. 
We already noted that \eqref{eqn:xIconstraintapp} and \eqref{eqn:BPS_fluxapp} are satisfied.
To prove \eqref{eqn:STUchargesapp}-\eqref{eq:Couzens_BHapp} are satisfied, we start by noting that \eqref{newtmcl}
imply the following inequalities are satisfied by the elementary symmetric polynomials of the fluxes $p^I$:
\begin{align}\label{eq:PElemSymmPolIneq}
&\hat{P}^{(2)} \leq \frac{3}{8} p_R^2 \,,\nn
&\hat{P}^{(3)} \leq \frac{4 (\hat{P}^{(2)})^2}{9 p_R} \leq
\frac{1}{6} p_R \hat{P}^{(2)} \leq \frac{1}{16} p_R^3 \,,\nn
&
\hat{P}^{(4)} \leq \frac{3 (\hat{P}^{(3)})^2}{8 \hat{P}^{(2)}} \leq \frac{\hat{P}^{(3)}\hat{P}^{(2)}}{6 p_R}\leq \frac{1}{36} (\hat{P}^{(2)})^2 \,.
\end{align}
We next prove some useful bounds for $s$. Using the definition for $s$ in \eqref{STU_defn_s}
and the last line of \eqref{eq:PElemSymmPolIneq}, we have:
\begin{align}\label{eq:sLowerBound1}
s^2 &= (\hat{P}^{(2)} + n_N n_S)^2 - 4 \hat{P}^{(4)} \geq (\hat{P}^{(2)} + n_N n_S)^2 - \frac{1}{9} (\hat{P}^{(2)})^2 \nn
&= \frac{8}{9} (\hat{P}^{(2)})^2 + 2n_N n_S \hat{P}^{(2)} + n_N^2 n_S^2 > n_N^2 n_S^2  \,.
\end{align}
We see that $s^2>0$ and therefore $s$ is indeed real, and furthermore $ s > n_N n_S $. In fact, we can derive another useful lower bound for $s$ as follows:
\begin{align}\label{eq:sLowerBound2}
s &= (\hat{P}^{(2)} + n_N n_S) \sqrt{1- \frac{4 \hat{P}^{(4)}}{(\hat{P}^{(2)} + n_N n_S)^2}} \nn
&\geq 
(\hat{P}^{(2)} + n_N n_S) \left[1- \frac{4 \hat{P}^{(4)}}{(\hat{P}^{(2)} + n_N n_S)^2}\right] 
= \hat{P}^{(2)} + n_N n_S - \frac{4 \hat{P}^{(4)}}{\hat{P}^{(2)} + n_N n_S} \,,
\end{align}
where we used the fact that $\sqrt{1-x}\geq 1-x$ for $0<x<1$.
We also have the following upper bound for $s$:
\begin{equation}\label{eq:sUpperBound}
s^2 \leq (\hat{P}^{(2)} + n_N n_S)^2 \quad \Rightarrow \quad
s \leq \hat{P}^{(2)} + n_N n_S \leq \frac{3}{8} p_R^2 + n_N n_S \,,
\end{equation}
where we used the first line of \eqref{eq:PElemSymmPolIneq}.
We also note another useful inequality:
\begin{equation}
\frac{\hat{P}^{(2)}}{\hat{P}^{(2)} + n_N n_S} < \frac{p_R}{n_N + n_S} = \frac{n_N - n_S}{n_N + n_S} < 1 \,.
\end{equation}
To see this note that $\frac{\hat{P}^{(2)}}{\hat{P}^{(2)} + n_N n_S} $ is a monotonically increasing function of $\hat{P}^{(2)}$ 
and since $\hat{P}^{(2)} \leq \frac{3}{8} p_R^2$ from
\eqref{eq:PElemSymmPolIneq} we have
\begin{equation}\label{eq:PP2AuxBound}
\frac{n_N + n_S}{p_R} \frac{\hat{P}^{(2)}}{\hat{P}^{(2)} + n_N n_S} \leq \frac{n_N + n_S}{p_R} \frac{\frac{3}{8} p_R^2}{\frac{3}{8}p_R^2 + n_N n_S} = \frac{n_N^2 - n_S^2}{n_N^2 + n_S^2 + \frac{2}{3} n_N n_S} < 1 \,.
\end{equation}

We now proceed with the proof. We start by showing that the expressions $A_1, A_2$ are both positive. Using the second line \eqref{eq:PElemSymmPolIneq}, we have:
\begin{align}
A_1 &\geq p_R (\hat{P}^{(2)} + n_N n_S) - \frac{1}{3} p_R \hat{P}^{(2)} + (n_N+n_S) s \nn
&= p_R \left( \frac{2}{3} \hat{P}^{(2)} + n_N n_S \right) + (n_N+n_S) s > 0 \,.
\end{align} 
For $A_2$, we use the lower bound \eqref{eq:sLowerBound2} as well as the third line of \eqref{eq:PElemSymmPolIneq} to obtain:
\begin{align}
A_2 &\geq -(n_N-n_S)(\hat{P}^{(2)} + n_N n_S) + 2\hat{P}^{(3)} + (n_N + n_S) \Bigg[ \hat{P}^{(2)} + n_N n_S - \frac{4 \hat{P}^{(4)}}{\hat{P}^{(2)} + n_N n_S} \Bigg] \nn
&= 2n_S (\hat{P}^{(2)} + n_N n_S) + 2\hat{P}^{(3)} - \frac{4(n_N+n_S) \hat{P}^{(4)}}{\hat{P}^{(2)} + n_N n_S} \nn
&\geq 2n_S (\hat{P}^{(2)} + n_N n_S) + 2\hat{P}^{(3)} \left[ 1- \frac{1}{3} \frac{n_N+n_S}{p_R} \frac{\hat{P}^{(2)}}{\hat{P}^{(2)} + n_N n_S} \right] \nn
&> 2n_S (\hat{P}^{(2)} + n_N n_S) + \frac{4}{3} \hat{P}^{(3)} > 0 \,,
\end{align}
where in the last line we used \eqref{eq:PP2AuxBound}. 
Since $A_1, A_2 >0 $, equation \eqref{eq:ConsChargeSymmPolRatios} implies that $\hat{\mathcal{E}}^{(k)}_N = \hat{\mathcal{E}}^{(k)}_S \, (k=1,2,3,4)$ as required.

Next we turn to proving the constraints \eqref{eq:XXIneqConstraints} using the expressions \eqref{eq:XIElemSymmPol}. For $\hat{X}^{(1)}_N, \hat{X}^{(1)}_S $ we have:
\begin{align}
s \hat{X}^{(1)}_N &= -2s + n_S(n_N+n_S) \leq -2n_N n_S + n_S (n_N+n_S) = n_S(n_S-n_N) < 0 \,,\nn
s \hat{X}^{(1)}_S &= -2s + n_N(n_N+n_S) \geq 
%-\frac{3}{4}(n_N-n_S)^2 - 2n_N n_S + n_N(n_N+n_S) \nn
%&= 
\frac{1}{4}(n_N-n_S) \left(n_N + 3n_S \right) >0 \,, 
\end{align}
where we used the bounds \eqref{eq:sLowerBound1} and \eqref{eq:sUpperBound}. 
Next, for $\hat{X}^{(2)}_N, \hat{X}^{(2)}_S $ we have:
\begin{align}
2s^2 \hat{X}^{(2)}_N &= n_S(n_N+3n_S)(\hat{P}^{(2)}+n_N n_S) + 2(n_N + 2n_S) \hat{P}^{(3)} - 3n_S(n_N+n_S)s + 2s^2  \nn 
&\geq n_S(n_N+3n_S)(\hat{P}^{(2)}+n_N n_S) + 2(n_N + 2n_S) \hat{P}^{(3)} - 3n_S (n_N+n_S) (\hat{P}^{(2)}+n_N n_S) + 2s^2 \nn
&= 2(\hat{P}^{(2)}+n_N n_S)\hat{P}^{(2)} + 2(n_N + 2n_S) \hat{P}^{(3)} - 8\hat{P}^{(4)} \nn
&\geq 2(\hat{P}^{(2)}+n_N n_S)\hat{P}^{(2)} + 2(n_N + 2n_S) \hat{P}^{(3)} - \frac{8}{36} (\hat{P}^{(2)})^2 \nn
&= \frac{16}{9} (\hat{P}^{(2)})^2 + 2n_N n_S \hat{P}^{(2)} + 2(n_N + 2n_S) \hat{P}^{(3)} > 0 \,, 
\end{align}
and
\begin{align}
2s^2 \hat{X}^{(2)}_S &= n_N(3n_N+n_S)(\hat{P}^{(2)}+n_N n_S) -2(2n_N+n_S) \hat{P}^{(3)} - 3n_N(n_N+n_S)s + 2s^2 \nn
&\geq n_N(3n_N+n_S)(\hat{P}^{(2)}+n_N n_S) - 2(2n_N+n_S) \hat{P}^{(3)} - 3n_N(n_N+n_S)(\hat{P}^{(2)}+n_N n_S) +2s^2 \nn
&= 2 (\hat{P}^{(2)}+n_N n_S) \hat{P}^{(2)} - 2(2n_N+n_S) \hat{P}^{(3)} - 8 \hat{P}^{(4)} \nn
&\geq 2 (\hat{P}^{(2)}+n_N n_S) \hat{P}^{(2)} - 2(2n_N+n_S) \frac{4 (\hat{P}^{(2)})^2}{9 p_R} - \frac{8}{36} (\hat{P}^{(2)})^2 \nn
&= 2 (\hat{P}^{(2)}+n_N n_S) \hat{P}^{(2)} \left[ 1 - \frac{4}{9} \frac{2n_N+n_S}{p_R} \frac{\hat{P}^{(2)}}{\hat{P}^{(2)} + n_N n_S} - \frac{1}{9} \frac{\hat{P}^{(2)}}{\hat{P}^{(2)} + n_N n_S} \right] \nn
&\geq 2 (\hat{P}^{(2)}+n_N n_S) \hat{P}^{(2)} \left[ 1 - \frac{8}{9} \frac{n_N+n_S}{p_R} \frac{\hat{P}^{(2)}}{\hat{P}^{(2)} + n_N n_S} - \frac{1}{9} \frac{\hat{P}^{(2)}}{\hat{P}^{(2)} + n_N n_S} \right] \nn
&>  2 (\hat{P}^{(2)}+n_N n_S)\hat{P}^{(2)} \left[ 1- \frac{8}{9} - \frac{1}{9} \right] = 0 \,,
\end{align}
where in the above we used the bound \eqref{eq:sUpperBound}, the second and third lines of \eqref{eq:PElemSymmPolIneq} and the bound \eqref{eq:PP2AuxBound}. 
For the  $\hat{X}^{(3)}_N, \hat{X}^{(3)}_S $ constraints we have (recall that $A_1, A_2 > 0$):
\begin{align}
-\frac{4s^3 \hat{X}^{(3)}_N}{A_1} &=  2 \hat{P}^{(3)} + (n_N+3n_S)(\hat{P}^{(2)}+n_N n_S -s) > 0 \,,\nn
\frac{4s^3 \hat{X}^{(3)}_S}{A_2} &= 2\hat{P}^{(3)}-(3n_N+n_S)(\hat{P}^{(2)}+n_N n_S -s)  
%\nn
%&
\geq 2\hat{P}^{(3)} - (3n_N+n_S) \frac{4 \hat{P}^{(4)}}{\hat{P}^{(2)} + n_N n_S} \nn
&\geq 2\hat{P}^{(3)} - (3n_N+n_S) \frac{4}{6p_R} \frac{\hat{P}^{(3)}\hat{P}^{(2)}}{\hat{P}^{(2)} + n_N n_S} 
%\nn
%&
= 2\hat{P}^{(3)} \left[ 1 - \frac{1}{3} \frac{3n_N+n_S}{p_R} \frac{\hat{P}^{(2)}}{\hat{P}^{(2)} + n_N n_S} \right] \nn
&\geq 2\hat{P}^{(3)} \left[ 1 - \frac{n_N+n_S}{p_R} \frac{\hat{P}^{(2)}}{\hat{P}^{(2)} + n_N n_S} \right] > 0 \,,
\end{align}
where we used the bounds \eqref{eq:sLowerBound2} and \eqref{eq:sUpperBound}, the last line of \eqref{eq:PElemSymmPolIneq} and the bound \eqref{eq:PP2AuxBound}. 
Finally, for $\hat{X}^{(4)}_N, \hat{X}^{(4)}_S $ we have:
\begin{align}
8s^4 \hat{X}^{(4)}_N &= A_1^2 (\hat{P}^{(2)}+n_N n_S -s) > 0 \,,\nn
8s^4 \hat{X}^{(4)}_S &= A_2^2 (\hat{P}^{(2)}+n_N n_S -s) > 0 \,,
\end{align}
where we again used the upper bound \eqref{eq:sUpperBound}.
We have thus proven \eqref{eq:XXIneqConstraints} and hence $ \left.(-1)^{t+1} x^I\right\vert_{N,S} > 0 $ in \eqref{STUexistcond}.

To conclude, we consider the black hole entropy expression \eqref{eq:SBHSymmPolExpr}. Since $A_1, A_2>0$, we can replace $|A_1|$ and $|A_2|$ by $A_1$ and $A_2$ respectively, and we obtain:
\begin{align}
S_{\textrm{BH}} &= \frac{\pi}{3n_N n_S (n_N+n_S) s} N^{3/2} \sqrt{\hat{P}^{(2)}+n_N n_S - s} \, \left( A_1 + A_2 \right) \nn
& =  
\frac{2\pi}{3n_N n_S}N^{3/2} \sqrt{\hat{P}^{(2)}+n_N n_S - s}
\,.
\end{align}
Thus, \eqref{eq:Couzens_BHapp} is established. It is also clear from \eqref{eq:sUpperBound} that this expression is indeed real and positive
and so the conditions in \eqref{STUexistcondapp} are all satisfied.

%%%%%%%%%%%%%%%
\section{Numerical hyperscalar solutions}\label{plotssols}

We have constructed various examples of $AdS_2\times \Sigma$ hyperscalar solutions by numerically solving the BPS equations. We have found that these only
exist in the anti-twist class. The system of BPS equations that we need to solve is given in 
section \ref{sec5}.
Instead of working in conformal gauge, we work in a gauge with $f=1$. The 
metric function $h$ is given by \eqref{eqn:BPSint} and the gauge fields are determined from \eqref{eq:fieldstrength}.
Our method for integrating the BPS equations is exactly analogous to that of appendix F of \cite{Arav:2025jee}.
In particular, we can fix all of the boundary conditions at the poles, apart from $\rho$. 
We then make a search over possible boundary conditions for $\rho$, as in \eqref{hsexppoles}, at one of the poles,
which after integrating the BPS equations then leads to a spindle solution with the correct boundary condition \eqref{hsexppoles} at the other pole.

\begin{figure}[h!]
	\centering
	\includegraphics[scale=0.54]{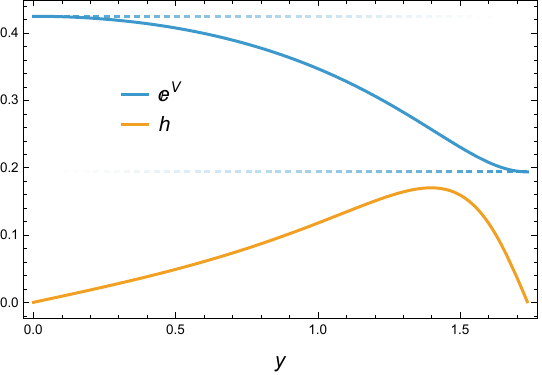}~
	\includegraphics[scale=0.54]{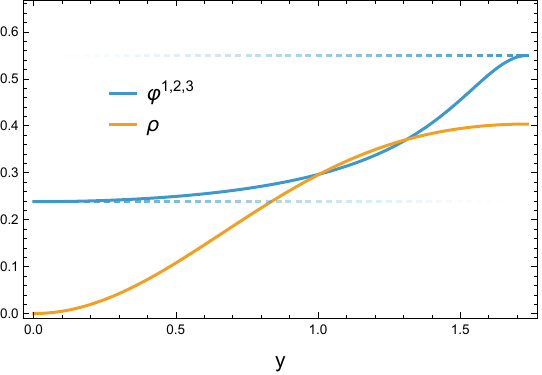}~
	\includegraphics[scale=0.54]{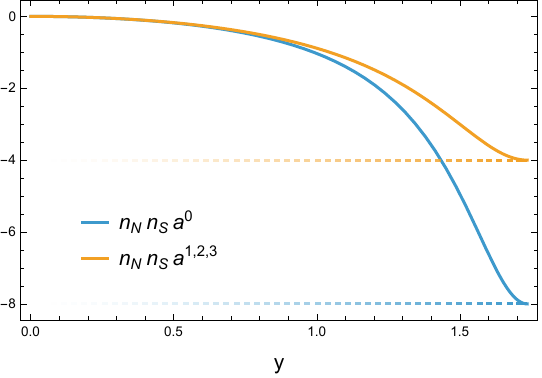}
	\caption{Metric, scalar functions and gauge fields for the solution with $(n_N, n_S)=(11,1)$ in table \ref{tablecoprimepbnonzero}.}
	\label{solnexample1}
	\end{figure}

\begin{figure}[h!]	
 \centering
		\includegraphics[scale=0.54]{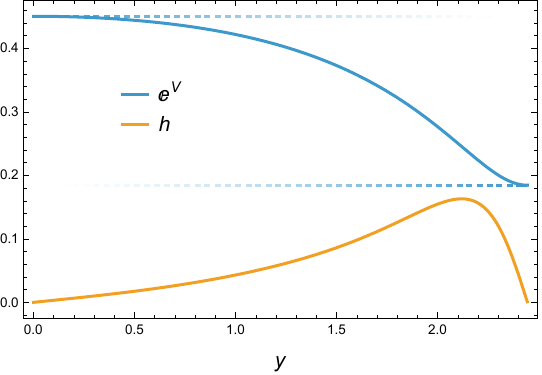}~
	\includegraphics[scale=0.54]{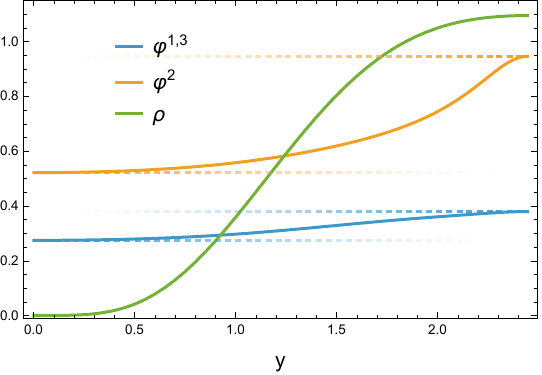}~
	\includegraphics[scale=0.54]{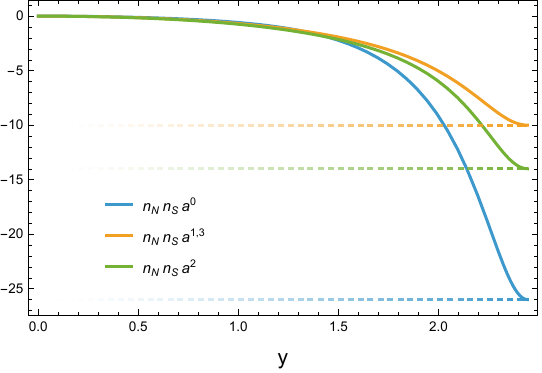}
	\caption{Metric, scalar functions and gauge fields for the solution with $(n_N, n_S)=(1,31)$ in table \ref{tablecoprimepbnonzero}.}
	\label{solnexample2}
\end{figure}

We have verified that hyperscalar solutions exist for many of the examples in 
tables \ref{pbzerotablecoprime}-\ref{noncoprime_pBneq0} (and more). Here we illustrate by plotting the metric functions
$e^V,h$, the scalars $\varphi_1,\varphi_2,\varphi_3$ and $\rho$, as well as the four gauge fields $a^I$, 
for two representative examples in figures \ref{solnexample1} and \ref{solnexample2}: they are both anti-twist solutions
as in \eqref{signchoicehyper}. The $N$ pole is taken to be located at $y=0$ and the location of the
$S$ pole is determined numerically. 
The behaviour of the function $h$ is consistent with the values of $(n_N,n_S)$ (recall \eqref{eqn:BPSint},\eqref{eqn:AdS2ansatz}, \eqref{eqn:sincosBC}). 
The dashed lines in the plots associated 
with the metric function $e^V$ as well as the scalars $\varphi_1,\varphi_2,\varphi_3$ are the boundary values
determined algebraically in studying the BPS equations, without having a solution. The gauge fields are plotted in a gauge
where $a^I=0$ at the $N$ pole; this then allows us to easily compare the behaviour
at the $S$ pole with the fluxes $p^I$, as indicated by dashed lines in the plots
(recall from \eqref{eqn:BPS_flux} that $p^I=\frac{1}{2} n_N n_S(a^I_S-a^I_N)=\frac{1}{2} n_N n_Sa^I_S$).
One can easily move to the gauge used in the text when discussing smooth
$S^1$ orbibundles, by suitably adding flat connections at the poles, parametrised by the
discrete orbibundle data $m_{N,S}^I$ and then gluing
the gauge fields at the equator of the spindle with a $U(1)$ gauge transformation
(recall \eqref{eq:circleconnection}, \eqref{eq:circlegdg}).

The case $(n_N,n_S)=(11,1)$ with $p^I=-(4,2,2,2)$ of table \ref{tablecoprimepbnonzero}
is presented in figure \ref{solnexample1}. 
This hyperscalar $AdS_2$ solution has $r_N=2$ and $r_S=0$; indeed we find
that $\rho$ vanishes at the $N$ pole quadratically, with $\rho \approx 0.496 y^2$. We find that the non-zero value
for $\rho$ at the $S$ pole is given by $\rho_S\approx 0.4034$. The south pole is located at $y_S = 1.736$.

The case $(n_N,n_S)=(31,1)$ with $p^I=-(13,5,7,5)$ of table \ref{tablecoprimepbnonzero}
is presented in figure \ref{solnexample2}.
This hyperscalar $AdS_2$ solution has $r_N=4$ and $r_S=0$; 
indeed we find that $\rho$ vanishes at the $N$ pole as $ \rho \approx 0.836 y^4$. We find that the non-zero value
for $\rho$ at the $S$ pole is given by $\rho_S\approx 1.095$. The south pole is located at $y_S = 2.444$.

%\bibliographystyle{utphys} 
%\bibliography{draft}{}

\providecommand{\href}[2]{#2}\begingroup\raggedright\endgroup

\end{document}